\documentclass[twocolumn]{aastex63}
\usepackage{amsmath}

\received{June 1, 2019}
\revised{January 10, 2019}
\accepted{\today}
\submitjournal{ApJ}

\shorttitle{Depletion of bright red giants}
\shortauthors{Zajacek et al.}
\graphicspath{{./}{figures/}}

\begin{document}

\title{Depletion of bright red giants in the Galactic center during its active phases}

\correspondingauthor{Michal Zaja\v{c}ek}
\email{zajacek@cft.edu.pl}

\author[0000-0001-6450-1187]{Michal Zaja\v{c}ek}
\affiliation{Center for Theoretical Physics, Polish Academy of Sciences, Al. Lotnikow 32/46, ´
02-668 Warsaw, Poland}

\author[0000-0001-7605-5786]{Anabella Araudo}
\affiliation{ELI Beamlines, Institute of Physics, Czech Academy of Sciences, CZ-25241 Doln\'i B\v{r}e\v{z}any, Czech Republic}
\affiliation{Astronomical Institute of the Czech Academy of Sciences,
Bo\v{c}n\'i II 1401, CZ-14100 Prague, Czech Republic}

\author[0000-0002-5760-0459]{Vladim\'ir Karas}
\affiliation{Astronomical Institute of the Czech Academy of Sciences,
Bo\v{c}n\'i II 1401, CZ-14100 Prague, Czech Republic}

\author[0000-0001-5848-4333]{Bo\.{z}ena Czerny}
\affiliation{Center for Theoretical Physics, Polish Academy of Sciences, Al.\ Lotnikow 32/46, ´
02-668 Warsaw, Poland}

\author[0000-0001-6049-3132]{Andreas Eckart}
\affiliation{I. Physikalisches Institut der Universit\"at zu K\"oln, Z\"ulpicher Strasse 77, D-50937 K\"oln, Germany}
\affiliation{Max-Planck-Institut f\"ur Radioastronomie (MPIfR), Auf dem H\"ugel 69, D-53121 Bonn, Germany}



\begin{abstract}

Observations in the near-infrared domain showed the presence of the flat core of bright late-type stars inside $\sim 0.5\,{\rm pc}$ from the Galactic center supermassive black hole (Sgr~A*), while young massive OB/Wolf-Rayet stars form a cusp. Several dynamical processes were proposed to explain this apparent paradox of the distribution of the Galactic center stellar populations. Given the mounting evidence about a significantly increased activity of Sgr A* during the past million years, we propose a scenario based on the interaction between the late-type giants and a nuclear jet, whose past existence and energetics can be inferred from the presence of $\gamma$-ray Fermi bubbles and bipolar radio bubbles. Extended, loose envelopes of red giant stars can be ablated by the jet with kinetic luminosity in the range of $L_{\rm j}\approx 10^{41}$--$10^{44}\,{\rm erg\,s^{-1}}$ within the inner $\sim 0.04\,{\rm pc}$ of Sgr~A* (S cluster region), which would lead to their infrared luminosity decrease after several thousand jet-star interactions. The ablation of the atmospheres of red giants is complemented by the process of tidal stripping that operates at distances of $\lesssim 1\,{\rm mpc}$, and by the direct mechanical interaction of stars with a clumpy disc at $\gtrsim 0.04\,{\rm pc}$, which can explain the flat density profile of bright late-type stars inside the inner half parsec from Sgr~A*.

\end{abstract}



\section{Introduction} \label{sec_intro}

The Galactic center supermassive black hole (hereafter SMBH) with the mass of $4.1\times 10^6\,M_{\odot}$ is located at the distance of $8.1\,{\rm kpc}$ \citep{2016ApJ...830...17B,2017ApJ...837...30G,2017ApJ...845...22P,2018A&A...615L..15G} and provides a unique laboratory for studying detailed dynamical processes and the mutual interaction between the nuclear star cluster (NSC) and a central massive black hole \citep{2005PhR...419...65A,2010RvMP...82.3121G,2017FoPh...47..553E,2020ApJ...896..100A} as well as with the multiphase gaseous-dusty circumnuclear medium \citep{1996ARA&A..34..645M,2017MNRAS.464.2090R}. The compact radio source Sgr A* associated with the SMBH is embedded in the Milky Way NSC, which is one of the densest stellar systems in the Galaxy \citep{2014CQGra..31x4007S}, and in addition, it is surrounded by an ionized, neutral, and molecular gas and dust \citep[see e.g.][and references therein]{2017A&A...603A..68M}.

The NSC consists of both late-type (red giants, supergiants and asymptotic giant branch stars) and early-type stars of O and B spectral classes \citep{1991ApJ...382L..19K,2009ApJ...703.1323D,2009A&A...499..483B,2018A&A...609A..26G}, which implies star-formation during the whole Galactic history, albeit most likely episodic \citep{2011ApJ...741..108P} with the star-formation peak at 10 Gyr, the minimum at 1-2 Gyr and a recent increase in the last few hundred million years.

In the innermost parsec of the Galactic center, there is a surprising abundance of young massive OB/Wolf-Rayet stars formed in-situ in the last 10 million years \citep{2003ApJ...586L.127G}. These young stars form an unrelaxed cusp-like distribution. On the other hand, former studies of the distribution of late-type stars showed that they exhibit a core-like distribution inside the inner $\sim 0.5\,{\rm pc}$, which has a projected flat or even decreasing profile towards the center \citep{2009ApJ...703.1323D,2009A&A...499..483B}. More recent studies provided a precise analysis of the distribution of late-type stars because of increasing their sensitivity towards larger magnitudes, i.e. fainter giants \citep{2018A&A...609A..26G,2019ApJ...872L..15H}. Using photometric number counts and diffuse light analysis, \citet{2018A&A...609A..26G} found that fainter late-type stars with magnitudes of $K\approx 18$ exhibit a cusp-like distribution within the sphere of influence of Sgr~A* with a 3D power-law exponent of $\gamma \simeq 1.43$. In comparison, there is an apparent lack of bright red giants with $K=12.5-16$ at the projected radii of $\lesssim 0.3\,{\rm pc}$ from Sgr~A*. \citet{2018A&A...609A..26G} estimate the number of missing giants to 100 for this distance range. The study of \citet{2019ApJ...872L..15H} also finds a cusp-like distribution for late-type stars with $K<17$ within $0.02-0.4\,{\rm pc}$. In agreement with \citet{2018A&A...609A..26G}, they found a core-like distribution for the brightest giants with $K<15.5$, although the number of missing giants appears to be lower than 100 according to their analysis. Although the surface brightness distribution of late-type stars
brighter than 15.5 magnitudes appears to be rather flat, already in Fig.~11 of \citet{2009A&A...499..483B} the inner point at $0.5''$ (where $1''\sim 0.04\,{\rm pc}$ at the Galactic center) of the distribution of the late-type stars as well as of the distribution of all the stars indicates the presence of a cusp.

In summary, there appears to be an internal mechanism within the NSC that preferentially depleted bright, large red giants on one hand, which has led to their apparent core-like distribution, and at the same time has been less efficient for early-type as well as fainter late-type stars on the other hand. Such a mechanism has altered either the spatial, luminosity, or the temperature distribution of the bright red-giant stars so that they effectively fall beyond the detection limit or they rather mimic younger, ``bluer'' stars. So far, mainly the following four mechanisms have been discussed to explain the apparent lack of bright red giants,
\begin{itemize}
    \item complete or partial tidal disruption of red giants by the SMBH \citep{1975Natur.254..295H,2014ApJ...788...99B,2020MNRAS.493L.120K} (envelope removal),
    \item envelope stripping by the collisions of red giants with the dense clumps within a self-gravitating accretion disc \citep{1996ApJ...470..237A,2014ApJ...781L..18A,2016ApJ...823..155K,2019arXiv191004774A} (envelope removal),
    \item collisions of red giants with field stars and compact remnants \citep{1989IAUS..136..543P,1993ApJ...408..496M,1996ApJ...472..153G,1999MNRAS.308..257B,2005PhR...419...65A,2005ApJ...624L..25D,2009MNRAS.393.1016D} (envelope removal),
    \item mass segregation effects: the dynamical effect of a secondary massive black hole \citep{2006MNRAS.372..174B,2006ApJ...648..890M,2006ApJ...641..319P,2007ApJ...656..879M,2008MNRAS.384..323L,2012ApJ...744...74G} or of an infalling massive cluster \citep{2003ApJ...597..312K,2009MNRAS.399..141E,2012ApJ...750..111A} or of stellar black holes \citep{1993ApJ...408..496M} (altered spatial distribution),
\end{itemize}
where in the parentheses we include the mechanism responsible for altering the population of late-type stars.
The importance of star--star and star--disc interactions was also analyzed generally for active galactic nuclei in terms of the effects on the accretion disk and Broad Line region structure as well as the NSC orbital distribution \citep{1994ApJ...434...46Z,1996ApJ...470..237A,2001A&A...376..686K,2002A&A...387..804V,2019arXiv190909645M,2016ApJ...823..155K}.

\begin{figure}[tbh]
    \centering
    \includegraphics[width=\columnwidth]{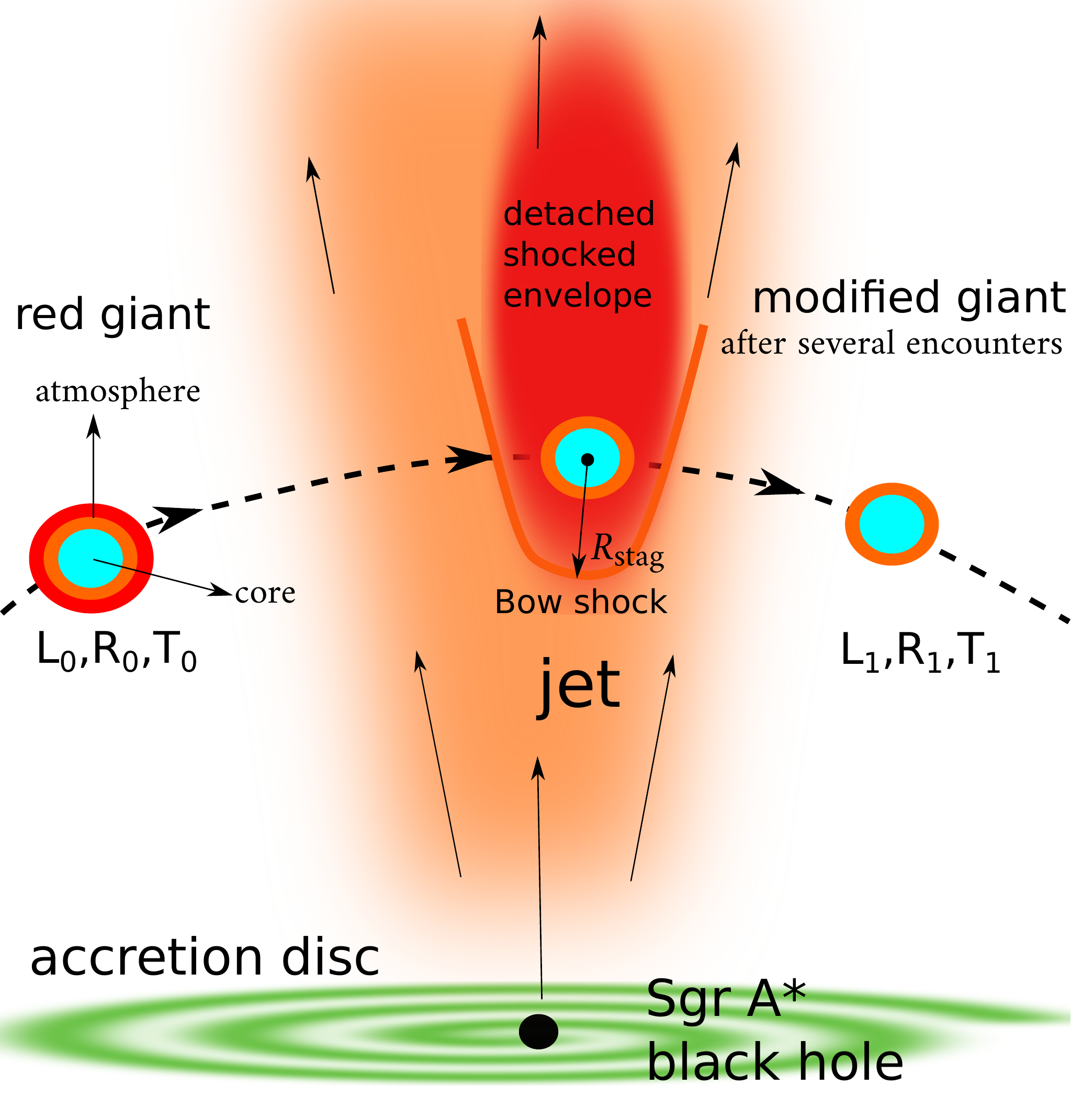}
    \caption{Illustration of the jet-red giant interaction in the vicinity of Sgr~A* during its active phase. The large, loosely bound envelope of the red giant (coloured as red and orange), which has an initial luminosity, radius, and temperature ($L_0, R_0, T_0$), is ablated by the jet ram pressure during several encounters since the lifetime of the jet $t_{\rm jet}\sim 0.5\,{\rm Myr}$ is much longer than the orbital timescale in the inner $\sim 0.5\,{\rm pc}$, $P_{\rm orb}\sim 1500(a/0.1\,{\rm pc})^{3/2}\,{\rm yr}$. After a few hundred encounters, star has modified parameters ($L_1, R_1, T_1$), which change the overall outlook of the giant in the near-infrared domain. Inspired by \citet{2012ApJ...749..119B} and \citet{2012A&A...539A..69B}.}
    \label{fig_jet_star}
\end{figure}

We propose here another mechanism based on the jet-star interactions \citep{2012ApJ...749..119B,2013MNRAS.436.3626A,2017bhns.work....1A}, which most likely coexisted with the mechanisms proposed above. In particular, the star--accretion disc collisions are expected to be accompanied by star--jet crossings during previous active phases of Sgr~A*. Since red-giant stars have typically large, loosely bound tenuous envelopes, dense compact cores, and slow winds with the terminal velocity $\lesssim 100\,{\rm km\,s^{-1}}$, they are in particular susceptible to mass removal in encounters with higher-pressure material \citep{2012ApJ...757..134M,2014ApJ...781L..18A,2016ApJ...823..155K}.\footnote{The estimate of the terminal wind velocity is given by the escape velocity 
$v_{\rm esc} = 62\,\left(\frac{M_{\star}}{1\,M_{\odot}}\right)^{1/2}\left(\frac{R_{\star}}{100\,R_{\odot}} \right)^{-1/2}\,{\rm km\,s^{-1}}$.} Therefore during the red giant--jet interactions, the jet ram pressure will remove the outer layers of the stellar envelope during the passage. We illustrate this 
idea in Fig.~\ref{fig_jet_star}. 

After several star--jet crossings, the atmosphere is removed similarly as for repetitive star--disc crossings \citep{2019arXiv191004774A} and the giant is modified in a way that it follows an evolutionary track in the Hertzsprung-Russell (HR) diagram approximately along the constant absolute magnitude towards higher effective temperatures. We show that this mechanism quite likely operated in the vicinity of Sgr~A* during its active Seyfert-like phase in the past few million years \citep{2019ApJ...886...45B} when the jet kinetic luminosity could have reached $\sim 10^{44}\,{\rm erg\,s^{-1}}$. In principle, even in the quiescent state, a tidal disruption event every $\sim 10^4$ years, which can be estimated for the Galactic center \citep{1999MNRAS.306...35S,2005PhR...419...65A}, can temporarily reactivate the jet of Sgr~A* and some of the bright red giants could be depleted during its existence. This makes the red giant--jet interaction in the Galactic center relevant and highly plausible in its recent history and the dynamical consequences can be inferred based on the so-far detected traces of the past active phase of Sgr~A* as well as the currently observed stellar density distribution. 

Previous jet-star interaction studies were focused on the emergent non-thermal radiation, in particular in the gamma-ray domain, and mass-loading and chemical enrichment of jets by stellar winds \citep[see e.g.][]{1994MNRAS.269..394K,2012ApJ...749..119B,2012A&A...539A..69B,2013MNRAS.436.3626A,2015ApJ...807..168B,2016A&A...591A..15D}. Here we focus on the effect of the jet on the stellar population. In most of the jetted active galactic nuclei, this is perhaps a secondary problem since stellar populations in the host bulge cannot be resolved out, i.e. one can only analyze the integrated starlight. In contrast, within the Galactic center NSC, one can not only disentangle late- and early-type stars, but it is also possible to study their distribution as well as kinematics of individual stars. Although in the current low-luminosity state there is not firm evidence for the presence of a relativistic jet, there are nowadays several multiwavelength signatures of the past active Seyfert-like state of Sgr~A* that occurred a few million years ago \citep{2019ApJ...886...45B,2019Natur.573..235H,2019Natur.567..347P}. However, even in the current quiescent state of Sgr~A*, studies by \citet{2012ApJ...758L..11Y} and \citet{2013ApJ...779..154L} indicate the existence of a low-surface-brightness pc-scale jet. In addition, the presence of the cometary-shaped infrared-excess bow-shock sources X3, X7 \citep{2010A&A...521A..13M} and recently X8 \citep{2019A&A...624A..97P} indicates that the star--outflow interaction is ongoing even in
a very low state of the Sgr~A* activity.
The morphology of these sources can be explained by the interaction with a strong accretion wind originating from Sgr~A* or with the collective wind of the cluster of young stars.

This paper proposes a new mechanism that could have affected the current population of bright late-type stars, namely their appearance as well as number counts in specific magnitude bins, in the Galactic center. We apply analytical and semi-analytical calculations to assess whether the potential past jet--star interactions could have had an effect on the stellar population in the sphere of influence of Sgr~A*. Although the analytical calculations introduce several simplifications, we show that the mechanism could have operated and the estimated number of affected red giants is in accordance with the up-to-date most sensitive studies \citep{2018A&A...609A..26G,2019ApJ...872L..15H}. A more detailed computational treatment including magnetohydrodynamic numerical simulations as well as a stellar evolution will be presented in our future studies.

The paper is structured as follows. In Section~\ref{sec_stagnation_radius}, we derive the stagnation radius, basic timescales, and the removed envelope mass for red-giant stars interacting with the jet of Sgr~A* during its past active phase. In Section~\ref{sec_NIR_domain}, we discuss the observational signatures in the near-infrared domain. Subsequently, in Section~\ref{sec_fraction_redgiant}, we estimate the number of red giants that could be affected by the jet interaction and visually depleted from the immediate vicinity of Sgr~A*. In Section~\ref{sec_discussion}, we discuss additional processes related to the red giant--jet interaction in the Galactic center. Finally, we summarize the main results and conclude with Section~\ref{sec_summary}.

\section{Derivation of jet-star stagnation radius and the jet-induced stellar mass-loss}
\label{sec_stagnation_radius}

The evidence for the active phase of Sgr~A* that is estimated to have occurred $4\pm 1\,{\rm Myr}$ ago is based on the X-ray/$\gamma$-ray bubbles with total energy content of $10^{56}-10^{57}\,{\rm erg}$ \citep{2019ApJ...886...45B}. The first evidence for the nuclear outburst was the kpc-scale 1.5 keV ROSAT X-ray emission that originated in the Galactic center \citep{2003ApJ...582..246B}. The X-ray structure coincides well with the more recently discovered Fermi $\gamma$-ray bubbles extending $50^{\circ}$ north and south of the Galactic plane at 1-100 GeV \citep[][]{2010ApJ...724.1044S}. The X-ray/$\gamma$-ray bubbles are energetically consistent with the nuclear AGN-like activity associated with the jet and/or disc-wind outflows with jet power $L_{\rm j}=2.3^{+5.1}_{-0.9}\times 10^{42}\,{\rm erg\,s^{-1}}$ and age $4.3^{+0.8}_{-1.4}\,{\rm Myr}$ \citep{2016ApJ...829....9M}. In comparison, the starburst origin of the Fermi bubbles is inconsistent with the bubble energetics by a factor of $\sim 100$ \citep{2003ApJ...582..246B}. On intermediate scales of hundreds of parsecs, the base of the Fermi bubbles coincides with the bipolar radio bubbles \citep{2019Natur.573..235H} as well as with the X-ray chimneys \citep{2019Natur.567..347P}.

Using hydrodynamic simulations, \citet{2012ApJ...756..181G} reproduce the basic radiative characteristics of the Fermi bubbles with the AGN jet duration of $\sim 0.1-0.5$ Myr, which corresponds to the jet luminosity $L_{\rm j} \approx 10^{56-57}\,{\rm erg}/(0.1-0.5\,{\rm Myr}) = 6.3 \times 10^{42}-3.2\times 10^{44}\,{\rm erg\,s^{-1}}$. The jet is dominated by the kinetic luminosity
$L_{\rm j}=\eta_{\rm j} L_{\rm acc}$, where $\eta_{\rm j}$ is the conversion efficiency from the accretion luminosity $L_{\rm acc}$ to the jet kinetic luminosity. The accretion luminosity is $L_{\rm acc} \lesssim L_{\rm Edd}$, where the Eddington luminosity is
%
\begin{equation}
   L_{\rm Edd}=5.03\times 10^{44}\left(\frac{M_{\bullet}}{4\times 10^6\,M_{\odot}} \right)\,{\rm erg\,s^{-1}}
   \label{eq_Ledd}
\end{equation}
and $\eta_{\rm j}<0.7$ for most radio sources  \citep{2008ApJ...685..828I}. This yields the maximum jet kinetic luminosity for Sgr~A* of $L_{\rm j}\approx 3.5 \times 10^{44}\,{\rm erg\,s^{-1}}$. 
We will consider $L_{\rm j}\approx 10^{41}-10^{44}\,{\rm erg\,s^{-1}}$, where the lower limit is given by the putative jet present in the current quiescent state \citep{2012ApJ...758L..11Y}, with the inferred kinetic luminosity of $L_{\rm min}\sim 1.2 \times 10^{41}\,{\rm erg\,s^{-1}}$, and the upper limit is given by the Eddington luminosity.  

We assume a conical jet with a half-opening angle  $\theta$ and  width  $R_{\rm j}=z\tan{\theta}$, where $z$ is the distance to Sgr~A*. The jet footpoint for Sgr~A* can be estimated to be located at $z_0\sim 2 \times 10^{-5} (M_{\bullet}/4\times 10^6\,M_{\odot})\,{\rm pc}=52 R_{\rm Schw}$ \citep{1999Natur.401..891J}, where $R_{\rm Schw}=2GM_{\bullet}/c^2=3.8\times 10^{-7}(M_{\bullet}/4\times 10^6\,M_{\odot})\,{\rm pc}$ is the Schwarzschild radius. Any red giant or supergiant with radius\footnote{The stellar radius is the sum of the core radius and the envelope radius, $R_{\star}=R_{\rm c}+R_{\rm env}$.} $R_{\star}$ and 
mass $m_{\star}$ is not expected to plunge below $z_0$
since the tidal disruption radius $r_{\rm t}=R_{\star}(2M_{\bullet}/m_{\star})^{1/3}$ \citep{1975Natur.254..295H,1988Natur.333..523R} is at least a factor of two larger,
\begin{equation}
  \frac{r_{\rm t}}{R_{\rm Schw}}=1\,20\,\left(\frac{R_{\star}}{10\,R_{\odot}}\right)\left(\frac{M_{\bullet}}{4\times 10^6\,M_{\odot}} \right)^\frac{1}{3} \left(\frac{m_{\star}}{M_{\odot}} \right)^{-\frac{1}{3}}.
  \label{eq_tidal_radius}
\end{equation} 
$R_{\star}=10\,R_{\odot}$ and  $m_{\star}=1\,M_{\odot}$ are typical intermediate values for the evolved late-type giants with extended envelopes \citep{2013degn.book.....M}. For numerical estimates, we consider the range of radii for red giants and supergiants, $R_{\star}\sim 4-1000\,R_{\odot}$, as indicated by the Hertzsprung-Russell diagram. These late-type stars have the large range of bolometric luminosities, $L_{\star}\sim 10-72000\,L_{\odot}$, and the temperature range of $T_{\star}\sim 5000-3000\,{\rm K}$, which corresponds to the spectral classes K and M, respectively. The K-band magnitude range is $K\sim 15.2$ mag for $R_{\star}=4\,R_{\odot}$, $T_{\star}=5000\,{\rm K}$ and $K\sim 4.4$ mag for $R_{\star}=1000\,R_{\star}$, $T_{\star}=3000\,{\rm K}$. More specifically, we focus on the late-type stars of $K=16\,{\rm mag}$ and brighter, which appear to form a core-like density distribution in the central $0.5$ pc. Using the isochrones obtained with the Parsec code \citep{2012MNRAS.427..127B}, these stars have $R_{\star}\sim 4\,R_{\odot}$ and larger for an age of 5 Gyr. The late-type stars that are completely absent in the S-cluster (inner $\sim 0.04\,{\rm pc}$) were inferred to have $R_{\star}=30\,R_{\odot}$ and larger \citep{2019ApJ...872L..15H}. Therefore, numerical estimates are typically scaled to $R_{\star}=30\,R_{\odot}$ unless otherwise indicated.

We will further focus on the region between the tidal radius of red giants and the outer edge of the S cluster, which approximately corresponds to the Bondi radius of the hot bremsstrahlung plasma \citep{2003ApJ...591..891B,2013Sci...341..981W},

\begin{equation}
   R_{\rm B}=\frac{GM_{\bullet}}{c_{\rm s}^2}=0.12\left(\frac{M_{\bullet}}{4\times 10^6\,M_{\odot}} \right)\left(\frac{T_{\rm p}}{10^7\,{\rm K}}\right)^{-1}\,{\rm pc}\,,
\end{equation}
which is two-three orders of magnitude larger than the tidal radius of red giants, $R_{\rm B}\sim 3.1 \times 10^5\,R_{\rm Schw}$. More generally speaking, the population of predominantly B-type stars lies within the innermost arcsecond ($\sim 0.04\,{\rm pc}$, S cluster), while the population of young massive OB/Wolf-Rayet stars stretches from $\sim 0.04\,{\rm pc}$ up to $\sim 0.5\,{\rm pc}$, a fraction of which forms a warped stellar disc \citep{2010RvMP...82.3121G}.   


The stellar wind ram pressure at a distance $r$ from the center of the star can be estimated as $P_{\rm sw}\sim \rho_{\rm w} v_{\rm w}^2=\dot{m}_{\rm w}v_{\rm w}/(4\pi r^2)$, where $\dot{m}_{\rm w}$ is the mass-loss rate and $v_{\rm w}$ is the terminal wind velocity. Using typical values for red giants with $\dot{m}_{\rm w}\approx 10^{-8}\,M_{\odot}\,{\rm yr^{-1}}$ and  $v_{\rm w}\approx 10\,{\rm km\,s^{-1}}$
\cite[e.g.][]{1987IAUS..122..307R,2016A&A...591A..15D}\footnote{For red giants, $\dot{m}_{\rm w}\approx 10^{-6}-10^{-9}\,{\rm M_{\odot}\,yr^{-1}}$  according to \citet{1987IAUS..122..307R}.}
%
\begin{align}
    P_{\rm sw}&=0.012\left(\frac{\dot{m}_{\rm w}}{10^{-8}\,M_{\odot}\,{\rm yr^{-1}}} \right)\left(\frac{v_{\rm w}}{10\,{\rm km\,s^{-1}}}\right) \times \notag\\
    & \times \left(\frac{r}{30\,R_{\odot}}\right)^{-2}\,{\rm erg\,cm^{-3}}\,.\label{eq_pressure_sw}
\end{align}
%
The  ram pressure 
of a relativistic jet with bulk motion Lorentz factor $\Gamma$ is $P_{\rm j}=\Gamma \rho_{\rm j} v_{\rm j}^2$, where 
the jet density is $\rho_{\rm j}=L_{\rm j}/[(\Gamma-1)c^2 \sigma_{\rm j} v_{\rm j}]$
and $\sigma_{\rm j}=\pi R_{\rm j}^2$ is the jet cross-sectional area. 
By assuming $v_{\rm j}\sim c$ and $\Gamma \sim 10$ 
the jet kinetic pressure for Sgr~A* can be written as
\begin{align}
    P_{\rm j} &\approx \frac{L_{\rm j}}{\sigma_{\rm j}c} = 
    \frac{L_{\rm j}}{\pi c z^2 \tan^2{\theta}}\notag\\
             &=0.014\left(\frac{L_{\rm j}}{10^{42}\,{\rm erg\,s^{-1}}} \right)\left(\frac{z}{0.04\,{\rm pc}} \right)^{-2}\,{\rm erg\,cm^{-3}}\,,\label{eq_jet_pressure}
\end{align}
where we have assumed
$\theta\simeq 12.5^{\circ}$ that corresponds to the jet sheath half-opening angle estimated for the current candidate jet of Sgr~A* \citep{2013ApJ...779..154L}.
Note that the innermost arcsecond ($z\sim 0.04$~pc) is also the outer radius of fast-moving  stars in the S cluster \citep{2010RvMP...82.3121G,2017FoPh...47..553E}.

\begin{figure}[tbh]
    \centering
    \includegraphics[width=\columnwidth]{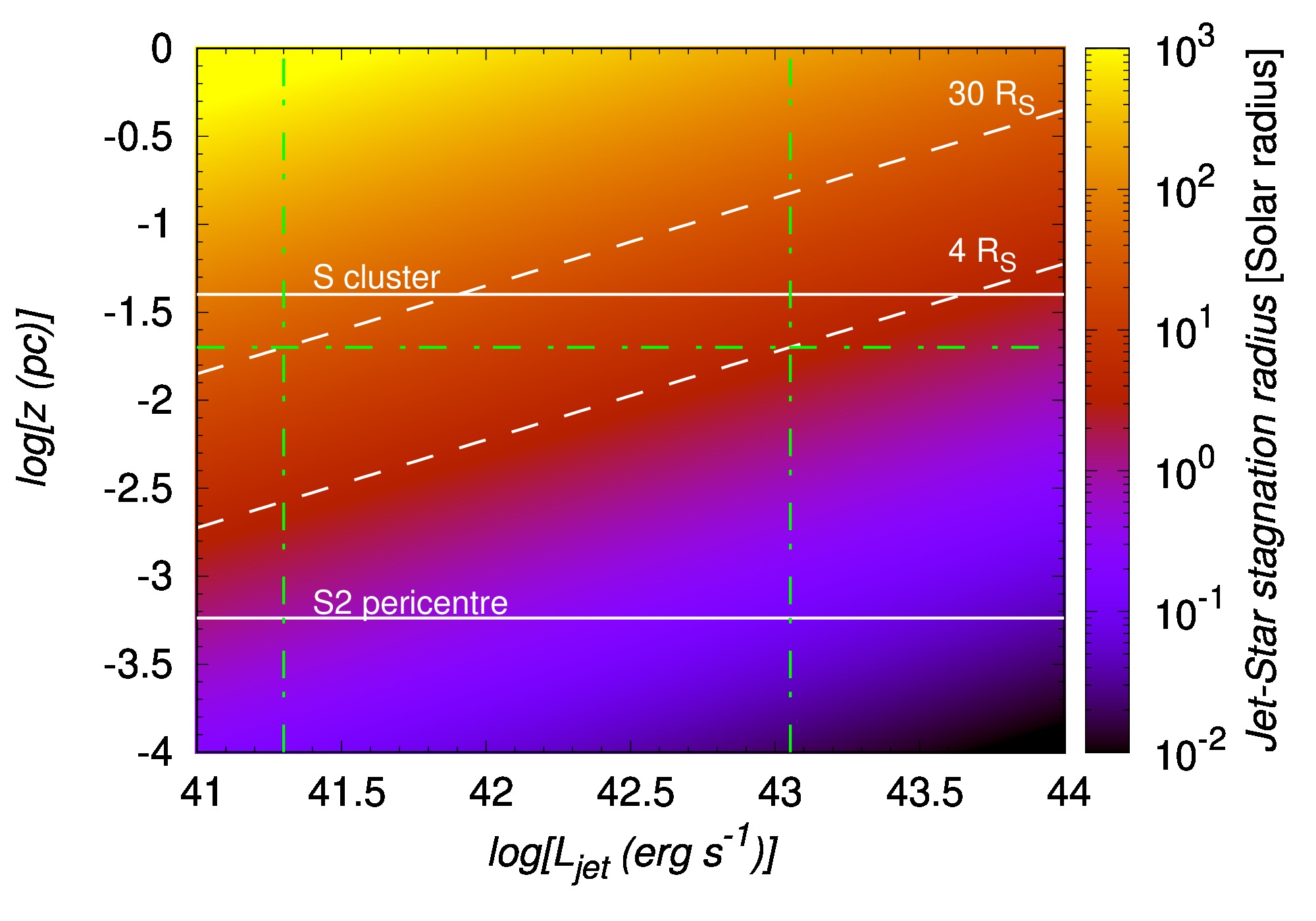}
    \caption{
   Stagnation radius $R_{\rm stag}/R_{\odot}$. The two horizontal white-solid lines indicate the radial extent of the S cluster between the S2 pericenter distance and the outer radius at $\sim 0.04\,{\rm pc}$. The two white-dashed lines stand for the atmosphere radius limits of late-type stars in the S cluster, 4$R_{\odot}$ and 30$R_{\odot}$ \citep{2019ApJ...872L..15H}. The  dot-dashed green lines indicate the jet luminosity limits that would yield the stellar atmosphere ablation at 30 and 4$R_{\odot}$ at $z=0.02$ pc.}
    \label{fig_stagnation_radius}
\end{figure}

By equating $P_{\rm sw}=P_{\rm j}$, we obtain the stagnation distance 
%
%
\begin{align}
  \frac{R_{\rm stag}}{R_{\odot}} &=  27\left(\frac{z}{0.04\,{\rm pc}}\right)\left(\frac{\dot{m}_{{\rm w}}}{10^{-8}\,M_{\odot}{\rm yr^{-1}}}\right)^\frac{1}{2} \times \notag \\
               & \times  \left(\frac{v_{\rm w}}{10\,{\rm km\,s^{-1}}}\right)^\frac{1}{2} \left(\frac{L_{\rm j}}{10^{42}\,{\rm erg\,s^{-1}}}\right)^{-\frac{1}{2}}\,
  \label{eq_stag_radius_values}
\end{align}
which characterizes by how much the red-giant envelope can be ablated by the jet in one encounter. Note that $R_{\rm stag}<R_{\star}$ for late-type giants and supergiants with $R_{\star}\sim 100-1000\,R_{\odot}$ (see Figure~\ref{fig_stagnation_radius}). 
The very tenuous wind of red giants cannot balance the jet ram pressure and therefore the jet plasma impacts on the stellar surface. 
As a consequence,  a fraction of the stellar envelope is removed as estimated by Eq.~\eqref{eq_stag_radius_values}.

Interestingly,  giant stars with $R_{\star}\gtrsim 30\,R_{\odot}$
appear to be missing in the S cluster. Only late-type stars  with $R_{\star}$ between $4$ and $30\,R_{\odot}$ (with absolute bolometric magnitudes between $-1.05$ and $3.32$, respectively, for the effective temperature of 4000 K) were
detected by \citet{2019ApJ...872L..15H} (see their figure~2).  
Stars with $4 \le R_{\star}/R_{\odot} \le 30$ within $0.02$~pc have $R_{\rm stag}/R_{\star} \sim 1$ when $2\times 10^{41} \le L_{\rm j}/{\rm erg\,s^{-1}} \le 1.1\times 10^{43}$, as it is
indicated in Fig.~\ref{fig_stagnation_radius}. This is in agreement with the estimated jet power $L_{\rm j}\sim 2.3\times 10^{42}\,{\rm erg\,s^{-1}}$ from X- and $\gamma$-ray bubbles  \citep{2016ApJ...829....9M}. Therefore, 
   an apparent lack of late-type giant stars with envelopes $R_{\star} \gtrsim 30\,R_{\odot}$ in the inner $\sim 0.04\,{\rm pc}$ of Galactic center could
   result from the jet-induced ablation of the stellar envelope during the last active phase of Sgr~A*, a few million years ago.

\begin{figure}[tbh]
    \centering
    \includegraphics[width=\columnwidth]{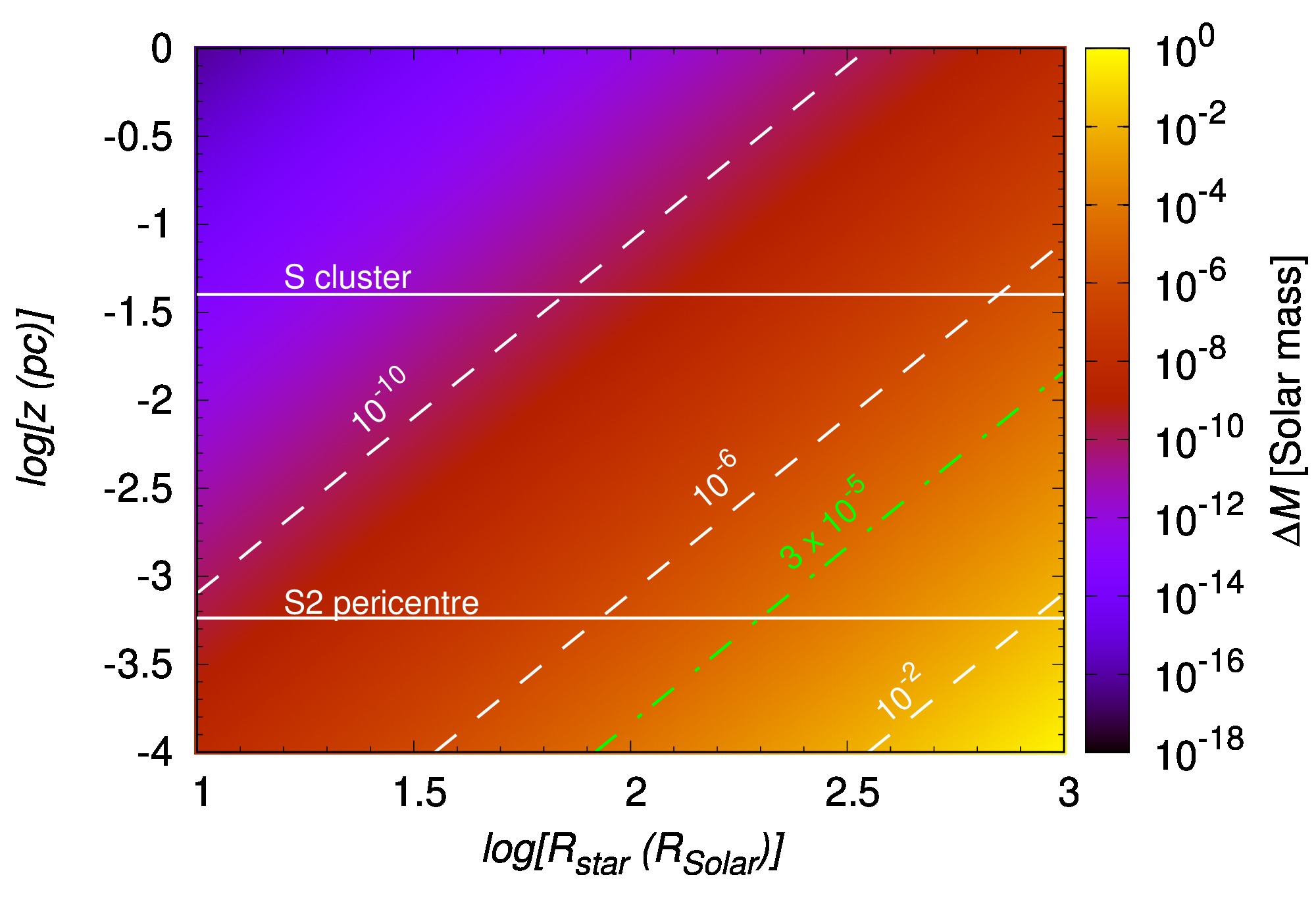}
    \caption{Mass removed from the red-giant envelope for a single jet--star interaction, $\Delta M$ in Eq.~(\ref{eq_mass_loss}), for the case with $L_{\rm j}=10^{42}\,{\rm erg\,s^{-1}}$,  $m_{\star}=1\,M_{\odot}$, and  $\theta=12.5^{\circ}$. The two horizontal lines indicate
    the radial extent of the S cluster between the S2 pericenter distance and the outer radius at $z\sim 0.04\,{\rm pc}$. 
    Dashed lines indicate 
    $\Delta M/M_{\odot}=10^{-10}$, $10^{-6}$, and $10^{-2}$. 
    The dot-dashed green line corresponds to $\Delta M=3\times 10^{-5}M_{\odot}$, which is equivalent to $\Delta M=6\times 10^{28}\,{\rm g}$ in equation~6 of \citet{2012ApJ...749..119B}.}
    \label{fig_mass_loss}
\end{figure}

\subsection{Basic timescales of the jet-star interaction}
\label{subsec_timescales}

The red giant, will enter the jet and not mix with its sheath layers on the surface if $v_{\rm orb}\gtrsim v_{\rm sc}$, where $v_{\rm orb}\sim (GM_{\bullet}/z)^{1/2}$ is the Keplerian orbital velocity of the star around Sgr~A* and $v_{\rm sc}\sim c(\Gamma \rho_{\rm j}/\rho_{\star})$ is the sound speed inside the shocked obstacle.
This condition can be written as
\begin{equation}
    \frac{\rho_{\star}}{\rho_{\rm j}}\gtrsim 5.2 \times 10^5 \left(\frac{z}{0.01\,{\rm pc}}\right)\left(\frac{\Gamma}{10}\right)\left(\frac{M_{\bullet}}{4\times 10^6\,M_{\odot}} \right)^{-1}\,,
    \label{eq_star_jet_density_ratio}
\end{equation}
which means that stellar atmosphere layers of the comparable density or greater than indicated by Eq.~(\ref{eq_star_jet_density_ratio}) will enter the jet and the less dense upper layers will mix with the jet surface layers.

Once inside, the bow shock is formed inside the jet on the very short timescale of $t_{\rm bs}\sim R_{\star}/c\sim 232 (R_{\star}/100\,R_{\odot})\,{\rm s}$. A shock also propagates through the red giant atmosphere on the shock-crossing or dynamical timescale, $t_{\rm d}\sim R_{\star}/v_{\rm sc}$, whose lower limit is imposed by the condition of penetration, $v_{\rm orb}\gtrsim v_{\rm sc}$, which leads to $t_{\rm d}\gtrsim R_{\star}/v_{\rm orb}=R_{\star}\sqrt{z/(GM_{\bullet})}\sim 53063 (R_{\star}/100\,R_{\odot}) (z/0.01\,{\rm pc})^{1/2} (M_{\bullet}/4\times 10^6\,M_{\odot})^{-1/2}\,{\rm s}$. The dynamical, shock-crossing time is at least $\sim 10$ times longer than the bow-shock formation time close to the footpoint of the jet, but the ratio becomes larger with the distance from Sgr~A* as $t_{\rm d}/t_{\rm bs}\gtrsim 229 (z/0.01\,{\rm pc})^{1/2} (M_{\bullet}/4\times 10^6)^{-1/2}$. 

The star-crossing time through the jet can be estimated as $t_{\star}\sim 2R_{\rm j}/v_{\rm orb}$. Using $R_{\rm j}=z\tan{\theta}$ and the condition $v_{\rm sc}\lesssim v_{\rm orb}$, we obtain
\begin{equation}
    \frac{t_{\star}}{t_{\rm d}}\lesssim \frac{2z\tan{\theta}}{R_{\star}}\sim 1966\,,
    \label{eq_ratio_star_crossing_dynamical_time}
\end{equation}
which implies that the shock propagates throughout the detached envelope, which is dragged by the jet and mixed with its material. Eventually, after several $t_{\rm d}$, the envelope material will reach the velocity of $v_{\rm j}\sim c$. Note that $t_{\star}/t_{\rm d}\sim 1$ when
$z\sim 10^{-5}\,{\rm pc}$, hence the removed envelope material should be dragged by the jet throughout the whole NSC.

The ablated red giant after the first crossing through the jet would first expand adiabatically to the original size on the thermal expansion timescale $t_{\rm exp}\sim R_{\star}/c_{\rm s} = R_{\star}\sqrt{\mu m_{\rm H}/(k_{\rm B}T_{\rm atm}})=0.2 (R_{\star}/100\,R_{\odot})(T_{\rm atm}/10^{4}\,{\rm K})^{-1/2}\,{\rm yr}$ because of the pressure of the warmer underlying layers as the star adjusts its size to reach a hydrodynamic equilibrium. This expansion timescale is shorter than the orbital timescale
\begin{align}
    P_{\rm orb} &= 2\pi \left(\frac{z^3}{GM_{\bullet}}\right)^{1/2}=\,\notag\\
                &= 47\left(\frac{z}{0.01\,{\rm pc}}\right)^{3/2}\left(\frac{M_{\bullet}}{4\times 10^6\,M_{\odot}}\right)^{-1/2}\,{\rm yr}.
\end{align}
\citet{2016ApJ...823..155K} infer a similar timescale for the envelope expansion using the hydrodynamic simulations of red giant--accretion clump collisions. According to their Fig.~7, the envelope expands to a larger size than the original stellar radius in $t_{\rm exp}\sim 1.5t_{\rm dyn}$ after the star emerges from the accretion clump, where $t_{\rm dyn}$ is a dynamical timescale of the star,
\begin{equation}
    t_{\rm dyn}\sim 0.32 \left(\frac{R_{\star}}{100\,R_{\odot}} \right)^{3/2}\left(\frac{m_{\star}}{1\,M_{\odot}} \right)^{-1/2}\,{\rm yr}\,,
    \label{eq_expansion_timescale}
\end{equation}
which leads to $t_{\rm exp}\sim 1.5\times 0.32\,{\rm yr}\approx 0.5\,{\rm yr}$.

The timescale of the thermal evolution of a star after the jet-star interaction is expressed by the Kelvin-Helmholtz (KH) or thermal timescale,
\begin{align}
    t_{\rm KH} & \approx \frac{Gm_{\star}^2}{R_{\star}L_{\star}}\notag\\
           &= 210 \left(\frac{m_{\star}}{1\,M_{\odot}} \right)^2 \left(\frac{R_{\star}}{100\,R_{\odot}} \right)^{-1} \left(\frac{L_{\star}}{1500\,L_{\odot}} \right)^{-1}\,{\rm yr}\,,
           \label{eq_kh_timescale}
\end{align}
where we estimated the stellar luminosity using $L_{\star}=4\pi R_{\star}^2 \sigma T_{\star}^4$ for $T_{\star}=3500-3700\,{\rm K}$ typical of red giants. For the whole range of stellar radii $R_{\star}\sim 4-1000\,R_{\odot}$ and stellar luminosities $L_{\star}\sim 10-7.2\times 10^3\,L_{\odot}$, $t_{\rm KH}$ differs considerably -- from $\sim 10^6\,$ yrs for the smallest giants to $\sim 0.43\,{\rm yr}$ for the largest ones.

Based on the comparison between the time between jet-star collisions $t_{\rm c}=P_{\rm orb}/2$ and the KH timescale, one can distinguish cool colliders when $t_{\rm c}\gtrsim t_{\rm KH}$, i.e. the star had enough time to radiate away the accumulated collisional heat and it cools down and shrinks before the next collision. For the case when $t_{\rm c} < t_{\rm KH}$, there is not enough time to radiate away the excess collisional heat and the star is warmer and larger at the time of the subsequent collision - these are so-called warm colliders. In the nuclear star cluster when the jet was active, there were both types of colliders with the approximate division given by $t_{\rm c}\approx t_{\rm KH}$, which leads to
\begin{align}
    z_{\rm c} & \approx \frac{G m_{\star}^{4/3}M_{\bullet}^{1/3}}{R_{\star}^{2/3}L_{\star}^{2/3}\pi^{2/3}}\notag\\
    & = 0.04\,\left(\frac{m_{\star}}{1\,M_{\odot}}\right)^{4/3}\left(\frac{M_{\bullet}}{4\times 10^6\,M_{\odot}}\right)^{1/3} \times \notag\\
    & \times \left(\frac{R_{\star}}{100\,R_{\odot}}\right)^{-2/3} \left(\frac{L_{\star}}{1500\,L_{\odot}}\right)^{-2/3}\,{\rm pc}.
    \label{eq_collider_division}
\end{align}

The length-scale $z_{\rm c}$ implies that red giants located inside the inner S cluster were collisionally heated up and bloated, which increased their mass removal during repetitive encounters with the jet. Stars orbiting at larger distances managed to cool down and shrink in size before the next collision, which has subsequently diminished their overall mass loss. However, note that Eq.~\eqref{eq_collider_division} is a function of stellar parameters $m_{\star}$, $R_{\star}$, and $L_{\star}$, hence $z_{\rm c}$ differs depending on the red-giant stage and its mass. For the smallest late-type stars with $R_{\star}\sim 4\,R_{\odot}$ and $L_{\odot}\sim 8.9\,L_{\odot}$, $z_{\rm c}\sim 10.5\,{\rm pc}$, therefore they can be classified as warm colliders throughout the nuclear star cluster. On the other hand, the late-type supergiants with $R_{\star}\sim 10^3\,R_{\odot}$ and $L_{\odot}\sim 7.2\times 10^4\,L_{\odot}$ have $z_{\rm c}\sim 0.7\,{\rm mpc}$, hence they can be classified as cool colliders beyond milliparsec distances.

\subsection{Jet-induced envelope removal}

The stellar evolution after a jet-star encounter is generally complicated given that the envelopes of red giants become bloated after the first passage through the jet. This is because of the pressure of lower, hotter layers and their subsequent nearly adiabatic expansion, which can make the red giant even larger and brighter \citep{2016ApJ...823..155K}. 
Note that the number of encounters $n_{\rm cross} =2 t_{\rm jet}/P_{\rm orb}$, where $t_{\rm jet}\sim 0.5$~Myr is the jet lifetime,  is typically $\gg 1$.
In particular,
%
\begin{equation}
    n_{\rm cross}  \sim 2\times 10^4  \left(\frac{t_{\rm jet}}{0.5\,{\rm Myr}}\right)\left(\frac{M_{\bullet}}{4\times10^6 \,M_{\odot}}\right)^\frac{1}{2}\left(\frac{z}{0.01\,{\rm pc}}\right)^{-\frac{3}{2}}.\label{eq_ncross}
\end{equation}
The number of encounters is also at least two orders of magnitude larger than the one expected from the star--accretion clump interaction investigated by \citet{2016ApJ...823..155K} and \citet{2019arXiv191004774A}. After the first passage, the bloated red giant has an even larger cross-section than before the encounter, which increases the mass removed during subsequent encounters \citep{1996ApJ...470..237A,2016ApJ...823..155K}. The jet--star interaction phase proceeds during the red-giant lifetime, which is $t_{\rm rg}\sim 10^8\,{\rm yr}$ \citep{2012ApJ...757..134M}. During the red giant phase, there are $n_{\rm orb}=t_{\rm rg}/P_{\rm orb}\sim 2.1\times 10^6(z/0.01\,{\rm pc})^{-3/2}$ orbits around Sgr~A*, out of which $n_{\rm cross}/n_{\rm orb}=2t_{\rm jet}/t_{\rm rg}\sim 1\%$ involve the interaction with the jet, assuming there was only one period of increased activity of Sgr~A* in the last 100 million years. This ensures that the repetitive jet-red giant interaction leads to a substantial mass-loss and the upper layer of the envelope is eventually removed.

The mass removal in a single passage due to the atmosphere ablation $\Delta M_1$ can be estimated through the balance of the jet ram force and the gravitational force acting on the envelope, i.e. $P_{\rm j}\pi R_{\star}^2 \simeq G\Delta M_1 m_{\star}/R_{\star}^2$, giving 
\citep{2012ApJ...749..119B}
\begin{align}
    \frac{\Delta M_1^{\rm max}}{M_{\odot}} &\approx 4 \times 10^{-10} \left(\frac{L_{\rm j}}{10^{42}\,{\rm erg\,s^{-1}}}\right)\left(\frac{R_{\star}}{100\,R_{\odot}}\right)^4 \times \notag \\
    & \times \left(\frac{z}{0.04\,{\rm pc}} \right)^{-2}\left(\frac{\theta}{0.22} \right)^{-2} \left(\frac{m_{\star}}{M_{\odot}} \right)^{-1}\,.
    \label{eq_mass_loss}
\end{align}
In Fig.~\ref{fig_mass_loss} we plot $\Delta M_1$.
Note that $\Delta M_1 \ll m_{\star}$ and, therefore, the orbital dynamics is 
not significantly affected by the {\rm single} passage through the jet. However, 
given that $\Delta M_1 \propto R_{\star}^4 z^{-2}$,
the mass-loss can be about one thousandth to one hundredth of the mass of a star for the largest giants on the asymptotic giant branch with $R_{\star}\sim 10^3\,R_{\odot}$ and distances an order of magnitude smaller than $z\simeq 0.04\,{\rm pc}$ 
(see the yellowish region in Fig.~\ref{fig_mass_loss}).
In fact, the value of $\Delta M_1=6 \times 10^{28}\,{\rm g}=3 \times 10^{-5}\,M_{\odot}$ discussed by \citet{2012ApJ...749..119B} for powerful blazars in connection to their very high-energy $\gamma$-ray emission can be reached in the Galactic center for red giants with radii $R_{\star}>200\,R_{\odot}$ at $z< 0.01\,{\rm pc}$. 
Such a large mass-loss with a certain momentum with respect to the star can already have an effect on the orbital dynamics, taking into account repetitive encounters of the red giant with the jet. In other words, the mass removal takes place at the expanse of the kinetic energy of the star, which has implications for the dynamics of the nuclear star cluster, see also \citet{2016ApJ...823..155K} for discussion. In addition, already for jets with a lower power corresponding to the active phase of the Galactic center, jet-red giant interactions can affect the short-term TeV emission in these sources. These effects are beyond the scope of the current paper and will be investigated in our future studies. 

The mass removal during a single jet encounter given by Eq.~\eqref{eq_mass_loss} can be considered as an upper limit since we assume that the cross-section of the star is given by its radius during the whole passage of the star through the jet, hence $\Delta M_1^{\rm max}\approx P_{\rm j}\pi R_{\star}^4/(Gm_{\star})$. However, this is only an approximation as realistically, during a few shock-crossing or dynamical timescales $t_{\rm d}$ the ram pressure of the jet will shape the red giant and its detached envelope into a comet-like structure, see Fig.~\ref{fig_jet_star}, for which the interaction cross-section is given by $R_{\rm stag}$ rather than by $R_{\star}$, which gives us a lower limit on the mass removal, $\Delta M_1^{\rm min}\approx P_{\rm j}\pi R_{\rm stag}^4/(Gm_{\star})\leq \Delta M_1^{\rm max}$. Using Eq.~\eqref{eq_stag_radius_values}, $\Delta M_1^{\rm min}$ can be expressed in terms of the basic parameters of the star and the jet and it can numerically be expressed by the same units as in Eqs.~\eqref{eq_mass_loss} and \eqref{eq_stag_radius_values},
\begin{align}
     \frac{\Delta M_1^{\rm min}}{M_{\odot}} &\approx  \frac{c(\dot{m}_{\rm w}v_{\rm w}z\tan{\theta})^2}{16Gm_{\star}L_{\rm j}}=\,\notag\\
     &= 2.1 \times 10^{-12} \left(\frac{L_{\rm j}}{10^{42}\,{\rm erg\,s^{-1}}} \right)^{-1} \left(\frac{m_{\star}}{M_{\odot}}\right)^{-1}\times \,\notag\\
     &\times \left(\frac{\dot{m}_{\rm w}}{10^{-8}\,M_{\odot}\,{\rm yr^{-1}}}\right)^2 \left(\frac{v_{\rm w}}{10\,{\rm km\,s^{-1}}}\right)^2\left(\frac{z}{0.04\,{\rm pc}}\right)^2\,, \label{eq_mass_loss_min_numerical}
\end{align}
where we adopted $\theta\sim 12.5^{\circ}$ as before. In comparison with $\Delta M_1^{\rm max}$ in Eq.~\eqref{eq_mass_loss}, which is proportional to $z^{-2}$, $\Delta M_1^{\rm min}$ in Eq.~\eqref{eq_mass_loss_min_numerical} increases as $z^2$. This implies that $\Delta M_1^{\rm min}\leq \Delta M_1^{\rm max}$ holds for $z\leq z_{\rm stag}$, where at $z_{\rm stag}$, $R_{\star}=R_{\rm stag}$. In other words, only at $z< z_{\rm stag}$ the mass removal due to the jet activity from the red giant atmosphere is possible, while at distances larger than $z_{\rm stag}$, the jet ablation is limited to the stellar-wind material, as is the case for the observed comet-shaped sources X3, X7, and X8 \citep{2010A&A...521A..13M,2019A&A...624A..97P}. From Eq.~\eqref{eq_stag_radius_values},  the relation for $z_{\rm stag}$ follows as,
\begin{align}
    z_{\rm stag} &\approx 0.15\,\left(\frac{R_{\star}}{100\,R_{\odot}} \right)\left(\frac{\theta}{0.22}\right)^{-1}\left(\frac{L_{\rm j}}{10^{42}\,{\rm erg\,s^{-1}}}\right)^{\frac{1}{2}}\times \,\notag\\
    &\times \left(\frac{\dot{m}_{\rm w}}{10^{-8}\,M_{\odot}\,{\rm yr^{-1}}} \right)^{-\frac{1}{2}}\left(\frac{v_{\rm w}}{10\,{\rm km\,s^{-1}}} \right)^{-\frac{1}{2}}\,{\rm pc}.\label{eq_stagnation_distance}
\end{align}

\begin{figure*}
    \centering
    \includegraphics[width=\columnwidth]{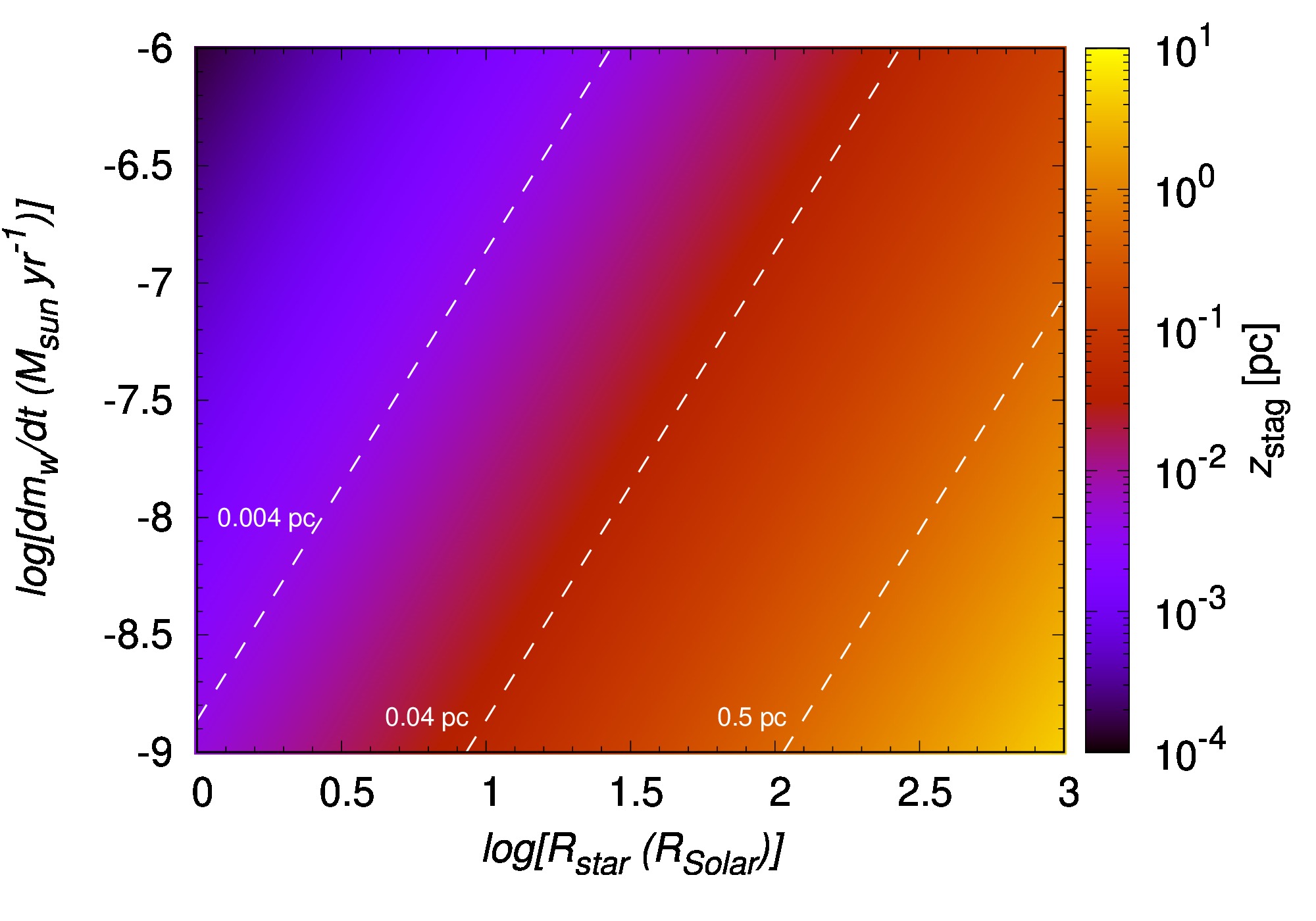}
    \includegraphics[width=\columnwidth]{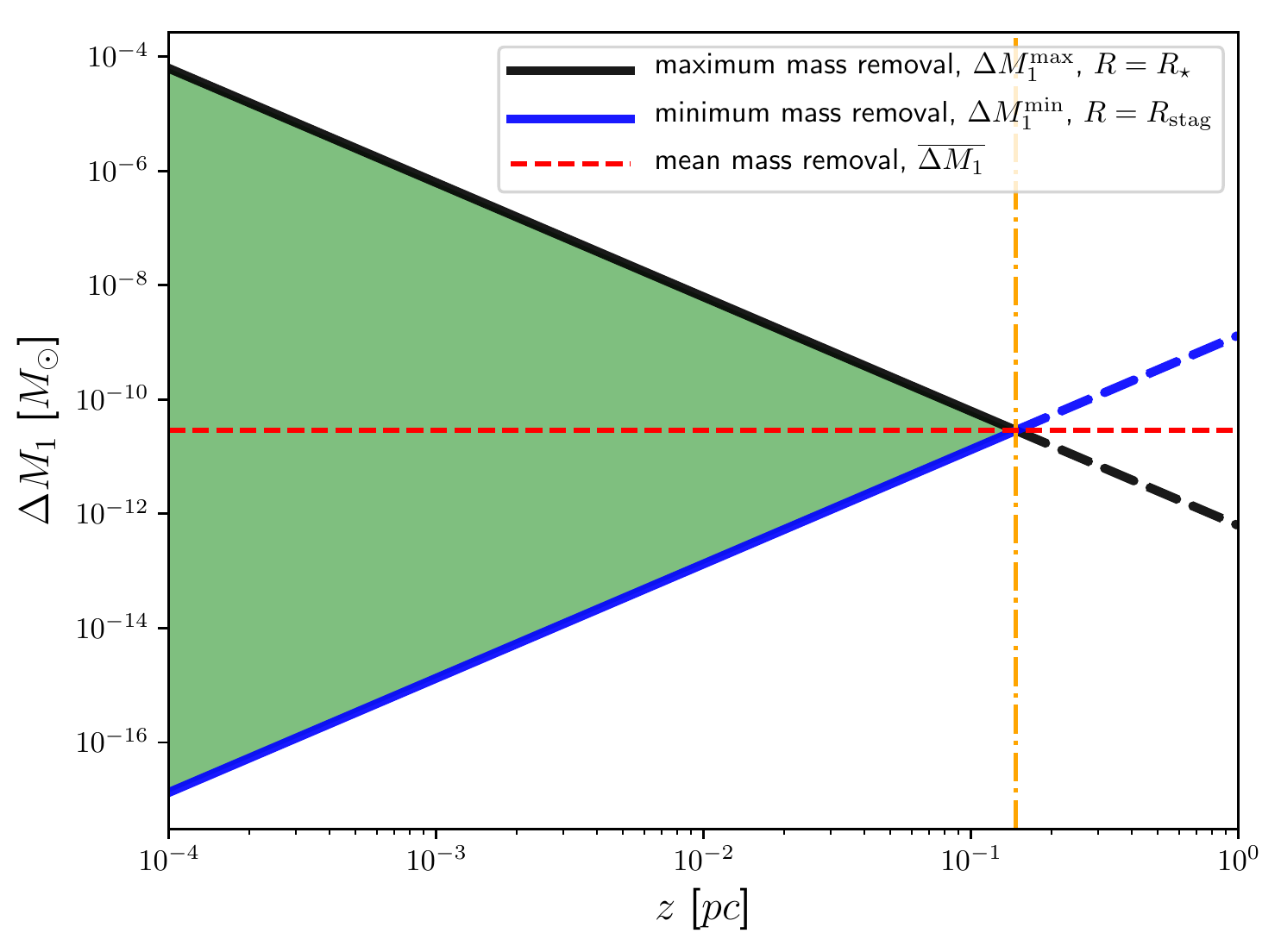}
    \caption{Mass removal during a single encounter of the red giant with the jet. {\bf Left panel: } The colour-coded distance in parsecs from Sgr~A* where $R_{\star}=R_{\rm stag}$ as a function of $\dot{m}_{\rm w}$ and $R_{\star}$. The dashed white lines mark $z_{\rm stag}$ equal to $0.004\,{\rm pc}$, $0.04\,{\rm pc}$, and $0.5\,{\rm pc}$ from the left to the right, respectively. {\bf Right panel:} An exemplary mass removal range $\Delta M_1^{\rm min}$--$\Delta M_1^{\rm max}$ (green-shaded region) for $R_{\star}=100\,R_{\odot}$, $\dot{m}_{\rm w}=10^{-8}\,M_{\odot}\,{\rm yr^{-1}}$, $v_{\rm w}=10\,{\rm km\,s^{-1}}$, $L_{\rm j}=10^{42}\,{\rm erg\,s^{-1}}$, and $\theta=12.5^{\circ}$. The dot-dashed vertical orange line marks $z_{\rm stag}$, see Eq.~\eqref{eq_stagnation_distance}, where $R_{\star}=R_{\rm stag}$. The red dashed line marks the mean mass removal $\overline{\Delta M_1}$, see Eq.~\eqref{eq_delta_mass_mean}, $\overline{\Delta M_1}=\Delta M_1^{\rm max}(z=z_{\rm stag})=\Delta M_1^{\rm min}(z=z_{\rm stag})$.}
    \label{fig_mass_removal_stagnation}
\end{figure*}

The dependence of $z_{\rm stag}$ on the stellar radius and the mass-loss rate is shown in Fig.~\ref{fig_mass_removal_stagnation} (left panel). It is apparent that the volume around the reactivated Sgr~A*, where the jet-ablation can occur for a particular red giant, depends considerably on $\dot{m}_{\rm w}$, which spans over three orders of magnitude depending on the evolutionary stage, $\dot{m}_{\rm w}\approx 10^{-9}-10^{-6}\,M_{\odot}\,{\rm yr^{-1}}$ \citep{1987IAUS..122..307R}. In particular, for red giants with $R_{\star}=10\,R_{\odot}$, $z_{\rm stag}$ shrinks from $0.047\,{\rm pc}$ to $0.0015\,{\rm pc}$ as the mass-loss rate increases from $\dot{m}_{\rm w}=10^{-9}\,M_{\odot}\,{\rm yr^{-1}}$ to $10^{-6}\,M_{\odot}\,{\rm yr^{-1}}$.

In Fig.~\ref{fig_mass_removal_stagnation} (right panel), we show an exemplary case for the mass removal range from the red giant atmosphere (red giant with the parameters of $R_{\star}=100\,R_{\odot}$, $\dot{m}_{\rm w}=10^{-8}\,M_{\odot}\,{\rm yr^{-1}}$, and $v_{\rm w}=10\,{\rm km\,s^{-1}}$) due to the single crossing through the jet with the luminosity of $L_{\rm j}=10^{42}\,{\rm erg\,s^{-1}}$ and the opening angle of $25^{\circ}$. Towards $z_{\rm stag}$, the mass removal due to a single encounter approaches the mean value of $\overline{\Delta M_1}=\Delta M_1^{\rm max}(z=z_{\rm stag})=\Delta M_1^{\rm min}(z=z_{\rm stag})$,
\begin{align}
    \frac{\overline{\Delta M_1}}{M_{\odot}} &= \frac{\dot{m}_{\rm w}v_{\rm w}R_{\star}^2}{4Gm_{\star}}=\,\notag\\
    &=2.9 \times 10^{-11} \left(\frac{\dot{m}_{\rm w}}{10^{-8}\,M_{\odot}\,{\rm yr^{-1}} }\right) \left(\frac{v_{\rm w}}{10\,{\rm km\,s^{-1}}}\right)\times\,\notag\\
    &\times \left(\frac{R_{\star}}{100\,R_{\odot}}\right)^2\left(\frac{m_{\star}}{1\,M_{\odot}}\right)^{-1}\,.\label{eq_delta_mass_mean}
\end{align}

\begin{figure}
    \centering
    \includegraphics[width=\columnwidth]{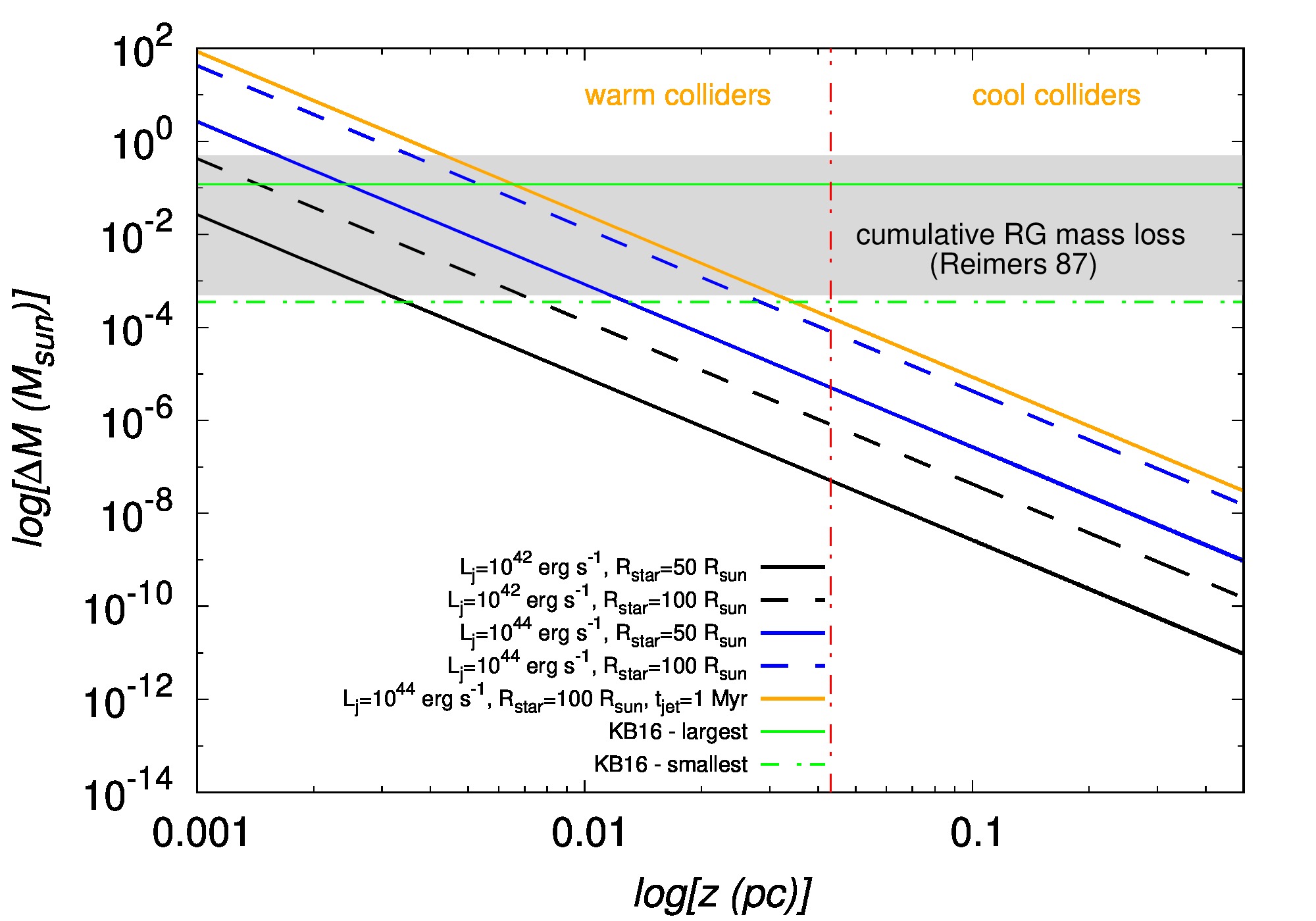}
    \caption{Cumulative mass removal $\Delta M$ due to the repetitive red giant--jet encounters as a function of the distance from Sgr~A*. We 
    fixed $L_{\rm j}=10^{42}$ and $10^{44}\,{\rm erg\,s^{-1}}$ and 
    $R_{\star}=50$ and $100\,R_{\odot}$. 
    In addition, we plot $\Delta M$
    for $L_{\rm j}=10^{44}\,{\rm erg\,s^{-1}}$, $R_{\star}=100\,R_{\odot}$, and a longer duration of the jet activity, $t_{\rm jet}=1\,{\rm Myr}$. For comparison, we also show the mass removal limits as inferred by \citet{2016ApJ...823..155K} for the red giant--clump collisions and the mass range as expected from the cumulative red giant (RG) mass loss due to stellar winds analyzed by \citet{1987IAUS..122..307R}. The red dot-dashed line marks the division between warm and cool colliders according to Eq.~\eqref{eq_collider_division}.}
    \label{fig_cumulative_mass_removal}
\end{figure}

Because of the estimated several thousands of red giant--jet encounters according to Eq.~\eqref{eq_ncross}, the cumulative mass loss from the red giant can be derived as 
$\Delta M\sim n_{\rm cross} \Delta M_1$, giving 
\begin{align}
  \frac{\Delta M}{M_{\odot}}    &\approx 10^{-4} \left(\frac{L_{\rm j}}{10^{42}\,{\rm erg\,s^{-1}}}\right)\left(\frac{R_{\star}}{100\,R_{\odot}}\right)^4 \left(\frac{z}{0.01\,{\rm pc}} \right)^{-\frac{7}{2}}\times \notag \\
    & \left(\frac{\theta}{0.22} \right)^{-2} \left(\frac{m_{\star}}{M_{\odot}} \right)^{-1}\left(\frac{t_{\rm jet}}{0.5\,{\rm Myr}}\right)\left(\frac{M_{\bullet}}{4\times10^6 \,M_{\odot}}\right)^\frac{1}{2}.
    \label{eq_total_mass_loss}
\end{align}
In Fig.~\ref{fig_cumulative_mass_removal}, we plot $\Delta M$ 
for $L_{\rm j}=10^{42}$ and $10^{44}\,{\rm erg\,s^{-1}}$, and 
$R_{\star}=50$ and $100\,R_{\odot}$. Additionally, we  plot $\Delta M$ 
for the longer jet lifetime of $t_{\rm jet}=1\,{\rm Myr}$ and $L_{\rm j}=10^{44}\,{\rm erg\,s^{-1}}$ and $R_{\star}=100\,R_{\odot}$, which can be considered as an upper limit of $\Delta M$ for a red giant orbiting Sgr~A*. We 
find that
$\Delta M$ 
within the S cluster is comparable to the mass removal 
inferred from red giant--clump collision simulations by \citet{2016ApJ...823..155K}. In Fig.~\ref{fig_cumulative_mass_removal}, we plot the upper and the lower limits of $\Delta M$ obtained by \citet{2016ApJ...823..155K}. Beyond $z=0.03\,{\rm pc}$, $\Delta M$ for the star--jet interaction is progressively smaller than $3.5\times 10^{-4}\,M_{\odot}$, which implies that the jet impact on the stellar evolution is the most profound for S cluster red giants. In this region, there also lies the division between the warm and the cool colliders as discussed in Subsection~\ref{subsec_timescales}, with warm colliders present inside $z_{\rm c}\sim 0.04\,{\rm pc}$, which is also marked in Fig.~\ref{fig_cumulative_mass_removal} with a vertical dot-dashed line. Since warm colliders are warmer and bigger, this further enhances the mass removal inside the S cluster. In addition, the mass removal due to the jet interaction is of a comparable order of magnitude as the mass loss expected from cool winds during the time interval of the active jet $\Delta M_{\rm w}\approx \dot{m}_{\rm w}t_{\rm jet}\sim 0.5-5\times 10^{-4}\,M_{\odot}$ when 
$\dot{m}_{\rm w}\approx 10^{-6}-10^{-9}\,{\rm M_{\odot}\,yr^{-1}}$ \citep{1987IAUS..122..307R}, as it is indicated by the shaded rectangle in Fig.~\ref{fig_cumulative_mass_removal}. This implies that the jet--star interaction perturbs the stellar evolution of passing red giants, in particular in the innermost parts of the nuclear stellar cluster.

Note that, on one hand, $\Delta M$ is supposed to be a lower limit since after the first passage through the jet, the giant is expected to expand to an even larger radius before the next encounter, which increases the mass removal efficiency \citep{2016ApJ...823..155K}. On the other hand, resonant relaxation 
of stellar orbits as well as a jet precession may change the frequency of the jet-star interactions
(see  Sections~\ref{sec_fraction_redgiant} and \ref{subsec_precession}) and therefore 
$n_{\rm cross}$ should be considered
as an upper limit of the number of encounters. Overall, $\Delta M$ in Eq.~(\ref{eq_total_mass_loss}) can be applied as an approximation for the total mass removal due to the giant-jet interactions.
Hence, the truncation of stellar envelopes of late-type stars by the jet during active phases of Sgr~A* appears to be efficient and complementary to other previously proposed processes, mainly tidal disruptions of giants and stellar collisions with other stars and/or the accretion disc.

\section{Missing red giants in the near-infrared domain}
\label{sec_NIR_domain}

Red giants are post-main-sequence evolutionary stages of stars with initial mass $0.5\, M_{\odot}\lesssim m_{\star} \lesssim 10 M_{\odot}$. These stars exhausted hydrogen supplies in their cores and the hydrogen fusion into helium continues in the shell. As a result, the mass of the helium core gradually increases and this is linked to the increase in the atmosphere radius as well as the luminosity. Stellar evolutionary models of red giants show that their atmosphere radius and the bolometric luminosity depend primarily on the mass of the helium core $m_{\rm c}$ as \citep{1971A&A....13..367R,1987ApJ...319..180J}
\begin{align}
    \frac{L_{\star}}{L_{\odot}} & \approx  \frac{10^{5.3}\mu_{\rm c}^6}{1+10^{0.4}\mu_{\rm c}^4+10^{0.5}\mu_{\rm c}^5}\,,\notag\\
    \frac{R_{\star}}{R_{\odot}} & \approx  \frac{3.7 \times 10^3 \mu_{\rm c}^4}{1+\mu_{\rm c}^3+1.75\mu_{\rm c}^4}\,,\label{eq_luminosity_radius}
\end{align}
where $\mu_{\rm c}\equiv m_{\rm c}/M_{\odot}$. This also holds for red supergiants with carbon-oxygen cores and burning hydrogen and helium in their shells \citep{1970AcA....20...47P}. 
In the red giant stage, the dominant energy source is the p-p process and hence the luminosity is mainly determined by the growth rate of the helium core $\dot{m}_{\rm c}$
\begin{equation}
    \frac{L_{\star}}{L_{\odot}}\simeq 1.02 \left(\frac{\dot{m}_{\rm c}}{10^{-11}\, M_{\odot}\,{\rm yr^{-1}}} \right)\,.
    \label{eq_luminosity_growth_rate}
\end{equation}


Eqs.~\eqref{eq_luminosity_radius} and \eqref{eq_luminosity_growth_rate} imply that the bolometric luminosity is not significantly affected by the jet-red giant interaction, since only the tenuous shell is ablated by the jet and the dense core is left untouched. Then the 
effective temperature $T_1$ of the ablated giant 
is
\begin{equation}
    \frac{T_1}{T_0}=\left(\frac{R_0}{R_1} \right)^\frac{1}{2}\,,
    \label{eq_temperature_bolometric}
\end{equation}
where $T_0$ is the original effective temperature and $R_0$ and $R_1$ are the atmosphere radii before and after the truncation, respectively. Here we implicitly assume that the red giant underwent $n_{\rm coll}$ interactions with the jet during the active phase given by Eq.~\eqref{eq_ncross}, which eventually leads to the decreased radius of $R_1\approx R_{\rm stag}$ according to Eq.~\eqref{eq_stag_radius_values}. The luminosity in the infrared domain between frequencies $\nu_1$ and $\nu_2$ can be expressed using the Rayleigh-Jeans approximation\footnote{Strictly speaking, for $T_{\star}=5000\,{\rm K}$, the condition $k_{\rm B}T_{\star}>h\nu$ applies for wavelengths longer than $2.9\,{\rm \mu m}$.} as $L_{\rm IR} \lesssim 8/3(\pi/c)^2 R_{\star}^2 k_{\rm B}T_{\star} (\nu_2^3-\nu_1^3) \propto R_{\star}^2 T_{\star}$ \citep[see][for a similar analysis]{2005PhR...419...65A}, which using Eq.~\eqref{eq_temperature_bolometric} leads to
\begin{equation}
    \frac{L_1}{L_0} \sim \left(\frac{R_1}{R_0} \right)^\frac{3}{2}\,.
    \label{eq_luminosity_ratio}
\end{equation}

For instance, the ablation of a red giant atmosphere from 120 to 30~$R_{\odot}$
would result in the increase of effective temperature by a factor of 2 
and a decrease by a factor of 8 in the IR luminosity or $\sim 2.26\,{\rm mag}$. The ablation of the envelope from 120 to 4$R_{\odot}$ 
would result in the decrease by as much as $\sim 5.5\,{\rm mag}$. The difference of 2-5 magnitudes can already affect the count rate of late-type stars in the near-infrared domain in the central arcsecond of the Galactic center.

As an exemplary case, we set up a simplified temporal evolution of a red giant using Eqs.~\eqref{eq_luminosity_radius} and \eqref{eq_luminosity_growth_rate}. We perform this calculation to estimate the potential difference in near-infrared magnitudes and the colour change for late-type stars before and after the active jet phase -- it does not represent realistic stellar evolution tracks, but can provide insight into the basic trends in the near-infrared magnitude evolution and the effective temperature. We evolve the stellar luminosity and the radius for an increasing core mass $\mu_{\rm c}=\mu_{\rm c 0}+\dot{\mu}_{\rm c} \mathrm{d}t$, where $\mu_{\rm c 0}=0.104$ and the time-step is $\mathrm{d}t=10^{4}\,{\rm yr}$. The overall evolution from $\mu_{\rm c 0}$ to $\mu_{\rm c}=0.55$ takes $\sim 8.41 \times 10^{9}\,{\rm yr}$, when neither the effect of stellar winds nor that of rotation is taken into account. The initial and the final core masses were chosen according to the limiting values for lighter stars, $m_{\star}<2\,M_{\odot}$, in which case $\mu_{\rm c}^{\rm min}\sim 0.1$ and $\mu_{\rm c}^{\rm max}\sim 0.5$. These are stars with degenerate helium cores and hydrogen burning shells \citep{1971A&A....13..367R}. The lower core-mass value of 0.1 also approximately corresponds to the Sch\"onberg-Chandrasekhar limit. The total duration is comparable to the time that stars of $1\,M_{\odot}$ spend on the giant and the asymptotic giant branches, which is of the order of $10^9\,{\rm yr}$ according to \citet{2012ApJ...757..134M}.

\begin{figure}[!ht]
    \centering
    \includegraphics[width=\columnwidth]{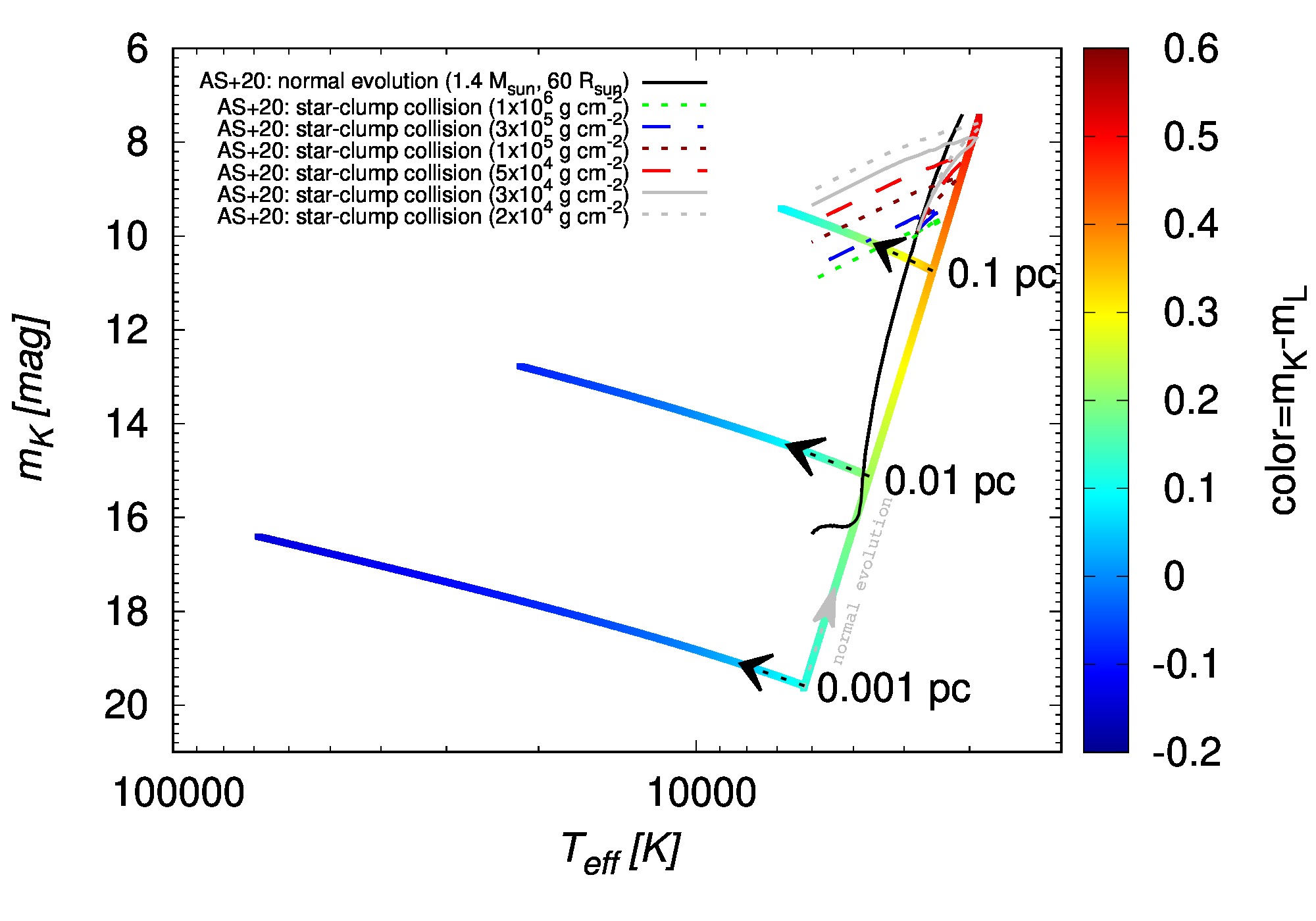}
    \caption{Near-infrared magnitude (K-band, $2.2\,{\rm \mu m}$, dereddened) and  temperature (or color index that is color-coded) of a red giant after crossing the jet $n_{\rm coll}$ times. We compare three temperature--$K_{\rm s}$-band magnitude curves affected by the series of collisions with the jet at $z=0.001$, $0.01$, and $0.1$~pc (marked by black arrows) with a normal, unaffected evolution (marked by a gray arrow). The evolution is driven by an increase in the helium core mass from $\mu_{\rm c}=0.104$ to $0.55\,M_{\odot}$ with the time-step of $10^4$ years. At each time-step we calculate $R_{\star}$ and
    $L_{\star}$ using Eqs.~\eqref{eq_luminosity_radius}. The overall evolution takes $\sim 8\times 10^{9}$ years. For comparison, we also show the evolutionary tracks for the normal evolution (black solid line) and the evolution affected by the star--clump collisions (lines according to the legend for different clump surface densities listed in parentheses) as calculated by \citet{2019arXiv191004774A} (abbreviated as AS+20) using the CESAM code \citep{2008Ap&SS.316...61M}.}
    \label{fig_evolution_tem_mag}
\end{figure}

To assess the observational effects of the star--jet collision in the near-infrared domain, we calculate the effective temperature at each step using $T_{\star}=T_{\odot}(L_{\star}/L_{\odot})^{1/4} (R_{\star}/R_{\odot})^{-1/2}$. Subsequently, we calculate the monochromatic flux density in K-band (2.2\,$\mu{\rm m}$) and L'-band (3.8\,$\mu{\rm m}$) using $F_{\nu}=(R_{\star}/d_{\rm GC})^2 \pi B_{\nu}(T_{\star})$, where $B_{\nu}(T_{\star})$ is the spectral brightness given by the Planck function at the given effective temperature. The corresponding magnitudes are calculated using $m_{K}=-2.5\log{(F_{K}/653\,{\rm Jy})}$ and $m_{L'}=-2.5\log{(F_{K}/253\,{\rm Jy})}$, from which the color index follows as CI$=m_{K}-m_{L'}$. 

For the analysis in this section as well as in Section~\ref{sec_surface_brightness}, we calculate intrinsic stellar magnitudes $m_{K}$ and $m_{L'}$. The calculated magnitudes and colours can then be compared to extinction-corrected magnitudes and the derived surface-brightness profiles of the nuclear star cluster, i.e. those corrected for the foreground extinction. To compare our results to observed magnitudes and derived surface profiles that are just corrected for the differential extinction but not for the foreground extinction \citep{2009A&A...499..483B,2019ApJ...872L..15H,2020arXiv200715950S}, it is necessary to increase the magnitudes using the corresponding mean extinction coefficients \citep[see e.g.,][]{2010A&A...511A..18S}, in particular $A_{K}=2.54 \pm 0.12$ mag and $A_{L'}=1.27 \pm 0.18$ mag.

\begin{table*}[tbh]
\caption{Parameters at the time of the jet-ablation and those of the final state of a red giant, which is evolved from the core mass of $\mu_{\rm c}=0.104$ to $0.55$. The upper three lines contain parameters of the ablated red giants orbiting at distances 0.001, 0.01, and 0.1 pc (first column). The second column lists the time, at which the ablation occurred with respect to the initial state of $\mu_{\rm c}=0.104$. The third column lists the truncation radius, when $R_{\star}=R_{\rm stag}$. The fourth column lists the effective temperature at the time of truncation. The fifth column lists the final effective temperature at the final stage of their evolution (when $\mu_{\rm c}=0.55$). The sixth and the seventh columns contain dereddened magnitudes at the ablation and the final times, respectively. The last column contains dereddened colour indices, $CI=m_{\rm K}-m_{\rm L}$, at the final state. The bottom line corresponds to the final state of a normal, unaffected evolution.}
    \centering
    \begin{tabular}{c|c|c|c|c|c|c|c}
    \hline
    \hline
    Ablation distance [pc] & Ablation time [yr] & $R_{\star}^{\rm abl}$ [$R_{\odot}$] & $T_{\star}^{\rm abl}$ [K]  & $T_{\star}^{\rm fin}$ [K] & $m_{K}^{\rm abl}$ [mag] & $m_{K}^{\rm fin}$ [mag] & $\text{CI}^{\rm fin}$\\
    \hline 
    0.001     & $3.0\times 10^8$  &$0.45$ & $6205$  & $68\,759$ & $19.6$ & $16.4$ & $-0.14$ \\
    0.01      & $8.0\times 10^9$   &$4.45$ & $4662$  &$21\,744$  & $15.2$  &$12.8$ & $-0.09$ \\
    0.1       & $8.4\times 10^9$  &$44.51$ & $3526$ &$6\,876$ & $10.7$  &$9.4$  & $-0.09$ \\
    \hline
    normal evolution & - &  $255.21$ (final radius) &  - & $2\,871$ & - & $7.47$ & $0.51$\\
    \hline 
    \end{tabular}
    \label{tab_redgiant_param}
\end{table*}

The imprint of the ablation of the stellar atmosphere by a jet, whose kinetic luminosity is fixed to $L_{\rm j}=2.3 \times 10^{42}\,{\rm erg\,s^{-1}}$, is modelled by assuming that the radius of an interacting red giant keeps evolving according to Eq.~\eqref{eq_luminosity_radius} when $R_{\star}<R_{\rm stag}$ at a given distance $z$ from Sgr~A*. After the stellar radius reaches the scale of the stagnation radius at a given distance $z$, we set $R_{\star}=R_{\rm stag}$ for the rest of the evolution which can by justified by the fact that the red giant propagates through the jet $n_{\rm coll}$-times and the envelope is removed after repetitive encounters. The expected number of encounters is $6\times 10^5$, $2\times 10^4$, and $632$ for 
$z =0.001\,{\rm pc}$, $0.01\,{\rm pc}$, and $0.1\,{\rm pc}$, respectively; see Eq.~\eqref{eq_ncross}. In Fig.~\ref{fig_evolution_tem_mag} we show three magnitude--effective temperature curves of the ablated red giants that underwent repetitive encounters with the jet at their orbital distances of $0.001$, $0.01$,
and $0.1\,{\rm pc}$ from Sgr~A*, which led to their truncation to a smaller radius close to the corresponding stagnation radius at a given distance. In addition, we compare the magnitude--temperature curves of ablated giants with an unaffected evolution. We list the stellar parameters at the time of the atmosphere truncation when $R_{\star}=R_{\rm stag}$ at the corresponding distance as well as the parameters for the final state of ablated giants in Table~\ref{tab_redgiant_param}. This is compared to an unaffected final state with the core mass of $0.55\,M_{\odot}$; see the bottom row of Table~\ref{tab_redgiant_param}. The basic signature of the jet-star interaction is that the star gets progressively warmer (with a bluer, more negative color index) and fainter in the near-infrared $K_{\rm s}$-band in comparison with the normal evolution without any atmosphere ablation. This trend is more apparent for red giants that are closer to Sgr~A* because of the smaller jet-star stagnation radius and hence a larger fraction of the stellar atmosphere that is removed.

Although we do not calculate stellar evolutionary tracks, only basic trends in terms of near-infrared magnitude and effective temperature, our results are consistent with those of \citet{2019arXiv191004774A} who calculated evolutionary tracks specifically for late-type stars ablated due to the red giant--accretion clump collisions. They show in their Fig. 2 that the collision affects the stellar evolution of a red giant in a way that after repetitive encounters it follows a track along a nearly constant absolute bolometric magnitude towards higher effective temperatures. The constant absolute bolometric magnitude or bolometric luminosity in combination with an increasing effective temperature results in the drop in the near-infrared luminosity, since $L_{\rm IR}/L_{\rm bol}\propto T_{\star}^{-3}$. We used their evolutionary tracks calculated using the CESAM code \citep{2008Ap&SS.316...61M} for estimating $K_{\rm s}$-band near-infrared magnitudes. These tracks are depicted in Fig.~\ref{fig_evolution_tem_mag} for the case of a normal evolution of a star with $M_{\star}=1.4\,M_{\odot}$ and $R_{\star}=60\,R_{\odot}$ (black solid line) and different collision cases for clumps with surface densities in the range $\Sigma=2 \times 10^4-10^6\,{\rm g\,cm^{-2}}$ (see the legend). Qualitatively, the perturbed stellar evolutionary tracks follow the temperature trends that we can also observe for giant--jet collisions: ablated giants move towards higher effective temperature. Also, they become fainter in the near-infrared domain in comparison with an unperturbed evolution. The main difference in comparison with the analysis of \citet{2019arXiv191004774A} is their trend towards larger magnitudes (stars become fainter), while we see a small gradual increase in brightness. This difference is due to our simplying assumption of a constant radius after the series of collisions with the jet, while in reality the radius should evolve, especially after the jet ceases to be active. Since $L_{\rm IR}\propto R_{\star}^2 T_{\star}$, the near-infrared luminosity grows linearly with increasing temperature for the fixed stellar radius. This motivates further exploration of the effect of star--jet collisions using a modified stellar evolutionary code.  

For asymptotic giant branch stars, an extreme transition from a red, cool luminous giant to a hot and faint white dwarf is possible when it is completely stripped off of its envelope. This was studied by \citet{2020MNRAS.493L.120K} for tidal stripping close to the SMBH, but cannot be excluded also for asymptotic giant-branch stars and jet collisions for a case when the giant star is at milliparsec separation from Sgr~A* and less, in which case the stagnation radius is typically a fraction of the Solar radius. In fact, an active jet can enlarge the volume around the SMBH where asymptotic giant-branch stars are turned into white dwarfs. Considering Eq.~\eqref{eq_stag_radius_values}, we can derive that in order for $R_{\rm stag}$ to be of the order of a white-dwarf radius $R_{\rm wd}\sim 0.01R_{\odot}$, the giant needs to orbit the SMBH at $z\approx 0.15\,{\rm mpc}$ so that the jet with $L_{\rm j}=10^{44}\,{\rm erg\,s^{-1}}$ can truncate it down to the white-dwarf size. The stellar interior that is not affected by tidal forces is characterized by the Hill radius
\begin{align}
    r_{\rm Hill} & \approx z\left(\frac{m_{\star}}{3M_{\bullet}} \right)^{\frac{1}{3}} = 29\left(\frac{z}{0.15\,{\rm mpc}}\right)\times \notag\\ & \times \left(\frac{m_{\star}}{1\,M_{\odot}}\right)^{\frac{1}{3}}\left(\frac{M_{\bullet}}{4\times 10^6\,M_{\odot}}\right)^{-\frac{1}{3}}\,R_{\odot}\,,     
\end{align}
from which we see that tidal forces alone will not truncate the giant down to its white-dwarf core since $r_{\rm Hill}>R_{\rm stag}$ at $z$.

In summary, the jet-star interaction could have affected the appearance of late-type giants in the central arcsecond by making them warmer or bluer in terms of a colour and hence fainter in the near-infrared domain. 

\section{Fraction of red giants interacting with the jet}
\label{sec_fraction_redgiant}

We estimate the number of late-type stars, i.e. stars that form a cusp, that could have passed and interacted with the jet during its estimated life-time of $\sim 0.1-0.5\,{\rm Myr}$ \citep{2012ApJ...756..181G}. For the density distribution of late-type stars in the inner $0.5\,{\rm pc}$, we adopt a cusp-like power-law distribution $n_{\rm RG}\approx n_0(z/z_{\rm 0})^{-\gamma}$, 
with $n_0\simeq 52\,{\rm pc^{-3}}$, $z_0\simeq 4.9\,{\rm pc}$, and $\gamma \simeq 1.43$ \citep{2018A&A...609A..26G}. The expected number of late-type stars within a certain distance $z_{\rm out}$ is
\begin{equation}
    N_{\star}(<z_{\rm out})=\int_0^{z_{\rm out}}n_{\rm RG}(z)4\pi z'^2\mathrm{d}z'=4\pi n_0 \frac{z_0^{\gamma} z_{\rm out}^{3-\gamma}}{3-\gamma}
\end{equation}
giving
$N_{\star}\approx 610$ and 25.8
inside 0.3 and 0.04~pc, respectively.
The number of stars inside the jet at any time is given by the jet covering factor $f_{\rm j}$ in a spherical volume $V_{\star} = (4/3)\pi z^3_{\rm out}$. By considering a conical jet and a counter-jet  with the total volume $V_{\rm j}=2/3\pi R_{\rm j}^2z_{\rm out}$, the covering factor
is $f_{\rm j} = V_{\rm j}/V_{\star} \sim (R_{\rm j}/z_{\rm out})^2$ and the number of red giants inside the jet is
$N_{\rm j}(<z_{\rm out})= f_{\rm j} N_{\star}(<z_{\rm out})\sim N_{\star}(<z_{\rm out})\tan^2(\theta)/2$.
Then the average number of red giants that are simultaneously inside the jet is $N_{\rm j}\approx 14.8$  and $0.62$ inside 0.3~pc and 0.04~pc, respectively.

\begin{figure}
    \centering
    \includegraphics[width=\columnwidth]{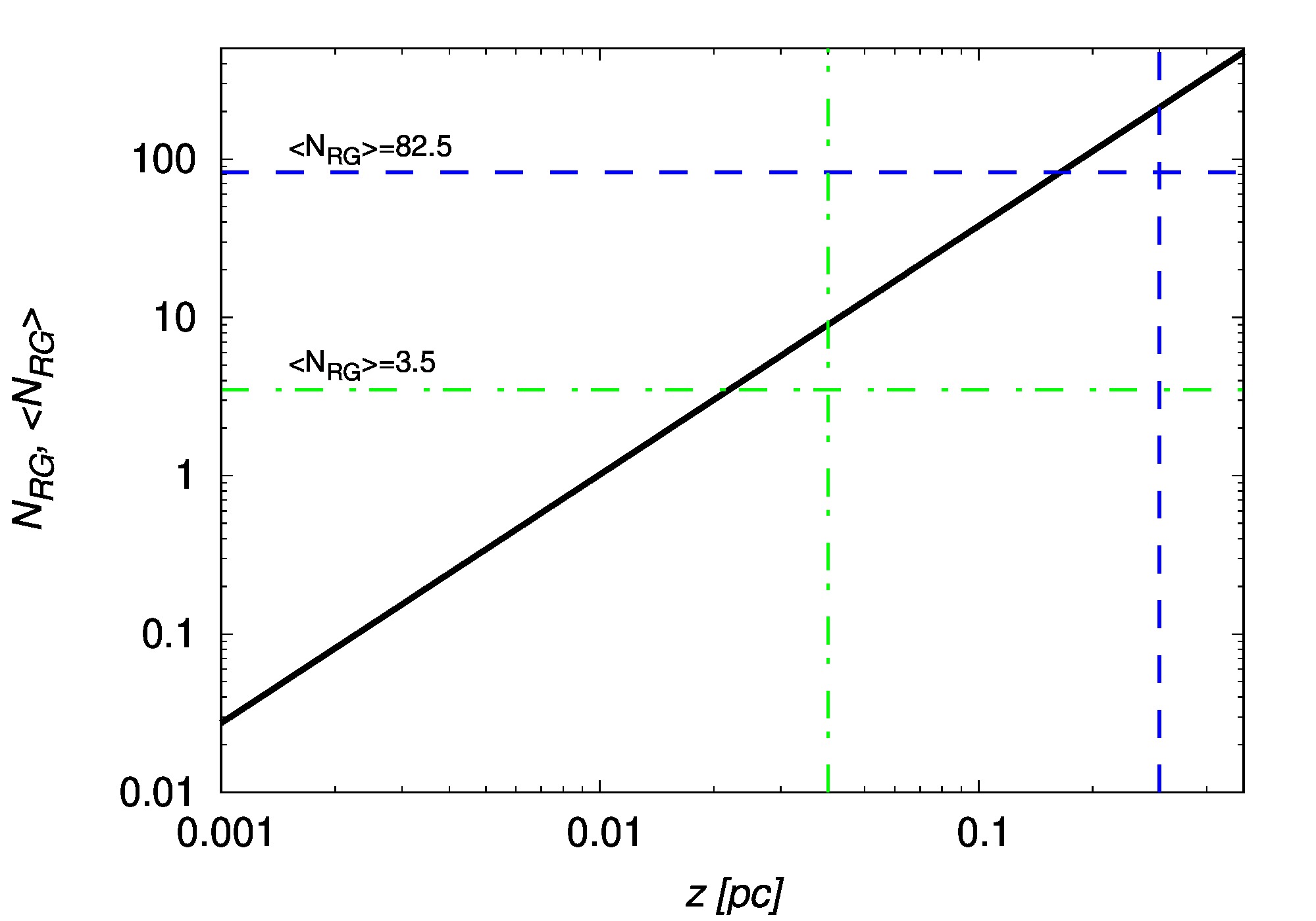}
    \caption{Number of red giants crossing the jet per orbital period (black solid line, see Eq.~\eqref{eq_Nrg}). The average number of red giant/jet interactions at
    $z\le 0.04$~pc and 0.3~pc are plotted in green dot-dashed and blue-dashed lines, respectively.
    }
    \label{fig_Nrg}
\end{figure}

To estimate the number of stars crossing the jet per an orbital timescale, we first calculate the jet-crossing rate per a unit of time. Since we focus on the region $z\lesssim 0.5\,{\rm pc}$, which is well inside the radius of influence of Sgr~A*, $r_{\rm h}\sim 2\,{\rm pc}$ \citep{2013degn.book.....M}, we approximate the stellar velocity dispersion by the local Keplerian velocity, $\sigma_{\star}\sim v_{\rm orb}=\sqrt{GM_{\bullet}/z}$; see e.g. \citet{2014ApJ...786..121S}. With the jet cross-section $S_{\rm jet}\approx zR_{\rm j}\sim z^2\tan{\theta}$, the number of late-type stars entering both the jet and the counter-jet per a unit of time is 
\begin{equation}
    \dot{N}_{\rm RG}\approx 2 n_{\rm RG}\sigma_{\star}S_{\rm jet} \simeq 2 n_0 z_0^{\gamma}\sqrt{GM_{\bullet}}\tan{\theta} z^{\frac{3}{2}-\gamma}.
    \label{eq_dotNrg}
\end{equation}
The number of red giants crossing the jet per orbital timescale is

\begin{equation}
    N_{\rm RG}=\dot{N}_{\rm RG} P_{\rm orb}=
    4 \pi n_0 z_0^{\gamma} \tan{\theta} z^{3-\gamma}\,,
    \label{eq_Nrg}
\end{equation}
where the orbital period in the sphere of influence of the SMBH follows from the third Keplerian law, $P_{\rm orb}=2\pi z^{3/2}/\sqrt{GM_{\bullet}}$. In Fig.~\ref{fig_Nrg} we plot $N_{\rm RG}$.
The average number of crossing giants per orbital period in the region with an outer radius $z_{\rm out}$ is
\begin{equation}
    \overline{N}_{\rm RG}=\frac{4\pi}{4-\gamma}n_0z_0^{\gamma} \tan{\theta} z_{\rm out}^{3-\gamma}\,.
    \label{eq_average_Nrg}
\end{equation}
In particular, $\overline{N}_{\rm RG}\simeq 3.5$ and $\overline{N}_{\rm RG}\simeq 82$ when
$z_{\rm out}=0.04\,{\rm pc}$ (S cluster) and $z_{\rm out}=0.3\,{\rm pc}$, respectively (see Fig.~\ref{fig_Nrg}).
%

In this case, $\overline{N}_{\rm RG}$ represents all late-type stars that cross the jet sheath per orbital period on average. The fraction of giants whose envelopes could have been stripped off by the jet can be estimated by comparing the radii of stars with the corresponding stagnation radius at a certain distance from Sgr~A*. The basic condition for the ablation is that $R_{\star}\gtrsim R_{\rm stag}$ at a given $z$ from Sgr~A*. In particular,  for $z=0.04\,{\rm pc}$ and the jet luminosity of $L_{\rm j}=10^{42}\,{\rm erg\,s^{-1}}$, the minimum stellar parameters for ablation are  $R_{\star}=27\,R_{\odot}$, $\mu_{\rm c}=0.3$, $L_{\star}=129\,L_{\odot}$, $T_{\star}=3743\,{\rm K}$, $m_{\rm abl}=11.7$ mag, where $m_{\rm abl}$ denotes the upper magnitude limit, below which stars are expected to be affected by the jet. Using the K-band luminosity function approximated by the power law, $\mathrm{d}\log{N}/\mathrm{d}m_{\rm K}=\beta$ with $\beta\simeq 0.3$ for late-type stars \citep{2009A&A...499..483B,2011ApJ...741..108P} between 12 and 18 mag, we can estimate the fraction of ablated stars as $\eta=N_{\rm abl}/N_{\rm tot}\times 100=10^{\beta(m_{\rm abl}-m_{\rm max})+2}\%$, where $m_{\rm max}$ is the limiting magnitude, which we set to 18 mag according to \citet{2011ApJ...741..108P}. Then for $z=0.04\,{\rm pc}$ and $L_{\rm j}=10^{42}\,{\rm erg\,s^{-1}}$ we get $\eta=1.27\%$. For the larger distance $z=0.5\,{\rm pc}$ and $L_{\rm j}=10^{42}\,{\rm erg\,s^{-1}}$, we obtain the minimum parameters of ablated stars as follows, $R_{\star}=338\,R_{\odot}$, $\mu_{\rm c}=0.6$, $L_{\star}=6081\,L_{\odot}$, $T_{\star}=2775\,{\rm K}$, $m_{\rm abl}=6.95$ mag with $\eta=0.05\%$. The limiting values and the percentage of ablated giants are quite sensitive to the jet luminosity. Increasing $L_{\rm j}$ to $10^{44}\,{\rm erg\,s^{-1}}$, we get $R_{\star}=2.7\,R_{\odot}$, $\mu_{\rm c}=0.16$, $L_{\star}=3.97\,L_{\odot}$, $T_{\star}=4960\,{\rm K}$, $m_{\rm abl}=16.1$ mag with $\eta=26.5\%$ for $z=0.04\,{\rm pc}$ and $R_{\star}=33.8\,R_{\odot}$, $\mu_{\rm c}=0.31$, $L_{\star}=181\,L_{\odot}$, $T_{\star}=3645\,{\rm K}$, $m_{\rm abl}=11.3$ mag with $\eta=0.95\%$ for $z=0.5\,{\rm pc}$. We summarize the relevant values in Table~\ref{tab_abl_fraction}. Although the fraction of affected late-type stars is small, it significantly affects brighter stars with smaller magnitudes -- stars brighter than 14 mag constitute $\sim 6.3\%$ of the total observed sample and stars brighter than 12 mag constitute only $1.6\%$. We explicitly show the change in the projected brightness distribution for brighter stars in Section~\ref{sec_surface_brightness}.

\begin{table*}[]
    \centering
    \caption{Limiting minimal stellar radii, maximum apparent K-band magnitudes (dereddened), and the fraction of ablated giants for two distances from Sgr~A* (0.04 pc and 0.5 pc) and two luminosities of its jet ($10^{42}\,{\rm erg\,s^{-1}}$ and $10^{44}\,{\rm erg\,s^{-1}}$).}
    \begin{tabular}{c|c|c}
    \hline
    \hline
    Distance & $L_{\rm j}=10^{42}\,{\rm erg\,s^{-1}}$ & $L_{\rm j}=10^{44}\,{\rm erg\,s^{-1}}$\\
    \hline
    $0.04\,{\rm pc}$     & $R_{\star}= 27\,R_{\odot}$, 
    $m_{\rm abl}=11.7$, $\eta=1.27\%$  & $R_{\star}= 2.7\,R_{\odot}$, $m_{\rm abl}=16.1$, $\eta=26.5\%$\\
    $0.5\,{\rm pc}$ &  $R_{\star}= 338\,R_{\odot}$, $m_{\rm abl}=6.95$, $\eta=0.05\%$    &  $R_{\star}= 33.8\,R_{\odot}$, $m_{\rm abl}=11.3$, $\eta=0.95\%$\\
    \hline 
    \end{tabular}
    \label{tab_abl_fraction}
\end{table*}

The jet-ablation could partially have contributed to the inferred 4-5 missing late-type giants in the region with $z_{\rm out}=0.04\,{\rm pc}$ \citep{2019ApJ...872L..15H} as well as to the $\sim 100$ missing late-type giants in the larger region with $z_{\rm out}=0.3\,{\rm pc}$ \citep{2018A&A...609A..26G}, especially for higher luminosities of the jet.

\begin{figure}
    \centering
    \includegraphics[width=\columnwidth]{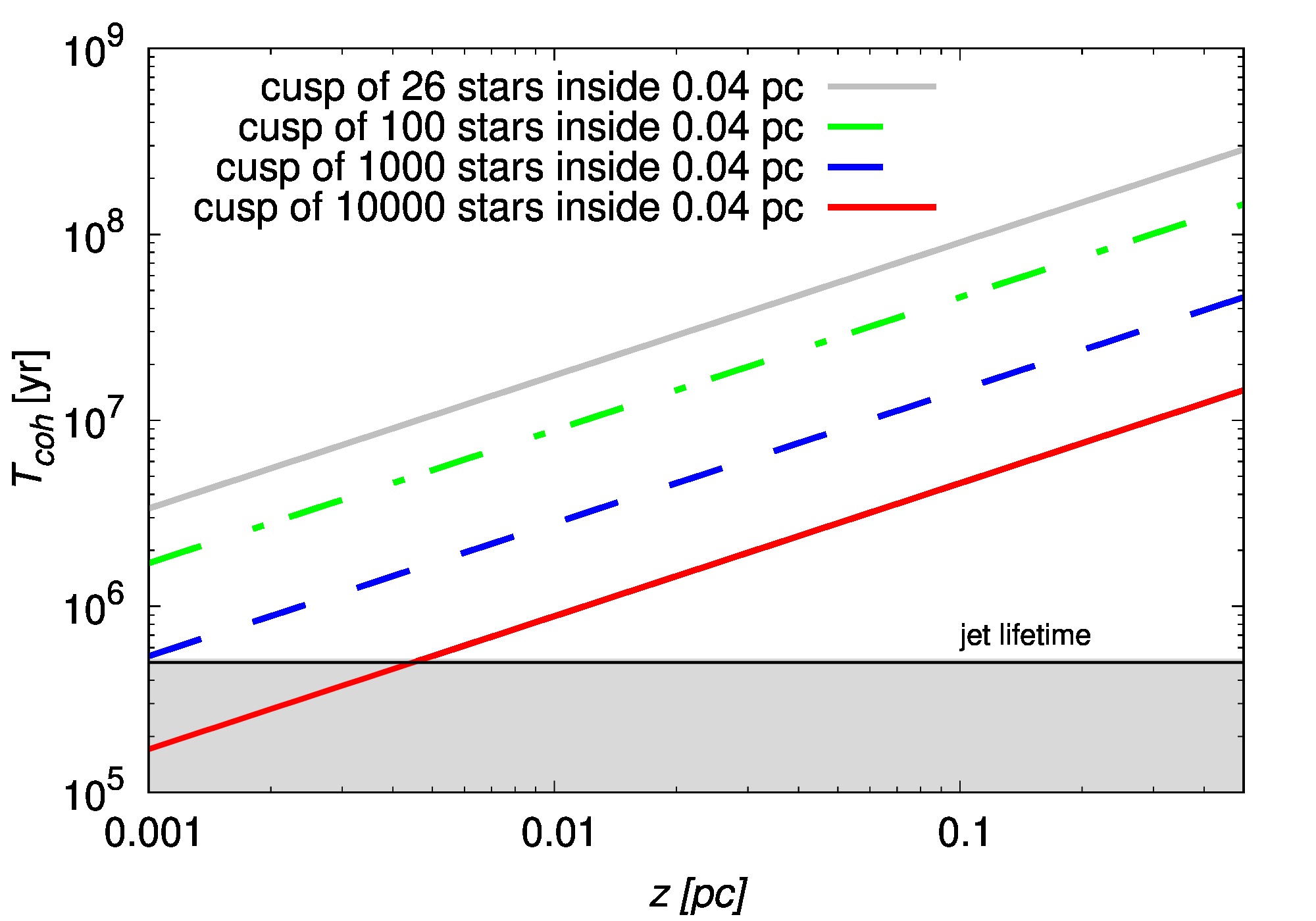}
    \caption{The coherence timescale $T_{\rm coh}$ for different number of stars in the S cluster according to the legend. The shaded area stands for the expected lifetime of the jet during the previous Seyfert-like activity of Sgr~A*.}
    \label{fig_coherence_timescale}
\end{figure}

The number of stars that can interact with the jet is increased via dynamical processes in the dense nuclear star cluster.
In particular,  the vector resonant relaxation (VRR)
changes the direction of the orbital angular momentum
\citep{2005PhR...419...65A,2013degn.book.....M}, and therefore
 stars that were not passing through the jet can do so on the resonant-relaxation timescale. More precisely, in the sphere of influence of the SMBH, stars move on Keplerian ellipses and the gravitational interactions between stars are correlated. Given the finite number of stars, there is a non-zero torque on a test star. During the time interval $\Delta t$, for which $P_{\rm orb}\ll \Delta t \ll T_{\rm coh}$ and  the coherence timescale
\begin{equation}
    T_{\rm coh}=\frac{P_{\rm orb}}{2\pi\sqrt{N_{\star}}} \frac{M_{\bullet}}{m_{\star}}
    = \sqrt{\frac{3-\gamma}{4\pi G n_0 z_0^{\gamma}}} \frac{M_{\bullet}^{\frac{1}{2}}}{m_{\star}}z^{\frac{\gamma}{2}}
    \label{eq_coherence_timescale}
\end{equation}
is inversely proportional to the square root of the number of enclosed stars, the angular momentum of a test star changes linearly with time.

The inclination of stellar orbits would change by $\sim \pi$ only when $T_{\rm coh}(z)\lesssim t_{\rm j}\sim 0.1-0.5\,{\rm Myr}$
and hence essentially all late-type stars could interact with the jet during its lifetime. The estimated number of enclosed stars in the S cluster ($z_{\rm out}\sim 0.04\,{\rm pc}$) is $26<N_{\star}<10\,000$, where the lower limit considers only late-type stars according to the analysis by \citet{2019ApJ...872L..15H} and the upper limit stands for all the stars including compact remnants. The upper limit is supposed to be closer to the actual number of stellar objects since the number of old neutron stars and stellar black holes in the central arcsecond could be of that order of magnitude \citep{1993ApJ...408..496M,2007MNRAS.377..897D,2018ApJS..235...26Z}. The total number of massive objects naturally affects the coherence timescale by more than an order of magnitude. In Fig.~\ref{fig_coherence_timescale} we plot $T_{\rm coh}$. We see that when $N_{\star}\sim 1000$ and more, $T_{\rm coh}$ is comparable to the lifetime of the jet in the inner parts of the S cluster. In summary, the coherent resonant relaxation makes the number of affected giants bigger and the estimates per orbital timescale
$\overline{N}_{\rm RG}$ can be considered as a lower limit. Another more hypothetical effect that can enlarge the number of affected giants is the jet precession (see Section~\ref{subsec_precession}).

The vector resonant relaxation can affect the number of encounters, $n_{\rm cross}$, see Eq.~\eqref{eq_ncross}. In case $T_{\rm coh}>t_{\rm jet}$, i.e. for a smaller number of enclosed objects ($N_{\star}\lesssim 100$), $n_{\rm cross}$ is still mainly determined by $t_{\rm jet}$, see Eq.~\eqref{eq_ncross}. However, then the mean number of interacting giants $\overline{N}_{\rm RG}$ is also not significantly enlarged. On the other hand, if $T_{\rm coh}\lesssim t_{\rm jet}$, then $n_{\rm cross}$ is reduced approximately by a factor of $2\theta/\pi$, which assumes that the angular momentum vector shifts linearly with time during $T_{\rm coh}$. In that sense, the interaction timescale with the jet is $t_{\rm int}\sim T_{\rm coh}(2\theta/\pi)$. Considering $T_{\rm coh}\sim t_{\rm jet}$, the number of crossings is
\begin{equation}
    n_{\rm cross}^{\rm RR}  \sim 2800 \left(\frac{T_{\rm coh}}{0.5\,{\rm Myr}}\right)\left(\frac{M_{\bullet}}{4\times10^6 \,M_{\odot}}\right)^\frac{1}{2}\left(\frac{z}{0.01\,{\rm pc}}\right)^{-\frac{3}{2}}\,,\label{eq_ncross_rr}
\end{equation}
which is smaller by an order of magnitude in comparison with $n_{\rm cross}$.


\section{Effect of jet-ablation on surface brightness profile of NSC}
\label{sec_surface_brightness}

\begin{figure*}
    \centering
    \includegraphics[width=\columnwidth]{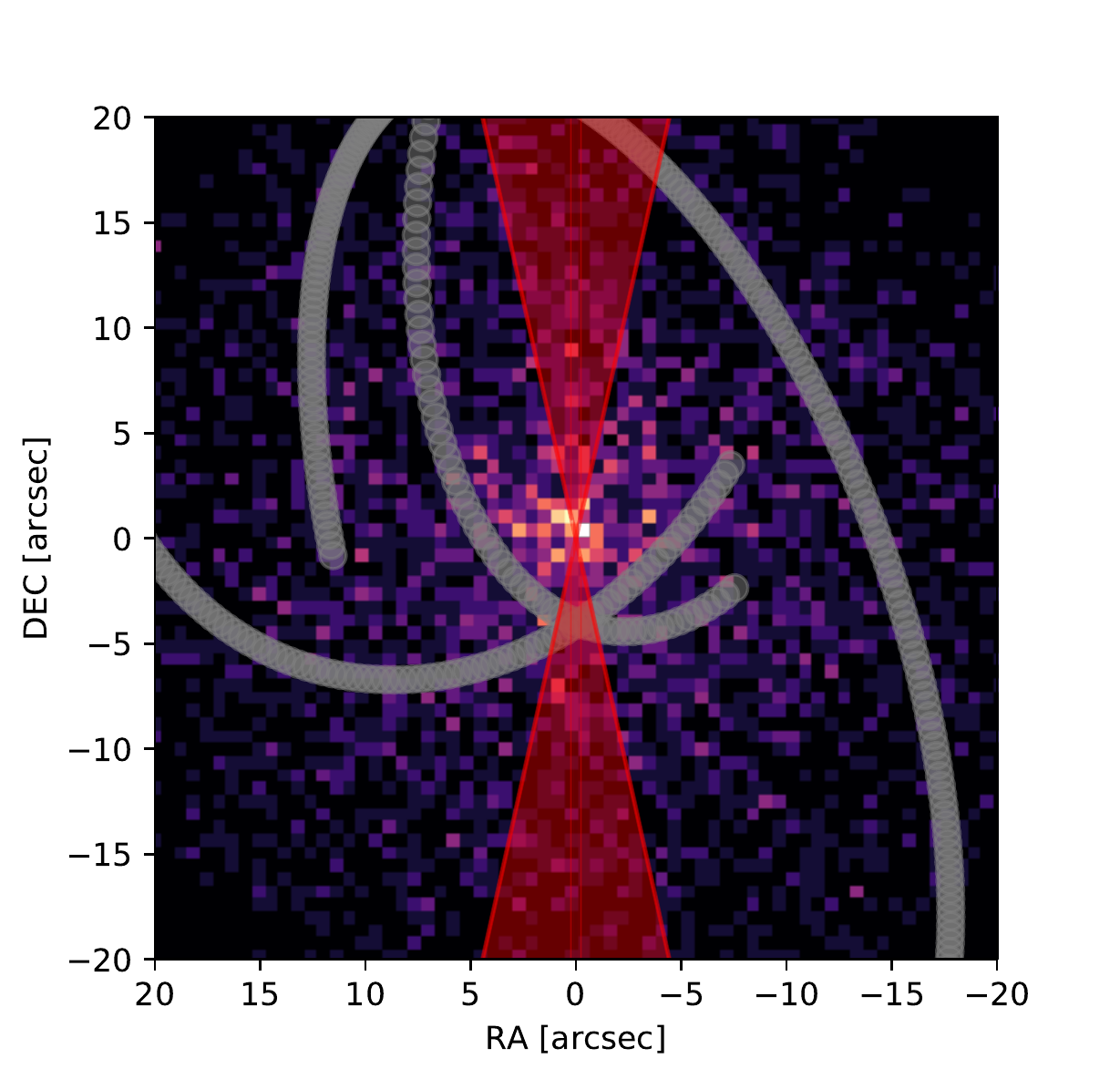}
    \includegraphics[width=\columnwidth]{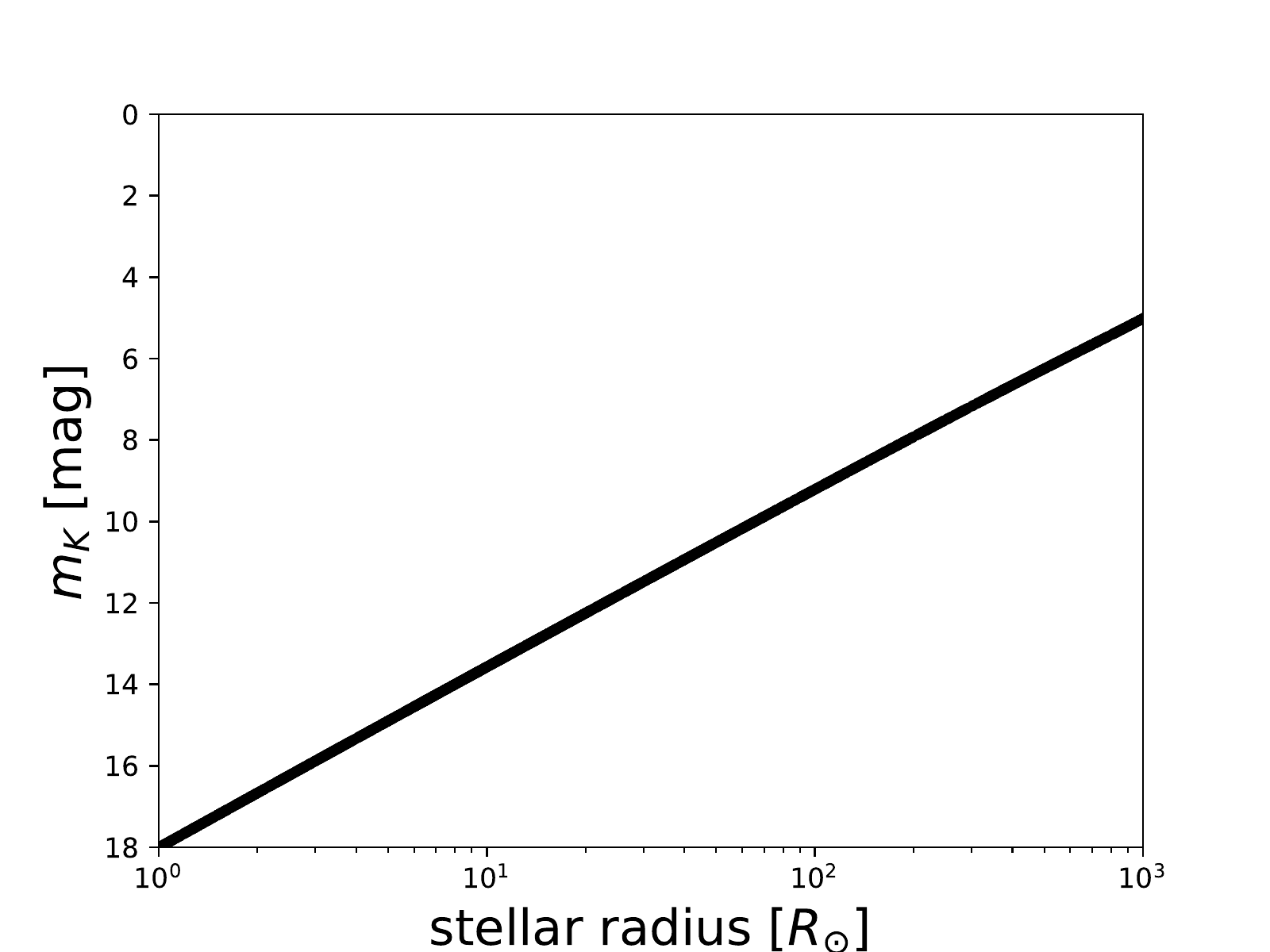}
    \caption{Initial properties of the Monte-Carlo generated mock NSC. {\bf Left panel:} The projected surface density distribution of 4000 stars with the illustrated active jet with the half-opening angle of $\theta=12.5^{\circ}$. The gray streamers illustrate the Minispiral arms according to the Keplerian model of \citet{2009ApJ...699..186Z}. {\bf Right panel:} The K-band magnitude (dereddened)--stellar radius relation for our generated NSC.}
    \label{fig_mk_radius_nsc}
\end{figure*}

To assess the observational signatures of the jet-ablation of late-type stars, we generate a mock spherical cluster of stars. Their initial spatial distribution follows $n_{\rm RG}=n_0(z/z_0)^{-\gamma}$ with $n_0=52\,{\rm pc^{-3}}$, $z_0=4.9\,{\rm pc}$, and $\gamma\sim 1.43$ \citep{2018A&A...609A..26G}. This spatial profile suggests that there are in total $\sim 4000$ late-type stars inside the inner one parsec, which we generate using the Monte Carlo approach to form a mock Nuclear Star Cluster (NSC), see Fig.~\ref{fig_mk_radius_nsc} (left panel) for illustration.

Each star is assigned its mass in the range from $0.08\,M_{\odot}$ to $100\,M_{\odot}$ following the Initial Mass Function (IMF) according to \citet{2001MNRAS.322..231K}, i.e.
\begin{equation}
    \zeta(m_{\star})=m_{\star}^{-\alpha}\,,\text{where}
    \begin{cases}
    \alpha=0.03\,, m_{\star}<0.08\,M_{\odot}\,,\\
    \alpha=1.3\,, 0.08\,M_{\odot}<m_{\star}<0.5\,M_{\odot}\,,\\
     \alpha=2.3\,, m_{\star}>0.5\,M_{\odot}\,.
    \end{cases}
\end{equation}
The Chabrier/Kroupa IMF is a good approximation for the observed mass distribution of the late-type NSC population \citep{2011ApJ...741..108P}.

In the next step, we assigned the core mass to each star of the mock cluster. Here we fix the ratio between the core mass and the stellar mass to $\mu_{\rm c}/m_{\star}=0.4$, which is in between the value inferred from the Sch\"onberg-Chandrasekhar limit\footnote{The Sch\"onberg-Chandrasekhar limit expresses the ratio between the isothermal core mass and the stellar mass, $m_{\rm ic}/m_{\star}\simeq 0.37(\mu_{\rm env}/\mu_{\rm ic})^2\sim 0.1$, where $\mu_{\rm env}$ and $\mu_{\rm ic}$ are mean molecular weights for the envelope and the isothermal core, respectively.} and the final phases of the stellar evolution, where the white-dwarf core constitutes most of the mass for the Solar-type stars. For more precise simulations, core masses from the stellar evolution of the NSC should be adopted, however, here we are interested in the first-order effects of the jet activity on the surface brightness distributions.

\begin{figure*}
    \centering
    \includegraphics[width=\columnwidth]{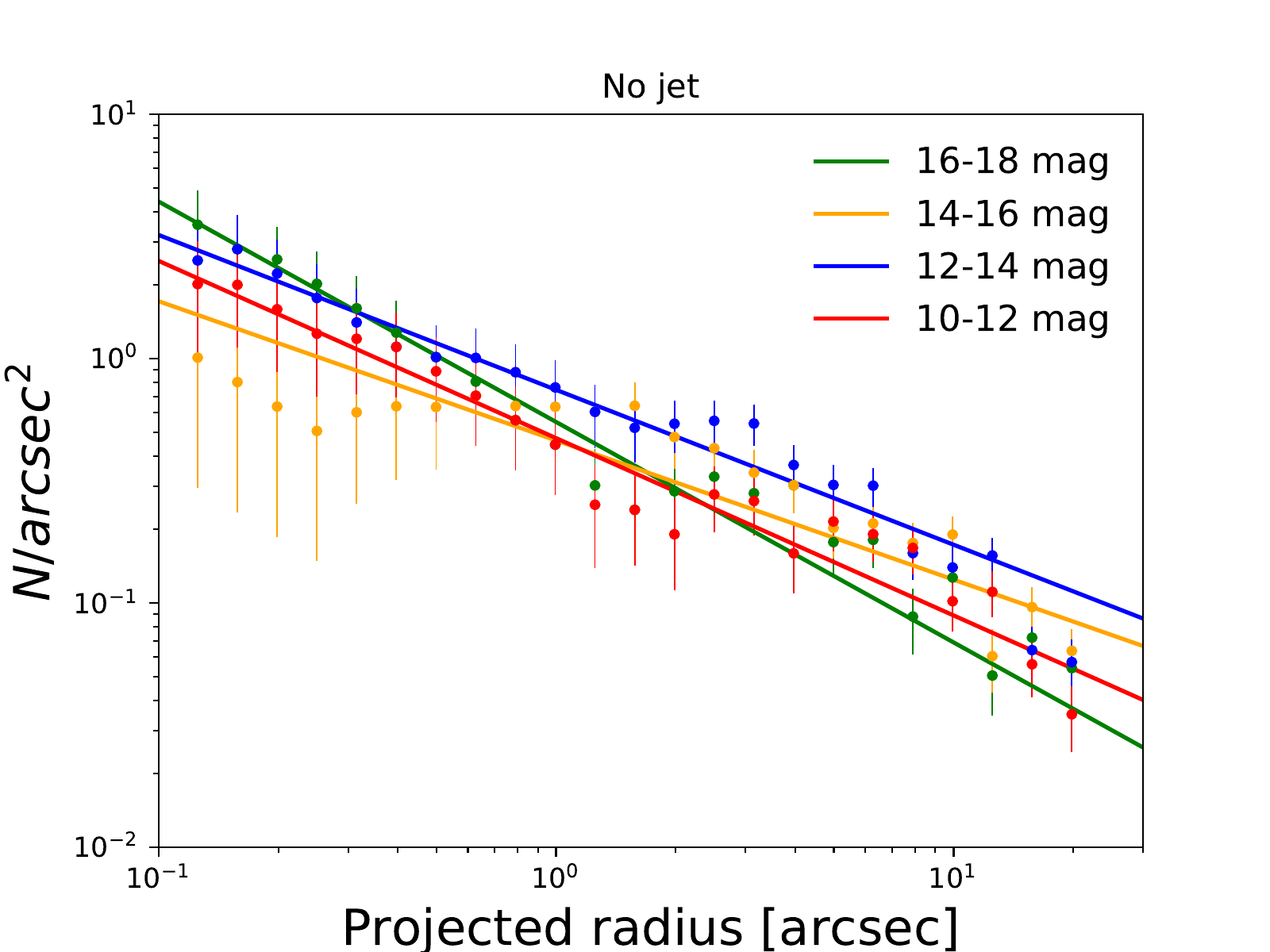}
    \includegraphics[width=\columnwidth]{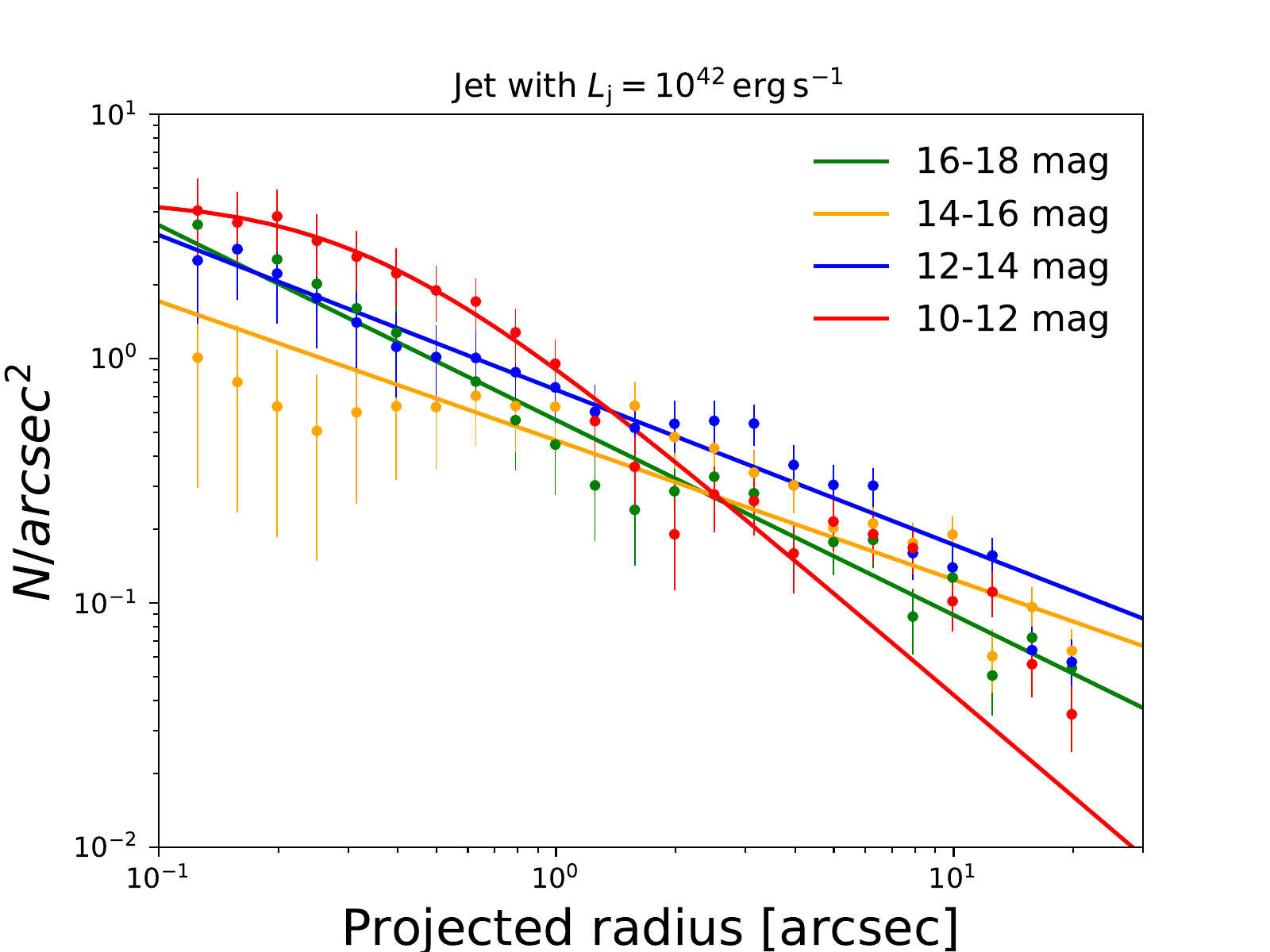}
    \includegraphics[width=\columnwidth]{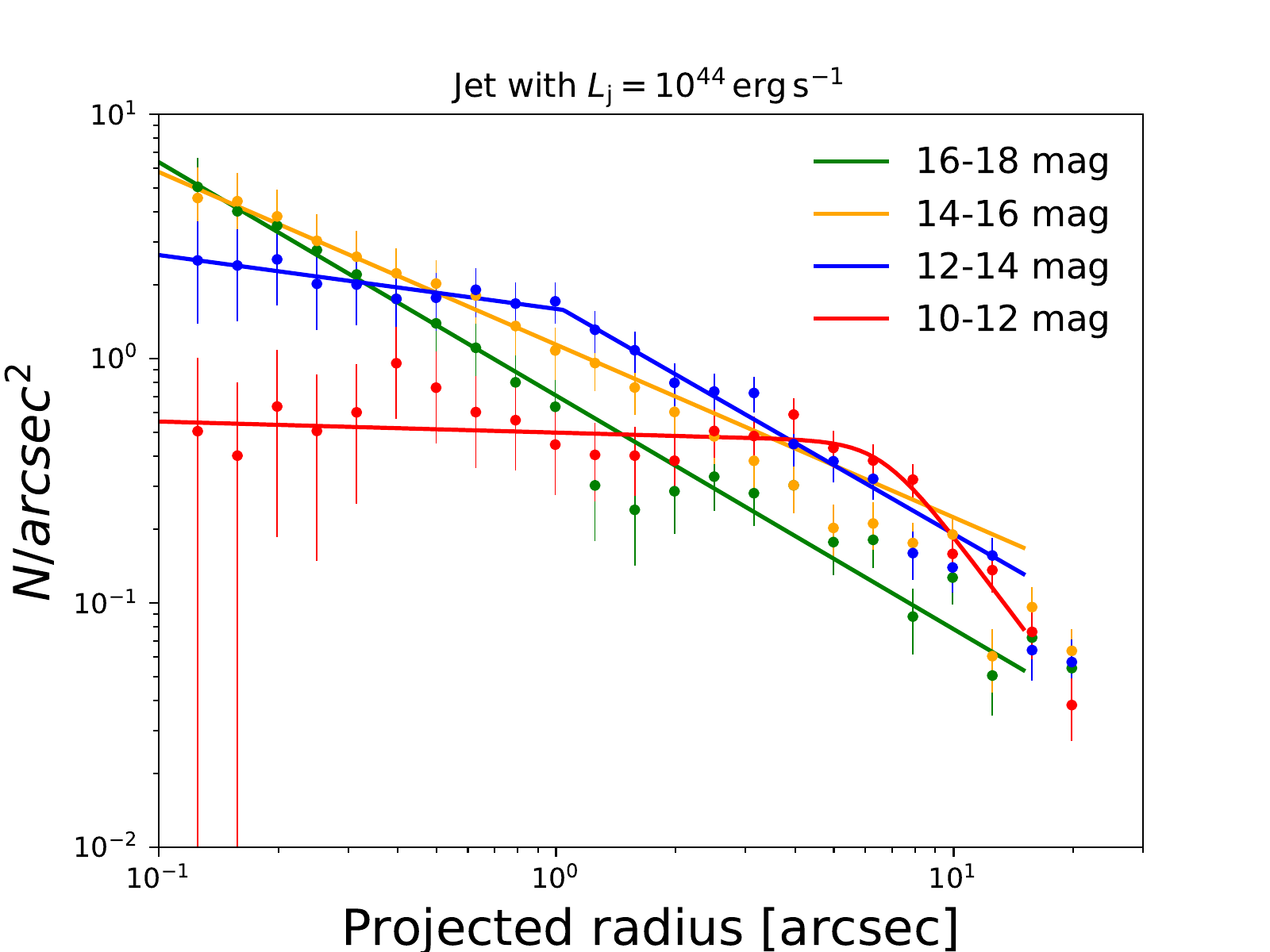}
    \caption{Potential effect of the jet activity on the surface distribution of the initially cuspy late-type NSC. The calculations assume that the coherent resonant relaxation timescale is comparable or shorter than the jet lifetime ($\sim 0.5\,{\rm Myr})$ within the S cluster region. In all the panels, the magnitude bins are dereddened. {\bf Top left panel:} The projected profile of an initial cusp of late-type stars with the surface slope of $\Gamma\sim 0.6-0.9$ across all magnitude bins (two-magnitude bins). {\bf Top right panel: } A modified projected profile for the jet luminosity of $10^{42}\,{\rm erg\,s^{-1}}$. The 10-12 mag surface profile flattens inside the inner arcsecond, while the fainter stars (12-18 mag) keep a cusp-like profile. {\bf Bottom panel:} A modified surface profile for the jet luminosity of $L_{\rm j}=10^{44}\,{\rm erg\,s^{-1}}$. The profile of the brightest stars (10-12 mag) flattens even more and decreases inside the inner arcsecond. The stars with 12-14 mag exhibit a flat profile as well in the inner arcsecond. The stars in the 14-16 mag and 16-18 mag bins keep the cuspy profile. For all three panels, the solid lines represent the single and the broken power-law function fits to the surface stellar distribution in four magnitude bins. The slopes and the break radii are listed in Table~\ref{tab_power_law}.}
    \label{fig_surface_distribution}
\end{figure*}

To construct the surface brightness profiles of the late-type population after the active jet phase in different magnitude bins, we followed these steps:
\begin{enumerate}
    \item We calculated $L_{\star}(\mu_{\rm c})$ and $R_{\star}(\mu_{\rm c})$ using Eq.~\ref{eq_luminosity_radius}.
    \item If the jet was set active with a certain luminosity $L_{\rm j}$, we compared $R_{\star}$ and $R_{\rm stag}$ for a given distance $z$ of the star. If $R_{\star}\geq R_{\rm stag}$, then we set $R_{\star}=R_{\rm stag}$. In this case we also implicitly assumed that at a given distance, all of the stars, for which $R_{\star}\geq R_{\rm stag}$, are eventually ablated by the jet, hence the resonant relaxation was assumed to be efficient and hence $T_{\rm coh}<t_{\rm jet}$. 
    \item We estimated the effective temperature of a star using $T_{\star}=T_{\odot}(L_{\star}/L_{\odot})^{1/4}(R_{\star}/R_{\odot})^{-1/2}$.
    \item From the Planck function $B_{\nu}(T_{\star})$ we calculated the monochromatic flux in the $K_{\rm s}$-band ($2.2\,\mu{\rm m}$) $F_{\nu}(R_{\star},T_{\star})=(R_{\star}/d_{\rm GC})^2\pi B_{\nu}(T_{\star})$ and the corresponding apparent magnitude $m_{\rm K}$ (dereddened). 
\end{enumerate}

The initial relation between the $K_{\rm s}$-band magnitude and the stellar radius is shown in Fig.~\ref{fig_mk_radius_nsc} in the right panel. We calculate the projected stellar density using the concentric annuli with the mean radius $R$ and the width of $\Delta R$, $\sigma_{\star}=N_{\star}/(2\pi R\Delta R)$, where $N_{\star}$ is a number of stars in an annulus. We estimate the uncertainty of the stellar number counts as $\sigma_{\rm N}\approx \sqrt{N_{\star}}$. Subsequently, we construct the surface stellar profiles in two-magnitude bins starting at $m_{\rm K}=18$ mag up to $m_{\rm K}=10$ mag, i.e. in total four bins; see Fig.~\ref{fig_surface_distribution}. In the top left panel of Fig.~\ref{fig_surface_distribution}, we plot the nominal projected distribution without considering the effect of jet. The brightness profile for all four magnitude bins can be approximated by simple power-law functions, $N(R)=N_0(R/R_0)^{-\Gamma}$, whose slopes are listed in Table~\ref{tab_power_law}. Hence, the initial cluster distribution is cusp-like. In the top right panel, we show the case with an active jet with the luminosity of $L_{\rm j}=10^{42}\,{\rm erg\,s^{-1}}$. We see that the profile for the brightest stars in the 10-12 mag bin becomes flat in the inner arcsecond and can be described as a broken power-law function, $N(R)=N_0(R/R_{\rm br})^{-\Gamma}[1+(R/R_{\rm br})^{\Delta}]^{(\Gamma-\Gamma_0)/\Delta}$, where $R_{\rm br}$ is a break radius, $\Gamma$ is a slope of the inner part, $\Gamma_0$ marks the slope of an outer part, and $\Delta$ denotes the sharpness of transition. The larger surface brightness values for this magnitude bin may be interpreted by an extra input of ablated giants with the initial magnitudes $m_{\rm K}<10$ mag that after the ablation fall into bins with a larger magnitude. Finally, we increase the jet luminosity to $L_{\rm j}=10^{44}\,{\rm erg\,s^{-1}}$, which makes the flattening of the brightest giants even more profound. The stars in the 12-14 mag bin also exhibit a flatter profile inside the inner arcsecond for this case, which shows the significance of the jet luminosity in affecting the observed surface profile of the NSC. We list the power-law slopes for both a simple and a broken power-law function and the break radii, where available, for all the magnitude bins and the three jet-activity cases in Table~\ref{tab_power_law}.

Fig.~\ref{fig_surface_distribution} demonstrates a potential signature of the jet activity on the surface profile of the NSC. It is important to study differential profiles, i.e. the distribution in different magnitude bins, since the lower-luminosity jet starts affecting the profile of bright stars (smaller magnitudes), while with an increasing jet luminosity, the fainter stars become affected as well, starting at smaller projected radii ($<1''$). Our Monte-Carlo simulation suggests that the active jet phase with $L_{\rm j}\lesssim 10^{44}\,{\rm erg\,s^{-1}}$ likely affected the late-type stars with $m_{\rm K} \leq 14$ mag that exhibit the flat profile inside the inner arcsecond. The fainter stars of $m_{\rm K}>14$ mag can still keep a cusp-like profile after the jet ceased its enhanced activity.

For better quantitative comparisons with observations, it is necessary to include the stars of different ages and hence different core masses. This is rather complex as there were recurrent star-formation episodes in the NSC, with $80\%$ of the stellar mass being formed 5 Gyr ago, the minimum in the star-formation rate close to 1 Gyr, and the renewed star-formation in the last 100-200 million years, although with the ten-times lower star-formation rate than at earlier episodes \citep{2011ApJ...741..108P}. These findings were confirmed by \citet{2020arXiv200715950S}, who estimate that 80$\%$ of stars formed 10 Gyr ago or earlier, then about 15$\%$ formed 3 Gyr ago, and the remaining fraction in the last 100 Myr. Furthermore, the dynamical effects such as the mass segregation and the relaxation processes could also have played a role in shaping the final observed profile of the NSC, and in addition we might expect other bright-giant depletion processes \citep{2005PhR...419...65A}. Despite these difficulties in comparing theoretical and observed profiles, the basic trend shown in Fig.~\ref{fig_surface_distribution}, in particular for $L_{\rm j}=10^{44}\,{\rm erg\,s^{-1}}$ (bottom panel), is consistent with the observational findings of \citet{2020arXiv200715950S}, who found cusp-like profiles for all magnitude bins apart from $m_{\rm K}=14.5-14$ mag bin (including foreground field extinction), which shows a flat/decreasing profile in their analysis. In our panels in Fig.~\ref{fig_surface_distribution}, this corresponds to the dereddened bins 10-12 and 12-14 mag, whose profiles become affected for $L_{\rm j}=10^{42}\,{\rm erg\,s^{-1}}$ starting from the projected radii below one arcsecond. However, \citet{2020arXiv200715950S} also note that precise surface profiles for late-type stars are difficult to construct due to the contamination at all magnitude bins by an unrelaxed population of young stars. \citet{2019ApJ...872L..15H} report a cusp-like profile for late-type stars of $m_{\rm K}<17$ mag (including foreground extinction), which approximately corresponds to our bin of 14-16 mag (yellow points) that maintains the cusp-like profile even for the largest jet luminosity (bottom panel). In conclusion, the expected trend of preferential depletion of bright late-type stars by the jet is confirmed.

\begin{table*}[tbh]
    \centering
     \caption{Summary of the power-law slopes (for both single and broken if relevant) for the four magnitude bins (two-magnitude intervals starting at 18 mag up 10 mag in the near-infrared $K_{\rm s}$-band; magnitudes are dereddened) and three jet-activity cases (no jet, $L_{\rm j}=10^{42}\,{\rm erg\,s^{-1}}$, $L_{\rm j}=10^{44}\,{\rm erg\,s^{-1}}$).}
    \begin{tabular}{c|c|c|c}
    \hline
    \hline
    Magnitude bin & No jet & Jet $L_{\rm j}=10^{42}\,{\rm erg\,s^{-1}}$ & Jet $L_{\rm j}=10^{44}\,{\rm erg\,s^{-1}}$\\
    \hline
    18-16 mag  & single: $\Gamma=0.9$   & single: $\Gamma=0.8$  & single: $\Gamma=1.0$   \\
    16-14 mag  & single: $\Gamma=0.6$  & single: $\Gamma=0.6$   & single: $\Gamma=0.7$   \\
    14-12 mag  & single: $\Gamma=0.6$   & single: $\Gamma=0.6$ & broken: $\Gamma=0.2$, $\Gamma_0=0.9$, $R_{\rm br}=1''$   \\
    12-10 mag  & single: $\Gamma=0.7$  & broken: $\Gamma=-0.08$, $\Gamma_0=1.4$, $R_{\rm br}=0.3''$   & broken: $\Gamma=0.04$, $\Gamma_0=2.2$, $R_{\rm br}=6.6''$  \\
    \hline
    \end{tabular}
    \label{tab_power_law}
\end{table*}

\section{Discussion}

We investigated the effects of a jet during an active phase of Sgr~A* in the last million years on the appearance of late-type giant stars with atmosphere radii more than 30$R_{\odot}$. We found that especially in the innermost arcsecond of the Galactic center (the S-cluster), the upper layers of the stellar envelope could be removed by the jet ram pressure. Hence, the jet-red giant interactions during the active phase of Sgr~A* could have contributed to the depletion of bright late-type stars. In other words, the atmosphere ablation by the jet would alter the red giant appearance in a way that would make them look bluer and fainter in the near-infrared bands (mainly K' and L bands), in which stars in the Galactic center region are generally monitored.
In the following, we outline several additional effects that could be associated with the jet/RG interaction.

\label{sec_discussion}

\subsection{Enlarging the number of affected stars by the jet precession}
\label{subsec_precession}
Jet precession is a phenomenon that accompanies the launching of jets during the evolution of galaxies and stellar binaries. It is caused by perturbations due to the misalignment of the accretion flow and the black hole spin, so-called Lense-Thirring precession, or by a secondary black hole. The jet precession was proposed to explain a long-term flux variability in radio galaxies, e.g. OJ 287 \citep{2018MNRAS.478.3199B}, 3C84 \citep{2019Galax...7...72B}, 3C279 \citep{1998ApJ...496..172A}, the neutrino emission from TXS 0506+056 \citep{2019A&A...630A.103B}, as well as in X-ray binaries \citep{2015A&A...574A.143M,2019Natur.569..374M}.  

For the Galactic center, the Lense-Thirring precession of the hot thick accretion flow was analyzed by \citet{2013MNRAS.432.2252D} in relation to the near-infrared and mm variability of Sgr~A*. This effect would also translate to the precession of the jet under the assumption it is coupled to the disc via the launching mechanism \citep[Blandford--Payne mechanism; see][]{1982MNRAS.199..883B}. The jet precession is also suggested by wide UV ionization cones with the opening angle of $60^{\circ}$ \citep{2019ApJ...886...45B}, which is larger by a factor of a few expected for the jet opening angle of $\sim 25^{\circ}$ \citep{2013ApJ...779..154L}. 

We estimate the factor by which the volume of the affected red giants is enlarged. We adopt the jet precession half-opening angle of $\Omega_{\rm p}\approx 30^{\circ}$ based on the scale of UV ionization cones \citep{2019ApJ...886...45B}. During the precession motion, the jet circumscribes a cone with the radius $R_{\rm p}=z \sin{\Omega_{\rm p}}$. The factor by which the volume at given distance $z$ enlarges is given by

\begin{equation}
    f_{\rm prec}=\frac{2\pi R_{\rm p}}{2R_{\rm j}}=\pi \frac{\sin{\Omega_{\rm p}}}{\tan{\theta}} \sim 7.1\,,
    \label{eq_factor_precession}
\end{equation}
when $\theta=12.5^{\circ}$ and $\Omega_{\rm p}=30^{\circ}$.
The number of affected late-type stars would then increase by the same factor to $N_{\rm j}\approx 105$ and 
$4.4$ within $0.3$ and $0.04\,{\rm pc}$, respectively,
which is comparable to the number of missing bright red giants at these scales -- 100 at $z<0.3\,{\rm pc}$ \citep{2018A&A...609A..26G} and 4 at $z<0.04\,{\rm pc}$ \citep{2019ApJ...872L..15H}. The factor derived in Eq.~\eqref{eq_factor_precession} should be treated as an upper limit on the volume enlargement since it assumes that the precession period is comparable to or less than the jet lifetime, but it could also be longer. 
On the other hand, larger volume means that the stars are affected correspondingly less (over a shorter period of time) by the jet action, which is spread in different directions with the precession duty cycle.


\subsection{High-energy particle acceleration and jet mass-loading due to jet/star interactions}

The detection of the Fermi bubbles in the GeV domain indicates the presence of relativistic particles emitting gamma-rays. \cite{2012ApJ...756..181G} considered that particles can be accelerated in the jet launching region or in the jet termination shocks. Note however that the coexistence of the jet with the dense NSC in the Galactic center makes jet/star interactions very likely. 
The NSC is composed of both early- and late-type stars. 
In the former case, the powerful wind of OB and Wolf-Rayet stars makes the stagnation distance $R_{\rm stag}>> R_{\star}$ and therefore a double bow-shock structure is formed \citep{2013MNRAS.436.3626A}.
In the latter case, the slow winds of low-mass stars cannot create a big bow shock around the stars, but a shock in the jet will be formed anyway. In both cases, particles can be accelerated through the Fermi I acceleration mechanism in the bow shocks \citep{1978MNRAS.182..147B}. Even when the interaction with massive stars is a better scenario to accelerate particles up to highest energies 
(given that the size of the acceleration region is $\sim R_{\rm stag}$), acceleration of particles up to GeV-energies is not difficult to achieve.

\cite{Barkov_10,Barkov_12} consider the interaction of AGN jets 
with red giant stars to explain the TeV emission in radiogalaxies and blazars. The mass stripped from the red giant forms clouds moving in the jet direction, see also our Fig.~\ref{fig_jet_star}. Particles are accelerated in the bow shock around the cloud formed by the pressure exerted by the jet from below. Another shock propagates into the cloud and as a consequence, it will heat up and expand. After a certain time, there will be a population of relativistic particles in the jet as well as a chemical enrichment by stellar envelopes
\citep{2017A&A...606A..40P}. These effects were previously not taken into account in the jet models of the Fermi bubbles.

\subsection{Chemically peculiar stars as remnant cores of ablated red giants}
\label{subsec_chemically_peculiar}

The past ablation of red giants by the jet would contribute to the apparent lack of late-type stars in the central region of the Galactic center. Another potential imprint of the past jet-red giant interaction would be the presence of chemically peculiar, high metallicity stars in the NSC. This can be predicted from the fact that as the jet ram pressure removes upper hydrogen- and helium-rich parts of the stellar atmosphere, the lower metal-rich parts as well as the denser core are exposed. In fact, two late-type stars at $\sim 0.5\,{\rm pc}$ from Sgr~A* were reported as having a super-solar metallicity \citep{2018ApJ...855L...5D} with an anomalous abundance of scandium, vanadium, and yttrium. A detailed modelling of the stellar evolution in combination with the treatment of the jet-star interaction is needed to confirm or exclude the previous interaction with the jet for these and similar candidate stars with super-solar metallicities.   

\subsection{A collimated jet or a broad-angle disc wind?}

In the current analysis, we took into account mainly highly collimated nuclear outflow -- jet -- with the small half-opening angle close to $\sim 10^{\circ}$. The analysis of the UV ionization cones by \citet{2019ApJ...886...45B} indicates a half-opening angle of $\sim 30^{\circ}$, which could be a signature of the jet precession as we discussed in Sec.~\ref{subsec_precession} or alternatively disc winds with a larger opening angle. In case the jet would be absent and the disc wind would be present with the larger half opening angle, the expected stagnation radius would be proportionally larger since $R_{\rm stag}\propto \tan{\theta}$. Assuming the same kinetic luminosity $L_{\rm j}=10^{42}\,{\rm erg\,s^{-1}}$ and the outflow velocity close to $c$ (ultrafast outflows), the ratio of the stagnation radii is $R_{\rm stag}^{\rm wind}/R_{\rm stag}^{\rm jet}=\tan{\theta_{\rm wind}}/\tan{\theta_{\rm jet}}\approx 2.6$, which yields for $R_{\rm stag}^{\rm wind}\sim 70\,R_{\odot}$ using Eq.~\eqref{eq_stag_radius_values}. The ablation effect would still take place but only for the largest red giants with $R_{\star}\gtrsim 100\,R_{\odot}$ at the outer radius of the S cluster. The stagnation radius of $30\,R_{\odot}$ would be reached at the distance of $z\sim 0.017\,{\rm pc}$ for the same stellar parameters as we assumed in Eq.~\eqref{eq_stag_radius_values} and a larger outflow half-opening angle of $30^{\circ}$. 

\subsection{Recurrent Seyfert-like activity and TDEs}

The X-ray/$\gamma$-ray Fermi bubbles were created during the increased Seyfert-like activity about $3.5 \pm 1$ Myr ago \citep{2019ApJ...886...45B} with total duration of $\sim 0.1-0.5$ Myr \citep{2012ApJ...756..181G}. Currently, it is unclear whether this increased activity is related to the star-formation event that led to the formation of massive OB/Wolf-Rayet stars a few million years ago, a fraction of which forms a stellar disc that is a remnant of a former massive gaseous disk \citep{2003ApJ...590L..33L}. In case the correlation between the episodic star-formation and the accretion activity exists, the Seyfert-like phase could occur every 100 million years based on the currently observed stellar populations in the Galactic center region \citep{2011ApJ...741..108P}. In the context of this work, this could provide a mechanism for the recurrent depletion of large red giants by the increased jet activity. However, the most relevant episode in terms of currently observed stellar populations is the most recent episode a few million years ago. There is an evidence for even more recent activity $\sim 400$ years ago, which is inferred from the X-ray reflections or propagating brightening of molecular clouds in the Central Molecular Zone \citep{1993ApJ...407..606S,1998MNRAS.297.1279S}. However, these repetitive events occur stochastically based on the presence of infalling clumps and lead to the increase by only several orders of magnitude and last for several years depending on the exact viscous timescale \citep{2013A&A...555A..97C}.

Another possibility of a recurrent launching of the jet is a tidal disruption event (TDE), which can occur in the Galactic center every $\sim 10^4$--$10^5$ years depending on the stellar type \citep{1999MNRAS.306...35S,2005PhR...419...65A,2015JHEAp...7..148K}. For completeness, we note that the jet is not always formed during the TDE \citep{2015JHEAp...7..148K}. However, for a few months to years, the TDE can trigger a jet activity similar to the Seyfert sources \citep{1975Natur.254..295H}. Due to the short duration of the TDE between several months to years given by the steep dependency of the luminosity on time, $L\propto t^{-5/3}$, the average number of interacting stars would be given by the estimates calculated in Section~\ref{sec_fraction_redgiant}. In general, the number of ablated red giants in the S cluster would be of the order of unity. The jet precession driven by the Lense-Thirring effect \citep{2006MNRAS.368.1196L} could enlarge this number depending on the precession period and $\Omega_{\rm p}$ (see Section~\ref{subsec_precession}). 

\subsection{Comparison with other mechanisms -- the region of efficiency}
\label{subsec_other_mechanisms}

The jet-induced alternation of the population of late-type stars is not necessarily an alternative to other proposed mechanism, listed in the introductory Section~\ref{sec_intro}. In reality, it could have co-existed simultaneously during the past few million years with other previously proposed mechanisms, in particular the tidal disruption of red-giant envelopes as well as direct star-disc interactions. This follows from the fact that these mechanisms have different length-scales of their efficiency, as we further outline in the paragraphs below.

First, the tidal disruption of red-giant envelopes takes place on the smallest scales -- less than one milliparsec from Sgr~A* -- as  given by the tidal radius, $r_{\rm t}\approx R_{\star}(2M_{\bullet}/m_{\star})^{1/3}$,
\begin{align}
    r_{\rm t} & \lesssim 6000 \left(\frac{R_{\star}}{30\,R_{\odot}} \right) \left(\frac{m_{\star}}{1\,M_{\odot}} \right)^{-\frac{1}{3}}\left(\frac{M_{\bullet}}{4\times 10^6\,M_{\odot}} \right)^{\frac{1}{3}}\,R_{\odot}\,\notag\\
    &=0.14\,{\rm mpc}=354 R_{\rm Schw}\,.
    \label{eq_tidal_radius_discussion}
\end{align}
The resonant relaxation process (discussed in Section~\ref{sec_fraction_redgiant}), in particular the scalar resonant relaxation, can cause an increase in orbital eccentricities and thus effectively induce the tidal disruption of giants as their orbital distance decreases below $r_{\rm t}$ close to the pericenter of their orbits. This could have contributed to the dearth of brighter red giants in the inner $\sim 0.1\,{\rm pc}$ \citep{2011ApJ...738...99M}, which is a larger scale than given by $r_{\rm t}$ but still smaller than the total extent of $0.5\,{\rm pc}$ of the red giant hole.

Then, the jet-induced mass removal is clearly the most efficient in the S-cluster region, $z\lesssim 0.04\,{\rm pc}$, according to the cumulative mass removal distance profile in Fig.~\ref{fig_cumulative_mass_removal}. Another way to constrain the region of the maximum efficiency is to use the relation in Eq.~\ref{eq_stag_radius_values} for the stagnation radius, from which we derive the distance $z$. It follows that for the maximum jet luminosity of $L_{\rm j}=10^{44}\,{\rm erg\,s^{-1}}$ and the stagnation radius range of $R_{\rm stag}=4-30\,R_{\odot}$, we obtain the distance range $z_{\rm jet}=0.06-0.44\,{\rm pc}$. Hence, within the S cluster, $z\lesssim 0.04\,{\rm pc}$, $R_{\rm stag}$ would be effectively below $4\,R_{\odot}$, which corresponds to $K\sim 16\,{\rm mag}$ stars for a typical age of 5 Gyr. Therefore, the jet luminosities close to the Eddington limit for Sgr~A* are required to truncate the atmospheres of late-type stars of $K\lesssim 16\,{\rm mag}$. For the moderate jet luminosity of $L_{\rm j}=10^{42}\,{\rm erg\,s^{-1}}$, the distance range decreases by an order of magnitude to $z_{\rm j}=0.006-0.04\,{\rm pc}$, hence only the brighter giants with $R_{\star}=30\,R_{\odot}$ ($K\sim 13.5\,{\rm mag}$) would be effectively truncated within the S cluster, while the smaller and fainter giants with $K\sim 16\,{\rm mag}$ would remain largely unaffected by the jet.

Finally, the star--clumpy disc collisions are the most efficient for the disc surface densities $\Sigma>10^{4}\,{\rm g\,cm^{-2}}$ typical of self-gravitating clumps \citep{2016ApJ...823..155K,2019arXiv191004774A}, which can form at larger distance scales where the condition for gravitational instability is met as given by the Toomre instability criterion \citep{2004ApJ...604L..45M}. In the Galactic center, this region likely corresponds to $0.04\lesssim z \lesssim 0.5\,{\rm pc}$, where the disk population of young massive stars is observed and they are believed to have formed in-situ in a massive gaseous disc \citep{2003ApJ...590L..33L}.

Hence, we speculate that the dearth of bright red giants for $z\lesssim 0.5\,{\rm pc}$ is due to the combination of the three processes -- tidal stripping, jet-induced atmosphere ablation, and star-disc interactions -- that operated the most efficiently at complementary length-scales up to $\sim 0.5\,{\rm pc}.$

\subsection{Observational signatures and falsifiability}

The jet-ablation mechanism likely operated in the central S cluster region ($<0.04\,{\rm pc}$) for stars of large atmosphere radii of $10\,R_{\odot}$, and especially for supergiants of $\sim 100\,R_{\odot}$ even at larger distances up to $\sim 0.1\,{\rm pc}$. However, since the enhanced jet activity and the resulting atmosphere ablation took place a few million years ago, any direct observational trace of the jet-star interaction is difficult to find. Here we list tentative observational signatures of the jet-ablation on the pre-existing cusp of late-type stars. Some of them have more possible interpretations due to other mechanisms operating simultaneously in the complex nuclear star cluster around Sgr~A*. The signatures proposed here can serve as a guideline towards confirming the jet activity and in particular the jet-star interaction in the central parsec. On the other hand, if other explanations become more likely, these can also serve as suitable falsifiability criteria for the jet-ablation theory. The signatures can be outlined as follows:
\begin{itemize}
    \item[(i)] Flattening of the density distribution for brighter late-type stars. This is a classical signature of the preferential bright late-type star depletion, which has more interpretations due to mechanisms, which likely operated on different scales; see Subsection~\ref{subsec_other_mechanisms}. Therefore, this signature should be treated with caution. However, we have shown in Section~\ref{sec_surface_brightness} that the jet active for a sufficiently long time can have an impact on the surface-brightness profile of the Nuclear Star Cluster when it reaches the kinetic luminosity at least $10^{42}\,{\rm erg\,s^{-1}}$; see also Section~\ref{sec_surface_brightness} and Fig.~\ref{fig_surface_distribution}. In particular, brighter giants with $m_{\rm K}<14$ mag could exhibit a flat profile due to the jet activity inside the inner arcsecond ($z\lesssim 0.04\,{\rm pc}$). Such a trend has also been recently reported by the high-sensitivity photometric analysis of \citet{2020arXiv200715950S}.
\item[(ii)] Detection of high metallicity stars. The jet-ablation of red giant and supergiant shells could reveal metal-rich deeper layers. A jet-ablation mechanism can be considered as one of the explanations for the occurrence of stars with anomalous metal concentrations in their atmospheres, as was found by \citet{2018ApJ...855L...5D}, see also Subsection~\ref{subsec_chemically_peculiar}.
    \item[(iii)] Cluster of remnant white dwarfs at millipasec scales. In relation to point (ii), an extreme case of jet-ablation could lead to the exposure of degenerate cores for asymptotic giant-branch (AGB) stars. This is analogous to the ablation by tidal stripping \citep{2020MNRAS.493L.120K}, however, the jet-ablation has a larger length-scale for Sgr~A*. As we derived in Section~\ref{sec_NIR_domain}, the AGB stars could be jet-ablated down to the white-dwarf cores for $z<0.015-0.15\,{\rm mpc}$ for $L_{\rm j}=10^{42}-10^{44}\,{\rm erg\,s^{-1}}$. The tidal stripping to the size of $10^{-2}\,R_{\odot}$ is only possible essentially below the event horizon. There has not been a direct observation of white dwarfs at such small distances from Sgr~A*, however, the hard X-ray flux peaking at Sgr~A* was hypothesized to originate in the cluster of accreting white dwarfs \citep{2015Natur.520..646P}.   
    \item[(iv)] Cusp of remnant blue OB stars. As we have shown in Section~\ref{sec_NIR_domain}, the late-type stars could be turned to blue stars of spectral type OB by the jet-ablation mechanism. In Table~\ref{tab_redgiant_param}, we show that the effective temperature could be of a few $\times 10^{4}\,{\rm K}$, and with the stagnation radius of $R_{\rm stag}=4.45\,R_{\odot}$, the K-band magnitude was estimated to reach $m_{\rm K}\sim 13$ mag, which is comparable to S2 star \citep[$m_{\rm K}\sim 14.1$ mag, ][]{2017ApJ...847..120H}. In this sense, we hypothesize that the fraction of S stars could be produced via the jet-ablation of older stars, however, the production rate within the S-cluster was of the order of unity, as we showed in Section~\ref{sec_fraction_redgiant}. Hence, the majority of B-type S stars was most likely formed in-situ in the circumnuclear gaseous material \citep{2016LNP...905..205M} and the current location and the kinematic structure of the S cluster are a result of different dynamical processes, most likely the Kozai-Lidov mechanism and the resonant relaxation \citep{2020ApJ...896..100A}.
     \item[(v)] Presence of bow-shock and comet-shaped sources. The presence of bow-shock sources X3 and X7 \citep{2010A&A...521A..13M} as well as X8 \citep{2019A&A...624A..97P} in the mini-cavity implies the interaction of these sources with a nuclear outflow in the current state of activity of Sgr~A*. The present nuclear outflow could be a signature of a low surface-brightness jet \citep{2020arXiv200804317Y}, which was much more luminous a few million years ago. The infrared-excess sources X3, X7, and X8 thus directly reveal the interaction of a nuclear outflow/jet with stars at the scale of $\sim 0.2\,{\rm pc}$.
    \item[(vi)] Non-spherical distribution of stars. In the broader context, the activity of the Seyfert-like jet at the scales of $\sim 0.1\,{\rm pc}$ could be imprinted in the non-spherical stellar structures, e.g. by affecting the distribution of denser star-forming material and its temperature via the jet feedback. Recently, \citet{2020ApJ...896..100A} revealed an x-shape structure of the S-cluster formed by two, nearly perpendicular stellar disks. Since S stars formed a few million years ago, their formation could be linked to the phase of higher accretion and the enhanced jet activity. Also, the x-structure implies that the resonant relaxation process (see Section~\ref{sec_fraction_redgiant}), in particular the vector resonant relaxation, is not as fast, otherwise the kinematic structure of the S cluster would be rather isotropic. However, the potential relation between the jet activity and the stellar kinematics needs to be verified via the detailed hydrodynamical/N-body or smooth-particle-hydrodynamics simulations.   
    \item[(vii)] Non-spherical distribution of ionization tracers around Sgr~A*. In a similar way as we discussed in point (vi) in terms of the non-spherical stellar distribution, the jet interaction with the surrounding gas could also be revealed via the non-isotropic distribution of ionization tracers. In particular, the shock-tracer line [FeIII] seems to be preferentially located in the direction of the mini-cavity \citep{2020A&A...634A..35P}, which suggests either the current or the past interaction of the gas with the high-velocity outflow/jet \citep{2020arXiv200804317Y}. On the scales larger than one parsec, the non-isotropic distribution is manifested by bipolar radio lobes \citep{2019Natur.573..235H} and X-ray \citep{2019Natur.567..347P} and $\gamma$-ray bubbles \citep{2010ApJ...724.1044S,2014ApJ...793...64A}. Recently, the analysis of the tilted, partially ionized disk in the inner Galaktic latitudes has shown that its optical line ratios are characteristic of low-ionization nuclear emission regions \citep[LINERs;][]{2020SciA....6.9711K}. The bipolar ionization structure is energetically in favor of the Seyfer-like jet activity rather than the star-formation event \citep{2019ApJ...886...45B}.
\end{itemize}

In summary, the jet-activity signs listed in points (i)--(vii) indicated the past enhanced activity of the jet and its interaction with the surrounding circumnuclear medium, including the nuclear star cluster. Although each of the above-mentioned points can have alternative explanations, the absence of all of these signatures would suggest that the jet did not operate in the past and our hypotheses would then be strongly disfavored. Future detailed observations by the Extremely Large Telescope (ELT) in combination with detailed numerical simulations of the jet-star interactions close to Sgr~A* will help to reveal the signatures of the current and the past jet--star interactions.

\section{Summary and conclusions}
\label{sec_summary}

We presented a novel scenario to explain the lack of bright red giants in the inner regions of the Galactic center in the sphere of influence of the currently quiescent, but previously active radio source Sgr~A*. Taking this increased activity into account, we focused on the effect of the jet on late-type stars at $\lesssim 0.3\,{\rm pc}$. By adopting the scenario of the recent active period 
of Sgr A$^{\star}$, we considered the interaction of red giants with a jet of a typical active Seyfert-like nucleus with the expected kinetic luminosity $L_{\rm j}=10^{41}-10^{44}\,{\rm erg\,s^{-1}}$. 
Given that red giants have a very slow wind, the jet can
significantly ablate the stellar envelope down to at least $\sim 30\,R_{\odot}$ within the S cluster ($z\lesssim 0.04\,{\rm pc}$) after repetitive encounters. Specifically, at $z=0.02\,{\rm pc}$, the stagnation radius is $4\leq R_{\rm stag}/R_{\odot}\leq 30$ for $2.0\times 10^{41}\leq L_{\rm j}/{\rm erg\,s^{-1}}\leq 1.1\times 10^{43}$. Hence, the higher luminosity end that corresponds to the less frequent events that formed the Fermi bubbles can ablate the stellar atmospheres of late-type giants by a factor of $\sim 7.5$ more than the more frequent, less energetic outbursts of Sgr~A*.

This truncation is accompanied by the removal of a large fraction of matter, reaching as much as $\Delta M_1\approx 3\times 10^{-5}\,M_{\odot}\approx 10 M_{\rm Earth}$ for red giants with radii $R_{\star}>200\,R_{\odot}$ at distances smaller than $0.01$ pc for a single encounter. After at least thousand of red giant--jet encounters, we expect the cumulative mass loss of at least $\Delta M\approx 10^{-4}\,M_{\odot}$ at the orbital distance of $0.01\,{\rm pc}$. This is comparable to the values inferred from red giant--accretion clump simulations. The proposed mechanism can thus help to explain the presence of late-type stars with the maximum atmosphere radius of $\sim 30 R_{\odot}$ within the S cluster as inferred from the near-infrared observations.

The reduction in the mass and radius of the red giant atmosphere after repetitive jet-star crossings will produce an estimated decrease in the near-infrared K-band magnitude by $1.9$, $5.3$, and $8.9$ magnitudes with respect to the normal evolution at $0.1$, $0.01$, and $0.001\,{\rm pc}$ from Sgr~A*, respectively. Simultaneously, the color index would decrease to negative values, i.e., the stars should appear bluer with a higher effective temperature. The mean expected number of red giant-jet crossings per orbital period is $3.5$ within $0.04$ pc, and $82.5$ within $0.3$ pc, respectively. For the jet kinetic luminosity of $10^{44}\,{\rm erg\,s^{-1}}$, $\sim 26.5\%$ of currently detectable late-type stars located at $z=0.04\,{\rm pc}$ (S cluster) with radii larger than $2.7\,R_{\odot}$ and K-band magnitudes smaller than $\sim 16$ mag could be affected by the jet ablation. The estimated numbers of interacting giants can be considered as lower limits since various dynamical effects, such as the coherent resonant relaxation within the nuclear star cluster as well as a potential jet precession would enlarge the number of affected giants.

Constructed surface-brightness profiles of the mock Nuclear Star Cluster affected by the jet with the luminosity of $10^{42}-10^{44}\,{\rm erg\,s^{-1}}$ show that profiles of mainly brighter late-type stars with $m_{\rm K}<14$ mag (dereddened, $<16$ mag with the foreground extinction included) are flattened by the jet inside the inner arcsecond ($\lesssim 0.04\,{\rm pc}$). Fainter stars keep the initially assumed cusp-like projected profile.

In summary, the interaction of red giants with the jet of Sgr~A* during its enhanced activity could contribute to the observed lack of bright red giants and hence affect their surface-brightness profile in the central parts of the nuclear star cluster. More likely, this mechanism operated in parallel with other previously proposed mechanisms, such as the star--disc interactions, star--star collisions, and tidal disruption events that have different spatial scales of efficiency. Detailed numerical computations of red giant--jet interactions in combination with a modified stellar evolution will help to verify our analytical estimates. 

\acknowledgments

We thank the referee for constructive comments that helped us to improve our manuscript.
MZ, BC and VK acknowledge the continued support by the National Science Center, Poland, grant No. 2017/26/A/ST9/00756 (Maestro 9), and the
Czech-Polish mobility program (MSMT 8J20PL037). AA and VK acknowledge the Czech Science Foundation under the grant GACR 20-19854S titled ``Particle Acceleration Studies in Astrophysical Jets'', and the European Space Agency PRODEX project eXTP in the Czech Republic. This work was carried out partly also within the Collaborative Research Center 956, sub-project [A02], funded by the Deutsche Forschungsgemeinschaft (DFG) project ID184018867. The work on this project was partially carried out during the short-term stay of MZ within the Polish-Czech bilateral exchange program supported by NAWA under the agreement PPN/BCZ/2019/1/00069. MZ also acknowledges the NAWA financial support under the agreement PPN/WYM/2019/1/00064 to perform a three-month exchange stay at the Astronomical Institute of the Czech Academy of Sciences in Prague. 



\begin{thebibliography}{}
\expandafter\ifx\csname natexlab\endcsname\relax\def\natexlab#1{#1}\fi
\providecommand{\url}[1]{\href{#1}{#1}}
\providecommand{\dodoi}[1]{doi:~\href{http://doi.org/#1}{\nolinkurl{#1}}}
\providecommand{\doeprint}[1]{\href{http://ascl.net/#1}{\nolinkurl{http://ascl.net/#1}}}
\providecommand{\doarXiv}[1]{\href{https://arxiv.org/abs/#1}{\nolinkurl{https://arxiv.org/abs/#1}}}

\bibitem[{{Abraham} \& {Carrara}(1998)}]{1998ApJ...496..172A}
{Abraham}, Z., \& {Carrara}, E.~A. 1998, \apj, 496, 172, \dodoi{10.1086/305387}

\bibitem[{{Ackermann} {et~al.}(2014){Ackermann}, {Albert}, {Atwood}, {Baldini},
  {Ballet}, {Barbiellini}, {Bastieri}, {Bellazzini}, {Bissaldi}, {Blandford},
  {Bloom}, {Bottacini}, {Brandt}, {Bregeon}, {Bruel}, {Buehler}, {Buson},
  {Caliandro}, {Cameron}, {Caragiulo}, {Caraveo}, {Cavazzuti}, {Cecchi},
  {Charles}, {Chekhtman}, {Chiang}, {Chiaro}, {Ciprini}, {Claus},
  {Cohen-Tanugi}, {Conrad}, {Cutini}, {D'Ammando}, {de Angelis}, {de Palma},
  {Dermer}, {Digel}, {Di Venere}, {Silva}, {Drell}, {Favuzzi}, {Ferrara},
  {Focke}, {Franckowiak}, {Fukazawa}, {Funk}, {Fusco}, {Gargano}, {Gasparrini},
  {Germani}, {Giglietto}, {Giordano}, {Giroletti}, {Godfrey}, {Gomez-Vargas},
  {Grenier}, {Guiriec}, {Hadasch}, {Harding}, {Hays}, {Hewitt}, {Hou},
  {Jogler}, {J{\'o}hannesson}, {Johnson}, {Johnson}, {Kamae}, {Kataoka},
  {Kn{\"o}dlseder}, {Kocevski}, {Kuss}, {Larsson}, {Latronico}, {Longo},
  {Loparco}, {Lovellette}, {Lubrano}, {Malyshev}, {Manfreda}, {Massaro},
  {Mayer}, {Mazziotta}, {McEnery}, {Michelson}, {Mitthumsiri}, {Mizuno},
  {Monzani}, {Morselli}, {Moskalenko}, {Murgia}, {Nemmen}, {Nuss}, {Ohsugi},
  {Omodei}, {Orienti}, {Orlando}, {Ormes}, {Paneque}, {Panetta}, {Perkins},
  {Pesce-Rollins}, {Petrosian}, {Piron}, {Pivato}, {Rain{\`o}}, {Rando},
  {Razzano}, {Razzaque}, {Reimer}, {Reimer}, {S{\'a}nchez-Conde}, {Schaal},
  {Schulz}, {Sgr{\`o}}, {Siskind}, {Spandre}, {Spinelli}, {Stawarz}, {Strong},
  {Suson}, {Tahara}, {Takahashi}, {Thayer}, {Tibaldo}, {Tinivella}, {Torres},
  {Tosti}, {Troja}, {Uchiyama}, {Vianello}, {Werner}, {Winer}, {Wood}, {Wood},
  \& {Zaharijas}}]{2014ApJ...793...64A}
{Ackermann}, M., {Albert}, A., {Atwood}, W.~B., {et~al.} 2014, \apj, 793, 64,
  \dodoi{10.1088/0004-637X/793/1/64}

\bibitem[{{Alexander}(2005)}]{2005PhR...419...65A}
{Alexander}, T. 2005, \physrep, 419, 65, \dodoi{10.1016/j.physrep.2005.08.002}

\bibitem[{{Ali} {et~al.}(2020){Ali}, {Paul}, {Eckart}, {Parsa}, {Zajacek},
  {Pei{\ss}ker}, {Subroweit}, {Valencia-S.}, {Thomkins}, \&
  {Witzel}}]{2020ApJ...896..100A}
{Ali}, B., {Paul}, D., {Eckart}, A., {et~al.} 2020, \apj, 896, 100,
  \dodoi{10.3847/1538-4357/ab93ae}

\bibitem[{{Amaro-Seoane} \& {Chen}(2014)}]{2014ApJ...781L..18A}
{Amaro-Seoane}, P., \& {Chen}, X. 2014, \apjl, 781, L18,
  \dodoi{10.1088/2041-8205/781/1/L18}

\bibitem[{{Amaro-Seoane} {et~al.}(2020){Amaro-Seoane}, {Chen}, {Sch{\"o}del},
  \& {Casanellas}}]{2019arXiv191004774A}
{Amaro-Seoane}, P., {Chen}, X., {Sch{\"o}del}, R., \& {Casanellas}, J. 2020,
  \mnras, 492, 250, \dodoi{10.1093/mnras/stz3507}

\bibitem[{{Antonini} {et~al.}(2012){Antonini}, {Capuzzo-Dolcetta},
  {Mastrobuono-Battisti}, \& {Merritt}}]{2012ApJ...750..111A}
{Antonini}, F., {Capuzzo-Dolcetta}, R., {Mastrobuono-Battisti}, A., \&
  {Merritt}, D. 2012, \apj, 750, 111, \dodoi{10.1088/0004-637X/750/2/111}

\bibitem[{{Araudo} {et~al.}(2013){Araudo}, {Bosch-Ramon}, \&
  {Romero}}]{2013MNRAS.436.3626A}
{Araudo}, A.~T., {Bosch-Ramon}, V., \& {Romero}, G.~E. 2013, \mnras, 436, 3626,
  \dodoi{10.1093/mnras/stt1840}

\bibitem[{{Araudo} \& {Karas}(2017)}]{2017bhns.work....1A}
{Araudo}, A.~T., \& {Karas}, V. 2017, in RAGtime 17-19: Workshops on Black
  Holes and Neutron Stars, 1--6

\bibitem[{{Armitage} {et~al.}(1996){Armitage}, {Zurek}, \&
  {Davies}}]{1996ApJ...470..237A}
{Armitage}, P.~J., {Zurek}, W.~H., \& {Davies}, M.~B. 1996, \apj, 470, 237,
  \dodoi{10.1086/177864}

\bibitem[{{Baganoff} {et~al.}(2003){Baganoff}, {Maeda}, {Morris}, {Bautz},
  {Brandt}, {Cui}, {Doty}, {Feigelson}, {Garmire}, {Pravdo}, {Ricker}, \&
  {Townsley}}]{2003ApJ...591..891B}
{Baganoff}, F.~K., {Maeda}, Y., {Morris}, M., {et~al.} 2003, \apj, 591, 891,
  \dodoi{10.1086/375145}

\bibitem[{{Bailey} \& {Davies}(1999)}]{1999MNRAS.308..257B}
{Bailey}, V.~C., \& {Davies}, M.~B. 1999, \mnras, 308, 257,
  \dodoi{10.1046/j.1365-8711.1999.02740.x}

\bibitem[{{Barkov} {et~al.}(2012{\natexlab{a}}){Barkov}, {Aharonian},
  {Bogovalov}, {Kelner}, \& {Khangulyan}}]{2012ApJ...749..119B}
{Barkov}, M.~V., {Aharonian}, F.~A., {Bogovalov}, S.~V., {Kelner}, S.~R., \&
  {Khangulyan}, D. 2012{\natexlab{a}}, \apj, 749, 119,
  \dodoi{10.1088/0004-637X/749/2/119}

\bibitem[{{Barkov} {et~al.}(2010){Barkov}, {Aharonian}, \&
  {Bosch-Ramon}}]{Barkov_10}
{Barkov}, M.~V., {Aharonian}, F.~A., \& {Bosch-Ramon}, V. 2010, \apj, 724,
  1517, \dodoi{10.1088/0004-637X/724/2/1517}

\bibitem[{{Barkov} {et~al.}(2012{\natexlab{b}}){Barkov}, {Bosch-Ramon}, \&
  {Aharonian}}]{Barkov_12}
{Barkov}, M.~V., {Bosch-Ramon}, V., \& {Aharonian}, F.~A. 2012{\natexlab{b}},
  \apj, 755, 170, \dodoi{10.1088/0004-637X/755/2/170}

\bibitem[{{Baumgardt} {et~al.}(2006){Baumgardt}, {Gualandris}, \& {Portegies
  Zwart}}]{2006MNRAS.372..174B}
{Baumgardt}, H., {Gualandris}, A., \& {Portegies Zwart}, S. 2006, \mnras, 372,
  174, \dodoi{10.1111/j.1365-2966.2006.10818.x}

\bibitem[{{Bednarek} \& {Banasi{\'n}ski}(2015)}]{2015ApJ...807..168B}
{Bednarek}, W., \& {Banasi{\'n}ski}, P. 2015, \apj, 807, 168,
  \dodoi{10.1088/0004-637X/807/2/168}

\bibitem[{{Bell}(1978)}]{1978MNRAS.182..147B}
{Bell}, A.~R. 1978, \mnras, 182, 147, \dodoi{10.1093/mnras/182.2.147}

\bibitem[{{Bland-Hawthorn} \& {Cohen}(2003)}]{2003ApJ...582..246B}
{Bland-Hawthorn}, J., \& {Cohen}, M. 2003, \apj, 582, 246,
  \dodoi{10.1086/344573}

\bibitem[{{Bland-Hawthorn} {et~al.}(2019){Bland-Hawthorn}, {Maloney},
  {Sutherland}, {Groves}, {Guglielmo}, {Hao Li}, {Curzons}, {Cecil}, \&
  {Fox}}]{2019ApJ...886...45B}
{Bland-Hawthorn}, J., {Maloney}, P.~R., {Sutherland}, R., {et~al.} 2019, \apj,
  886, 45, \dodoi{10.3847/1538-4357/ab44c8}

\bibitem[{{Blandford} \& {Payne}(1982)}]{1982MNRAS.199..883B}
{Blandford}, R.~D., \& {Payne}, D.~G. 1982, \mnras, 199, 883,
  \dodoi{10.1093/mnras/199.4.883}

\bibitem[{{Boehle} {et~al.}(2016){Boehle}, {Ghez}, {Sch{\"o}del}, {Meyer},
  {Yelda}, {Albers}, {Martinez}, {Becklin}, {Do}, {Lu}, {Matthews}, {Morris},
  {Sitarski}, \& {Witzel}}]{2016ApJ...830...17B}
{Boehle}, A., {Ghez}, A.~M., {Sch{\"o}del}, R., {et~al.} 2016, \apj, 830, 17,
  \dodoi{10.3847/0004-637X/830/1/17}

\bibitem[{{Bogdanovi{\'c}} {et~al.}(2014){Bogdanovi{\'c}}, {Cheng}, \&
  {Amaro-Seoane}}]{2014ApJ...788...99B}
{Bogdanovi{\'c}}, T., {Cheng}, R.~M., \& {Amaro-Seoane}, P. 2014, \apj, 788,
  99, \dodoi{10.1088/0004-637X/788/2/99}

\bibitem[{{Bosch-Ramon} {et~al.}(2012){Bosch-Ramon}, {Perucho}, \&
  {Barkov}}]{2012A&A...539A..69B}
{Bosch-Ramon}, V., {Perucho}, M., \& {Barkov}, M.~V. 2012, \aap, 539, A69,
  \dodoi{10.1051/0004-6361/201118622}

\bibitem[{{Bressan} {et~al.}(2012){Bressan}, {Marigo}, {Girardi}, {Salasnich},
  {Dal Cero}, {Rubele}, \& {Nanni}}]{2012MNRAS.427..127B}
{Bressan}, A., {Marigo}, P., {Girardi}, L., {et~al.} 2012, \mnras, 427, 127,
  \dodoi{10.1111/j.1365-2966.2012.21948.x}

\bibitem[{{Britzen} {et~al.}(2019{\natexlab{a}}){Britzen}, {Fendt},
  {Zaja{\v{c}}ek}, {Jaron}, {Pashchenko}, {Aller}, \&
  {Aller}}]{2019Galax...7...72B}
{Britzen}, S., {Fendt}, C., {Zaja{\v{c}}ek}, M., {et~al.} 2019{\natexlab{a}},
  Galaxies, 7, 72, \dodoi{10.3390/galaxies7030072}

\bibitem[{{Britzen} {et~al.}(2018){Britzen}, {Fendt}, {Witzel}, {Qian},
  {Pashchenko}, {Kurtanidze}, {Zajacek}, {Martinez}, {Karas}, {Aller}, {Aller},
  {Eckart}, {Nilsson}, {Ar{\'e}valo}, {Cuadra}, {Subroweit}, \&
  {Witzel}}]{2018MNRAS.478.3199B}
{Britzen}, S., {Fendt}, C., {Witzel}, G., {et~al.} 2018, \mnras, 478, 3199,
  \dodoi{10.1093/mnras/sty1026}

\bibitem[{{Britzen} {et~al.}(2019{\natexlab{b}}){Britzen}, {Fendt},
  {B{\"o}ttcher}, {Zaja{\v{c}}ek}, {Jaron}, {Pashchenko}, {Araudo}, {Karas}, \&
  {Kurtanidze}}]{2019A&A...630A.103B}
{Britzen}, S., {Fendt}, C., {B{\"o}ttcher}, M., {et~al.} 2019{\natexlab{b}},
  \aap, 630, A103, \dodoi{10.1051/0004-6361/201935422}

\bibitem[{{Buchholz} {et~al.}(2009){Buchholz}, {Sch{\"o}del}, \&
  {Eckart}}]{2009A&A...499..483B}
{Buchholz}, R.~M., {Sch{\"o}del}, R., \& {Eckart}, A. 2009, \aap, 499, 483,
  \dodoi{10.1051/0004-6361/200811497}

\bibitem[{{Czerny} {et~al.}(2013){Czerny}, {Kunneriath}, {Karas}, \&
  {Das}}]{2013A&A...555A..97C}
{Czerny}, B., {Kunneriath}, D., {Karas}, V., \& {Das}, T.~K. 2013, \aap, 555,
  A97, \dodoi{10.1051/0004-6361/201118124}

\bibitem[{{Dale} {et~al.}(2009){Dale}, {Davies}, {Church}, \&
  {Freitag}}]{2009MNRAS.393.1016D}
{Dale}, J.~E., {Davies}, M.~B., {Church}, R.~P., \& {Freitag}, M. 2009, \mnras,
  393, 1016, \dodoi{10.1111/j.1365-2966.2008.14254.x}

\bibitem[{{Davies} \& {King}(2005)}]{2005ApJ...624L..25D}
{Davies}, M.~B., \& {King}, A. 2005, \apjl, 624, L25, \dodoi{10.1086/430308}

\bibitem[{{de la Cita} {et~al.}(2016){de la Cita}, {Bosch-Ramon},
  {Paredes-Fortuny}, {Khangulyan}, \& {Perucho}}]{2016A&A...591A..15D}
{de la Cita}, V.~M., {Bosch-Ramon}, V., {Paredes-Fortuny}, X., {Khangulyan},
  D., \& {Perucho}, M. 2016, \aap, 591, A15,
  \dodoi{10.1051/0004-6361/201527084}

\bibitem[{{Deegan} \& {Nayakshin}(2007)}]{2007MNRAS.377..897D}
{Deegan}, P., \& {Nayakshin}, S. 2007, \mnras, 377, 897,
  \dodoi{10.1111/j.1365-2966.2007.11659.x}

\bibitem[{{Dexter} \& {Fragile}(2013)}]{2013MNRAS.432.2252D}
{Dexter}, J., \& {Fragile}, P.~C. 2013, \mnras, 432, 2252,
  \dodoi{10.1093/mnras/stt583}

\bibitem[{{Do} {et~al.}(2009){Do}, {Ghez}, {Morris}, {Lu}, {Matthews}, {Yelda},
  \& {Larkin}}]{2009ApJ...703.1323D}
{Do}, T., {Ghez}, A.~M., {Morris}, M.~R., {et~al.} 2009, \apj, 703, 1323,
  \dodoi{10.1088/0004-637X/703/2/1323}

\bibitem[{{Do} {et~al.}(2018){Do}, {Kerzendorf}, {Konopacky}, {Marcinik},
  {Ghez}, {Lu}, \& {Morris}}]{2018ApJ...855L...5D}
{Do}, T., {Kerzendorf}, W., {Konopacky}, Q., {et~al.} 2018, \apjl, 855, L5,
  \dodoi{10.3847/2041-8213/aaaec3}

\bibitem[{{Eckart} {et~al.}(2017){Eckart}, {H{\"u}ttemann}, {Kiefer},
  {Britzen}, {Zaja{\v{c}}ek}, {L{\"a}mmerzahl}, {St{\"o}ckler}, {Valencia-S},
  {Karas}, \& {Garc{\'\i}a-Mar{\'\i}n}}]{2017FoPh...47..553E}
{Eckart}, A., {H{\"u}ttemann}, A., {Kiefer}, C., {et~al.} 2017, Foundations of
  Physics, 47, 553, \dodoi{10.1007/s10701-017-0079-2}

\bibitem[{{Ernst} {et~al.}(2009){Ernst}, {Just}, \&
  {Spurzem}}]{2009MNRAS.399..141E}
{Ernst}, A., {Just}, A., \& {Spurzem}, R. 2009, \mnras, 399, 141,
  \dodoi{10.1111/j.1365-2966.2009.15305.x}

\bibitem[{{Gallego-Cano} {et~al.}(2018){Gallego-Cano}, {Sch{\"o}del}, {Dong},
  {Nogueras-Lara}, {Gallego-Calvente}, {Amaro-Seoane}, \&
  {Baumgardt}}]{2018A&A...609A..26G}
{Gallego-Cano}, E., {Sch{\"o}del}, R., {Dong}, H., {et~al.} 2018, \aap, 609,
  A26, \dodoi{10.1051/0004-6361/201730451}

\bibitem[{{Genzel} {et~al.}(2010){Genzel}, {Eisenhauer}, \&
  {Gillessen}}]{2010RvMP...82.3121G}
{Genzel}, R., {Eisenhauer}, F., \& {Gillessen}, S. 2010, Reviews of Modern
  Physics, 82, 3121, \dodoi{10.1103/RevModPhys.82.3121}

\bibitem[{{Genzel} {et~al.}(1996){Genzel}, {Thatte}, {Krabbe}, {Kroker}, \&
  {Tacconi-Garman}}]{1996ApJ...472..153G}
{Genzel}, R., {Thatte}, N., {Krabbe}, A., {Kroker}, H., \& {Tacconi-Garman},
  L.~E. 1996, \apj, 472, 153, \dodoi{10.1086/178051}

\bibitem[{{Ghez} {et~al.}(2003){Ghez}, {Duch{\^e}ne}, {Matthews}, {Hornstein},
  {Tanner}, {Larkin}, {Morris}, {Becklin}, {Salim}, {Kremenek}, {Thompson},
  {Soifer}, {Neugebauer}, \& {McLean}}]{2003ApJ...586L.127G}
{Ghez}, A.~M., {Duch{\^e}ne}, G., {Matthews}, K., {et~al.} 2003, \apjl, 586,
  L127, \dodoi{10.1086/374804}

\bibitem[{{Gillessen} {et~al.}(2017){Gillessen}, {Plewa}, {Eisenhauer}, {Sari},
  {Waisberg}, {Habibi}, {Pfuhl}, {George}, {Dexter}, {von Fellenberg}, {Ott},
  \& {Genzel}}]{2017ApJ...837...30G}
{Gillessen}, S., {Plewa}, P.~M., {Eisenhauer}, F., {et~al.} 2017, \apj, 837,
  30, \dodoi{10.3847/1538-4357/aa5c41}

\bibitem[{{Gravity Collaboration} {et~al.}(2018){Gravity Collaboration},
  {Abuter}, {Amorim}, {Anugu}, {Baub{\"o}ck}, {Benisty}, {Berger}, {Blind},
  {Bonnet}, {Brandner}, {Buron}, {Collin}, {Chapron}, {Cl{\'e}net}, {Coud{\'e}
  Du Foresto}, {de Zeeuw}, {Deen}, {Delplancke-Str{\"o}bele}, {Dembet},
  {Dexter}, {Duvert}, {Eckart}, {Eisenhauer}, {Finger}, {F{\"o}rster
  Schreiber}, {F{\'e}dou}, {Garcia}, {Garcia Lopez}, {Gao}, {Gendron},
  {Genzel}, {Gillessen}, {Gordo}, {Habibi}, {Haubois}, {Haug}, {Hau{\ss}mann},
  {Henning}, {Hippler}, {Horrobin}, {Hubert}, {Hubin}, {Jimenez Rosales},
  {Jochum}, {Jocou}, {Kaufer}, {Kellner}, {Kendrew}, {Kervella}, {Kok},
  {Kulas}, {Lacour}, {Lapeyr{\`e}re}, {Lazareff}, {Le Bouquin}, {L{\'e}na},
  {Lippa}, {Lenzen}, {M{\'e}rand}, {M{\"u}ler}, {Neumann}, {Ott}, {Palanca},
  {Paumard}, {Pasquini}, {Perraut}, {Perrin}, {Pfuhl}, {Plewa}, {Rabien},
  {Ram{\'\i}rez}, {Ramos}, {Rau}, {Rodr{\'\i}guez-Coira}, {Rohloff}, {Rousset},
  {Sanchez-Bermudez}, {Scheithauer}, {Sch{\"o}ller}, {Schuler}, {Spyromilio},
  {Straub}, {Straubmeier}, {Sturm}, {Tacconi}, {Tristram}, {Vincent}, {von
  Fellenberg}, {Wank}, {Waisberg}, {Widmann}, {Wieprecht}, {Wiest},
  {Wiezorrek}, {Woillez}, {Yazici}, {Ziegler}, \& {Zins}}]{2018A&A...615L..15G}
{Gravity Collaboration}, {Abuter}, R., {Amorim}, A., {et~al.} 2018, \aap, 615,
  L15, \dodoi{10.1051/0004-6361/201833718}

\bibitem[{{Gualandris} \& {Merritt}(2012)}]{2012ApJ...744...74G}
{Gualandris}, A., \& {Merritt}, D. 2012, \apj, 744, 74,
  \dodoi{10.1088/0004-637X/744/1/74}

\bibitem[{{Guo} \& {Mathews}(2012)}]{2012ApJ...756..181G}
{Guo}, F., \& {Mathews}, W.~G. 2012, \apj, 756, 181,
  \dodoi{10.1088/0004-637X/756/2/181}

\bibitem[{{Habibi} {et~al.}(2017){Habibi}, {Gillessen}, {Martins},
  {Eisenhauer}, {Plewa}, {Pfuhl}, {George}, {Dexter}, {Waisberg}, {Ott}, {von
  Fellenberg}, {Baub{\"o}ck}, {Jimenez-Rosales}, \&
  {Genzel}}]{2017ApJ...847..120H}
{Habibi}, M., {Gillessen}, S., {Martins}, F., {et~al.} 2017, \apj, 847, 120,
  \dodoi{10.3847/1538-4357/aa876f}

\bibitem[{{Habibi} {et~al.}(2019){Habibi}, {Gillessen}, {Pfuhl}, {Eisenhauer},
  {Plewa}, {von Fellenberg}, {Widmann}, {Ott}, {Gao}, {Waisberg},
  {Baub{\"o}ck}, {Jimenez-Rosales}, {Dexter}, {de Zeeuw}, \&
  {Genzel}}]{2019ApJ...872L..15H}
{Habibi}, M., {Gillessen}, S., {Pfuhl}, O., {et~al.} 2019, \apjl, 872, L15,
  \dodoi{10.3847/2041-8213/ab03cf}

\bibitem[{{Heywood} {et~al.}(2019){Heywood}, {Camilo}, {Cotton}, {Yusef-Zadeh},
  {Abbott}, {Adam}, {Aldera}, {Bauermeister}, {Booth}, {Botha}, {Botha},
  {Brederode}, {Brits}, {Buchner}, {Burger}, {Chalmers}, {Cheetham}, {de
  Villiers}, {Dikgale-Mahlakoana}, {du Toit}, {Esterhuyse}, {Fanaroff},
  {Foley}, {Fourie}, {Gamatham}, {Goedhart}, {Gounden}, {Hlakola}, {Hoek},
  {Hokwana}, {Horn}, {Horrell}, {Hugo}, {Isaacson}, {Jonas}, {Jordaan},
  {Joubert}, {J{\'o}zsa}, {Julie}, {Kapp}, {Kenyon}, {Kotz{\'e}}, {Kriel},
  {Kusel}, {Lehmensiek}, {Liebenberg}, {Loots}, {Lord}, {Lunsky}, {Macfarlane},
  {Magnus}, {Magozore}, {Mahgoub}, {Main}, {Malan}, {Malgas}, {Manley},
  {Maree}, {Merry}, {Millenaar}, {Mnyandu}, {Moeng}, {Monama}, {Mphego}, {New},
  {Ngcebetsha}, {Oozeer}, {Otto}, {Passmoor}, {Patel}, {Peens-Hough},
  {Perkins}, {Ratcliffe}, {Renil}, {Rust}, {Salie}, {Schwardt}, {Serylak},
  {Siebrits}, {Sirothia}, {Smirnov}, {Sofeya}, {Swart}, {Tasse}, {Taylor},
  {Theron}, {Thorat}, {Tiplady}, {Tshongweni}, {van Balla}, {van der Byl}, {van
  der Merwe}, {van Dyk}, {Van Rooyen}, {Van Tonder}, {Van Wyk}, {Wallace},
  {Welz}, \& {Williams}}]{2019Natur.573..235H}
{Heywood}, I., {Camilo}, F., {Cotton}, W.~D., {et~al.} 2019, \nat, 573, 235,
  \dodoi{10.1038/s41586-019-1532-5}

\bibitem[{{Hills}(1975)}]{1975Natur.254..295H}
{Hills}, J.~G. 1975, \nat, 254, 295, \dodoi{10.1038/254295a0}

\bibitem[{{Ito} {et~al.}(2008){Ito}, {Kino}, {Kawakatu}, {Isobe}, \&
  {Yamada}}]{2008ApJ...685..828I}
{Ito}, H., {Kino}, M., {Kawakatu}, N., {Isobe}, N., \& {Yamada}, S. 2008, \apj,
  685, 828, \dodoi{10.1086/591036}

\bibitem[{{Joss} {et~al.}(1987){Joss}, {Rappaport}, \&
  {Lewis}}]{1987ApJ...319..180J}
{Joss}, P.~C., {Rappaport}, S., \& {Lewis}, W. 1987, \apj, 319, 180,
  \dodoi{10.1086/165443}

\bibitem[{{Junor} {et~al.}(1999){Junor}, {Biretta}, \&
  {Livio}}]{1999Natur.401..891J}
{Junor}, W., {Biretta}, J.~A., \& {Livio}, M. 1999, \nat, 401, 891,
  \dodoi{10.1038/44780}

\bibitem[{{Karas} \& {{\v{S}}ubr}(2001)}]{2001A&A...376..686K}
{Karas}, V., \& {{\v{S}}ubr}, L. 2001, \aap, 376, 686,
  \dodoi{10.1051/0004-6361:20011009}

\bibitem[{{Kieffer} \& {Bogdanovi{\'c}}(2016)}]{2016ApJ...823..155K}
{Kieffer}, T.~F., \& {Bogdanovi{\'c}}, T. 2016, \apj, 823, 155,
  \dodoi{10.3847/0004-637X/823/2/155}

\bibitem[{{Kim} \& {Morris}(2003)}]{2003ApJ...597..312K}
{Kim}, S.~S., \& {Morris}, M. 2003, \apj, 597, 312, \dodoi{10.1086/378347}

\bibitem[{{King}(2020)}]{2020MNRAS.493L.120K}
{King}, A. 2020, \mnras, 493, L120, \dodoi{10.1093/mnrasl/slaa020}

\bibitem[{{Komissarov}(1994)}]{1994MNRAS.269..394K}
{Komissarov}, S.~S. 1994, \mnras, 269, 394, \dodoi{10.1093/mnras/269.2.394}

\bibitem[{{Komossa}(2015)}]{2015JHEAp...7..148K}
{Komossa}, S. 2015, Journal of High Energy Astrophysics, 7, 148,
  \dodoi{10.1016/j.jheap.2015.04.006}

\bibitem[{{Krabbe} {et~al.}(1991){Krabbe}, {Genzel}, {Drapatz}, \&
  {Rotaciuc}}]{1991ApJ...382L..19K}
{Krabbe}, A., {Genzel}, R., {Drapatz}, S., \& {Rotaciuc}, V. 1991, \apjl, 382,
  L19, \dodoi{10.1086/186204}

\bibitem[{{Krishnarao} {et~al.}(2020){Krishnarao}, {Benjamin}, \&
  {Haffner}}]{2020SciA....6.9711K}
{Krishnarao}, D., {Benjamin}, R.~A., \& {Haffner}, L.~M. 2020, Science
  Advances, 6, 9711, \dodoi{10.1126/sciadv.aay9711}

\bibitem[{{Kroupa}(2001)}]{2001MNRAS.322..231K}
{Kroupa}, P. 2001, \mnras, 322, 231, \dodoi{10.1046/j.1365-8711.2001.04022.x}

\bibitem[{{Levin} \& {Beloborodov}(2003)}]{2003ApJ...590L..33L}
{Levin}, Y., \& {Beloborodov}, A.~M. 2003, \apjl, 590, L33,
  \dodoi{10.1086/376675}

\bibitem[{{Li} {et~al.}(2013){Li}, {Morris}, \&
  {Baganoff}}]{2013ApJ...779..154L}
{Li}, Z., {Morris}, M.~R., \& {Baganoff}, F.~K. 2013, \apj, 779, 154,
  \dodoi{10.1088/0004-637X/779/2/154}

\bibitem[{{L{\"o}ckmann} \& {Baumgardt}(2008)}]{2008MNRAS.384..323L}
{L{\"o}ckmann}, U., \& {Baumgardt}, H. 2008, \mnras, 384, 323,
  \dodoi{10.1111/j.1365-2966.2007.12699.x}

\bibitem[{{Lodato} \& {Pringle}(2006)}]{2006MNRAS.368.1196L}
{Lodato}, G., \& {Pringle}, J.~E. 2006, \mnras, 368, 1196,
  \dodoi{10.1111/j.1365-2966.2006.10194.x}

\bibitem[{{MacLeod} {et~al.}(2012){MacLeod}, {Guillochon}, \&
  {Ramirez-Ruiz}}]{2012ApJ...757..134M}
{MacLeod}, M., {Guillochon}, J., \& {Ramirez-Ruiz}, E. 2012, \apj, 757, 134,
  \dodoi{10.1088/0004-637X/757/2/134}

\bibitem[{{MacLeod} \& {Lin}(2019)}]{2019arXiv190909645M}
{MacLeod}, M., \& {Lin}, D. N.~C. 2019, arXiv e-prints, arXiv:1909.09645.
\newblock \doarXiv{1909.09645}

\bibitem[{{Madigan} {et~al.}(2011){Madigan}, {Hopman}, \&
  {Levin}}]{2011ApJ...738...99M}
{Madigan}, A.-M., {Hopman}, C., \& {Levin}, Y. 2011, \apj, 738, 99,
  \dodoi{10.1088/0004-637X/738/1/99}

\bibitem[{{Mapelli} \& {Gualandris}(2016)}]{2016LNP...905..205M}
{Mapelli}, M., \& {Gualandris}, A. 2016, {Star Formation and Dynamics in the
  Galactic Centre}, ed. F.~{Haardt}, V.~{Gorini}, U.~{Moschella}, A.~{Treves},
  \& M.~{Colpi}, Vol. 905, 205, \dodoi{10.1007/978-3-319-19416-5_6}

\bibitem[{{Matsubayashi} {et~al.}(2007){Matsubayashi}, {Makino}, \&
  {Ebisuzaki}}]{2007ApJ...656..879M}
{Matsubayashi}, T., {Makino}, J., \& {Ebisuzaki}, T. 2007, \apj, 656, 879,
  \dodoi{10.1086/510344}

\bibitem[{{Merritt}(2013)}]{2013degn.book.....M}
{Merritt}, D. 2013, {Dynamics and Evolution of Galactic Nuclei (Princeton:
  Princeton University Press)}

\bibitem[{{Merritt} \& {Szell}(2006)}]{2006ApJ...648..890M}
{Merritt}, D., \& {Szell}, A. 2006, \apj, 648, 890, \dodoi{10.1086/506010}

\bibitem[{{Miller} \& {Bregman}(2016)}]{2016ApJ...829....9M}
{Miller}, M.~J., \& {Bregman}, J.~N. 2016, \apj, 829, 9,
  \dodoi{10.3847/0004-637X/829/1/9}

\bibitem[{{Miller-Jones} {et~al.}(2019){Miller-Jones}, {Tetarenko}, {Sivakoff},
  {Middleton}, {Altamirano}, {Anderson}, {Belloni}, {Fender}, {Jonker},
  {K{\"o}rding}, {Krimm}, {Maitra}, {Markoff}, {Migliari}, {Mooley}, {Rupen},
  {Russell}, {Russell}, {Sarazin}, {Soria}, \& {Tudose}}]{2019Natur.569..374M}
{Miller-Jones}, J. C.~A., {Tetarenko}, A.~J., {Sivakoff}, G.~R., {et~al.} 2019,
  \nat, 569, 374, \dodoi{10.1038/s41586-019-1152-0}

\bibitem[{{Milosavljevi{\'c}} \& {Loeb}(2004)}]{2004ApJ...604L..45M}
{Milosavljevi{\'c}}, M., \& {Loeb}, A. 2004, \apjl, 604, L45,
  \dodoi{10.1086/383467}

\bibitem[{{Monceau-Baroux} {et~al.}(2015){Monceau-Baroux}, {Porth}, {Meliani},
  \& {Keppens}}]{2015A&A...574A.143M}
{Monceau-Baroux}, R., {Porth}, O., {Meliani}, Z., \& {Keppens}, R. 2015, \aap,
  574, A143, \dodoi{10.1051/0004-6361/201425015}

\bibitem[{{Morel} \& {Lebreton}(2008)}]{2008Ap&SS.316...61M}
{Morel}, P., \& {Lebreton}, Y. 2008, \apss, 316, 61,
  \dodoi{10.1007/s10509-007-9663-9}

\bibitem[{{Morris}(1993)}]{1993ApJ...408..496M}
{Morris}, M. 1993, \apj, 408, 496, \dodoi{10.1086/172607}

\bibitem[{{Morris} \& {Serabyn}(1996)}]{1996ARA&A..34..645M}
{Morris}, M., \& {Serabyn}, E. 1996, \araa, 34, 645,
  \dodoi{10.1146/annurev.astro.34.1.645}

\bibitem[{{Moser} {et~al.}(2017){Moser}, {S{\'a}nchez-Monge}, {Eckart},
  {Requena-Torres}, {Garc{\'\i}a-Marin}, {Kunneriath}, {Zensus}, {Britzen},
  {Sabha}, {Shahzamanian}, {Borkar}, \& {Fischer}}]{2017A&A...603A..68M}
{Moser}, L., {S{\'a}nchez-Monge}, {\'A}., {Eckart}, A., {et~al.} 2017, \aap,
  603, A68, \dodoi{10.1051/0004-6361/201628385}

\bibitem[{{Mu{\v{z}}i{\'c}} {et~al.}(2010){Mu{\v{z}}i{\'c}}, {Eckart},
  {Sch{\"o}del}, {Buchholz}, {Zamaninasab}, \& {Witzel}}]{2010A&A...521A..13M}
{Mu{\v{z}}i{\'c}}, K., {Eckart}, A., {Sch{\"o}del}, R., {et~al.} 2010, \aap,
  521, A13, \dodoi{10.1051/0004-6361/200913087}

\bibitem[{{Paczy{\'n}ski}(1970)}]{1970AcA....20...47P}
{Paczy{\'n}ski}, B. 1970, \actaa, 20, 47

\bibitem[{{Parsa} {et~al.}(2017){Parsa}, {Eckart}, {Shahzamanian}, {Karas},
  {Zaja{\v{c}}ek}, {Zensus}, \& {Straubmeier}}]{2017ApJ...845...22P}
{Parsa}, M., {Eckart}, A., {Shahzamanian}, B., {et~al.} 2017, \apj, 845, 22,
  \dodoi{10.3847/1538-4357/aa7bf0}

\bibitem[{{Pei{\ss}ker} {et~al.}(2020){Pei{\ss}ker}, {Hosseini},
  {Zaja{\v{c}}ek}, {Eckart}, {Saalfeld}, {Valencia-S.}, {Parsa}, \&
  {Karas}}]{2020A&A...634A..35P}
{Pei{\ss}ker}, F., {Hosseini}, S.~E., {Zaja{\v{c}}ek}, M., {et~al.} 2020, \aap,
  634, A35, \dodoi{10.1051/0004-6361/201935953}

\bibitem[{{Pei{\ss}ker} {et~al.}(2019){Pei{\ss}ker}, {Zaja{\v{c}}ek}, {Eckart},
  {Sabha}, {Shahzamanian}, \& {Parsa}}]{2019A&A...624A..97P}
{Pei{\ss}ker}, F., {Zaja{\v{c}}ek}, M., {Eckart}, A., {et~al.} 2019, \aap, 624,
  A97, \dodoi{10.1051/0004-6361/201834947}

\bibitem[{{Perez} {et~al.}(2015){Perez}, {Hailey}, {Bauer}, {Krivonos}, {Mori},
  {Baganoff}, {Barri{\`e}re}, {Boggs}, {Christensen}, {Craig}, {Grefenstette},
  {Grindlay}, {Harrison}, {Hong}, {Madsen}, {Nynka}, {Stern}, {Tomsick}, {Wik},
  {Zhang}, {Zhang}, \& {Zoglauer}}]{2015Natur.520..646P}
{Perez}, K., {Hailey}, C.~J., {Bauer}, F.~E., {et~al.} 2015, \nat, 520, 646,
  \dodoi{10.1038/nature14353}

\bibitem[{{Perucho} {et~al.}(2017){Perucho}, {Bosch-Ramon}, \&
  {Barkov}}]{2017A&A...606A..40P}
{Perucho}, M., {Bosch-Ramon}, V., \& {Barkov}, M.~V. 2017, \aap, 606, A40,
  \dodoi{10.1051/0004-6361/201630117}

\bibitem[{{Pfuhl} {et~al.}(2011){Pfuhl}, {Fritz}, {Zilka}, {Maness},
  {Eisenhauer}, {Genzel}, {Gillessen}, {Ott}, {Dodds-Eden}, \&
  {Sternberg}}]{2011ApJ...741..108P}
{Pfuhl}, O., {Fritz}, T.~K., {Zilka}, M., {et~al.} 2011, \apj, 741, 108,
  \dodoi{10.1088/0004-637X/741/2/108}

\bibitem[{{Phinney}(1989)}]{1989IAUS..136..543P}
{Phinney}, E.~S. 1989, in IAU Symposium, Vol. 136, The Center of the Galaxy,
  ed. M.~{Morris}, 543

\bibitem[{{Ponti} {et~al.}(2019){Ponti}, {Hofmann}, {Churazov}, {Morris},
  {Haberl}, {Nandra}, {Terrier}, {Clavel}, \& {Goldwurm}}]{2019Natur.567..347P}
{Ponti}, G., {Hofmann}, F., {Churazov}, E., {et~al.} 2019, \nat, 567, 347,
  \dodoi{10.1038/s41586-019-1009-6}

\bibitem[{{Portegies Zwart} {et~al.}(2006){Portegies Zwart}, {Baumgardt},
  {McMillan}, {Makino}, {Hut}, \& {Ebisuzaki}}]{2006ApJ...641..319P}
{Portegies Zwart}, S.~F., {Baumgardt}, H., {McMillan}, S. L.~W., {et~al.} 2006,
  \apj, 641, 319, \dodoi{10.1086/500361}

\bibitem[{{Rees}(1988)}]{1988Natur.333..523R}
{Rees}, M.~J. 1988, \nat, 333, 523, \dodoi{10.1038/333523a0}

\bibitem[{{Refsdal} \& {Weigert}(1971)}]{1971A&A....13..367R}
{Refsdal}, S., \& {Weigert}, A. 1971, \aap, 13, 367

\bibitem[{{Reimers}(1987)}]{1987IAUS..122..307R}
{Reimers}, D. 1987, in IAU Symposium, Vol. 122, Circumstellar Matter, ed.
  I.~{Appenzeller} \& C.~{Jordan}, 307--318

\bibitem[{{R{\'o}{\.z}a{\'n}ska} {et~al.}(2017){R{\'o}{\.z}a{\'n}ska},
  {Kunneriath}, {Czerny}, {Adhikari}, \& {Karas}}]{2017MNRAS.464.2090R}
{R{\'o}{\.z}a{\'n}ska}, A., {Kunneriath}, D., {Czerny}, B., {Adhikari}, T.~P.,
  \& {Karas}, V. 2017, \mnras, 464, 2090, \dodoi{10.1093/mnras/stw2460}

\bibitem[{{Sch{\"o}del} {et~al.}(2014){Sch{\"o}del}, {Feldmeier}, {Neumayer},
  {Meyer}, \& {Yelda}}]{2014CQGra..31x4007S}
{Sch{\"o}del}, R., {Feldmeier}, A., {Neumayer}, N., {Meyer}, L., \& {Yelda}, S.
  2014, Classical and Quantum Gravity, 31, 244007,
  \dodoi{10.1088/0264-9381/31/24/244007}

\bibitem[{{Sch{\"o}del} {et~al.}(2010){Sch{\"o}del}, {Najarro}, {Muzic}, \&
  {Eckart}}]{2010A&A...511A..18S}
{Sch{\"o}del}, R., {Najarro}, F., {Muzic}, K., \& {Eckart}, A. 2010, \aap, 511,
  A18, \dodoi{10.1051/0004-6361/200913183}

\bibitem[{{Sch{\"o}del} {et~al.}(2020){Sch{\"o}del}, {Nogueras-Lara},
  {Gallego-Cano}, {Shahzamanian}, {Gallego-Calvente}, \&
  {Gardini}}]{2020arXiv200715950S}
{Sch{\"o}del}, R., {Nogueras-Lara}, F., {Gallego-Cano}, E., {et~al.} 2020,
  arXiv e-prints, arXiv:2007.15950.
\newblock \doarXiv{2007.15950}

\bibitem[{{Su} {et~al.}(2010){Su}, {Slatyer}, \&
  {Finkbeiner}}]{2010ApJ...724.1044S}
{Su}, M., {Slatyer}, T.~R., \& {Finkbeiner}, D.~P. 2010, \apj, 724, 1044,
  \dodoi{10.1088/0004-637X/724/2/1044}

\bibitem[{{Sunyaev} \& {Churazov}(1998)}]{1998MNRAS.297.1279S}
{Sunyaev}, R., \& {Churazov}, E. 1998, \mnras, 297, 1279,
  \dodoi{10.1046/j.1365-8711.1998.01684.x}

\bibitem[{{Sunyaev} {et~al.}(1993){Sunyaev}, {Markevitch}, \&
  {Pavlinsky}}]{1993ApJ...407..606S}
{Sunyaev}, R.~A., {Markevitch}, M., \& {Pavlinsky}, M. 1993, \apj, 407, 606,
  \dodoi{10.1086/172542}

\bibitem[{{Syer} \& {Ulmer}(1999)}]{1999MNRAS.306...35S}
{Syer}, D., \& {Ulmer}, A. 1999, \mnras, 306, 35,
  \dodoi{10.1046/j.1365-8711.1999.02445.x}

\bibitem[{{Vilkoviskij} \& {Czerny}(2002)}]{2002A&A...387..804V}
{Vilkoviskij}, E.~Y., \& {Czerny}, B. 2002, \aap, 387, 804,
  \dodoi{10.1051/0004-6361:20020255}

\bibitem[{{{\v{S}}ubr} \& {Haas}(2014)}]{2014ApJ...786..121S}
{{\v{S}}ubr}, L., \& {Haas}, J. 2014, \apj, 786, 121,
  \dodoi{10.1088/0004-637X/786/2/121}

\bibitem[{{Wang} {et~al.}(2013){Wang}, {Nowak}, {Markoff}, {Baganoff},
  {Nayakshin}, {Yuan}, {Cuadra}, {Davis}, {Dexter}, {Fabian}, {Grosso},
  {Haggard}, {Houck}, {Ji}, {Li}, {Neilsen}, {Porquet}, {Ripple}, \&
  {Shcherbakov}}]{2013Sci...341..981W}
{Wang}, Q.~D., {Nowak}, M.~A., {Markoff}, S.~B., {et~al.} 2013, Science, 341,
  981, \dodoi{10.1126/science.1240755}

\bibitem[{{Yusef-Zadeh} {et~al.}(2020){Yusef-Zadeh}, {Royster}, {Wardle},
  {Cotton}, {Kunneriath}, {Heywood}, \& {Michail}}]{2020arXiv200804317Y}
{Yusef-Zadeh}, F., {Royster}, M., {Wardle}, M., {et~al.} 2020, arXiv e-prints,
  arXiv:2008.04317.
\newblock \doarXiv{2008.04317}

\bibitem[{{Yusef-Zadeh} {et~al.}(2012){Yusef-Zadeh}, {Arendt}, {Bushouse},
  {Cotton}, {Haggard}, {Pound}, {Roberts}, {Royster}, \&
  {Wardle}}]{2012ApJ...758L..11Y}
{Yusef-Zadeh}, F., {Arendt}, R., {Bushouse}, H., {et~al.} 2012, \apjl, 758,
  L11, \dodoi{10.1088/2041-8205/758/1/L11}

\bibitem[{{Zhao} {et~al.}(2009){Zhao}, {Morris}, {Goss}, \&
  {An}}]{2009ApJ...699..186Z}
{Zhao}, J.-H., {Morris}, M.~R., {Goss}, W.~M., \& {An}, T. 2009, \apj, 699,
  186, \dodoi{10.1088/0004-637X/699/1/186}

\bibitem[{{Zhu} {et~al.}(2018){Zhu}, {Li}, \& {Morris}}]{2018ApJS..235...26Z}
{Zhu}, Z., {Li}, Z., \& {Morris}, M.~R. 2018, \apjs, 235, 26,
  \dodoi{10.3847/1538-4365/aab14f}

\bibitem[{{Zurek} {et~al.}(1994){Zurek}, {Siemiginowska}, \&
  {Colgate}}]{1994ApJ...434...46Z}
{Zurek}, W.~H., {Siemiginowska}, A., \& {Colgate}, S.~A. 1994, \apj, 434, 46,
  \dodoi{10.1086/174703}

\end{thebibliography}



\end{document}